\documentclass[aps,prl,twocolumn,showpacs,10pt,superscriptaddress,floatfix,longbibliography]{revtex4-2}

\usepackage{mathptmx}
\usepackage{graphicx}
\usepackage{subfigure}
\usepackage{calrsfs}
\DeclareMathAlphabet{\pazocal}{OMS}{zplm}{m}{n}
\usepackage{amsmath}
\usepackage{amsfonts}
\usepackage{amssymb}
\usepackage{mathrsfs}
\usepackage{bm}
\usepackage{subfigure}
\usepackage{xcolor}
\usepackage{braket}
\usepackage[colorlinks=true,linkcolor=blue,anchorcolor=red,citecolor=blue, urlcolor=blue]{hyperref}\usepackage{ulem}
\usepackage[resetlabels,labeled]{multibib}

\newcommand{\be}{\begin{equation}}
\newcommand{\ee}{\end{equation}}
\newcommand{\bea}{\begin{eqnarray}}
\newcommand{\eea}{\end{eqnarray}}
\newcommand{\bse}{\begin{subequations}}
\newcommand{\ese}{\end{subequations}}

\newlength{\seplinewidth}
\newlength{\seplinesep}
\colorlet{sepline}{orange}

\begin{document}
\title{Massive Dirac fermions in moir\'e superlattices: a route towards topological flat minibands}
\author{Ying Su}
\affiliation{Theoretical Division, T-4 and CNLS, Los Alamos National Laboratory, Los Alamos, New Mexico 87545, USA}
\affiliation{Department of Physics, The University of Texas at Dallas, Richardson, Texas 75080, USA}

\author{Heqiu Li}

\affiliation{Department of Physics, University of Michigan, Ann Arbor, Michigan 48109, USA}
\affiliation{Department of Physics, University of Toronto, Toronto, Ontario, Canada}

\author{Chuanwei Zhang}

\affiliation{Department of Physics, The University of Texas at Dallas, Richardson, Texas 75080, USA}

\author{Kai Sun}

\affiliation{Department of Physics, University of Michigan, Ann Arbor, Michigan 48109, USA}

\author{Shi-Zeng Lin}

\affiliation{Theoretical Division, T-4 and CNLS, Los Alamos National Laboratory, Los Alamos, New Mexico 87545, USA}

\begin{abstract}
We demonstrate a generic mechanism to realize topological flat minibands by confining massive Dirac fermions in a periodic moir\'e potential, which can be achieved in a heterobilayer of transition metal dichalcogenides. We show that the topological phase can be protected by the symmetry of moir\'e potential and survive to arbitrarily large Dirac band gap. We take the MoTe$_2$/WSe$_2$ heterobilayer as an example and find that the topological phase can be driven by a vertical electric field. By projecting the Coulomb interaction onto the topological fat minibands, we identify a correlated Chern insulator at half filling and a quantum valley-spin Hall insulator at full filling which explains  the topological states observed in the MoTe$_2$/WSe$_2$ in experiment. Our work clarifies the importance of Dirac structure for the topological minibands and
unveils a general strategy to design topological moir\'e materials.
\end{abstract}
\date{\today}
\maketitle

{\it Introduction.---}Electrons confined by periodic potential in a crystal can behave very differently from a free particle. 
Perhaps the most prominent example is graphene in which the time-reversal, inversion, and three-fold rotation symmetries together stabilize a pair of Dirac cones at Brillouin zone corners \cite{RevModPhys.81.109}. The massless Dirac fermions can acquire a mass when the time-reversal or inversion symmetry is broken, like in transition metal dichalcogenide (TMD) \cite{PhysRevLett.105.136805,Di_coupled_2012}. The nontrivial topological properties associated with the exotic quasiparticles enable novel quantum effects such as the Klein tunneling \cite{katsnelson2006chiral}, valley Hall effect \cite{PhysRevLett.99.236809,mak2014valley}, and valley-selective circular dichroism \cite{PhysRevB.77.235406,Di_coupled_2012,cao2012valley} in these 2D materials which are considered candidates for the next-generation microelectronics.

When overlapping these 2D materials, the moir\'e superlattices (MSL) formed by misalignment open a new possibility to confine the Dirac fermions in a periodic moir\'e potential generated by interlayer hybridization and lattice corrugation. Recently, the topological flat minibands identified in twisted multilayer graphene \cite{Sharpe605,zhang2019nearly,Song2019All,Zhang2019Twisted,lee2019theory,serlin2020intrinsic,stepanov2020competing,Xie2020Nature,Bultinck2020Mechanism,Wu2020Collective,Su2020Current,nuckolls2020strongly,choi2021correlation,wu2021chern,das2021symmetry,park2021flavour,wang2021topological,pierce2021unconventional,he2021chirality}, ABC-stacked-trilayer graphene/hBN heterostructure \cite{Chittari2019Gate,Zhang2019Bridging,chen2020tunable}, and TMD homobilayer \cite{Wu2019Topological} have evoked great interest because the interplay between electronic correlation and nontrivial topology can stabilize exotic quantum states including unconventional superconductivity \cite{cao_unconventional_2018,Yankowitz1059,chen2019signatures,lu2019superconductors,stepanov2020untying,arora2020superconductivity,saito2020independent,cao2021nematicity,park2021tunable,hao2021electric,cao2021pauli,kim2021spectroscopic,Xu2018topological,Guo2018Pairing,Wu2018Theory,Fidrysiak2018Unconventional,Su2018Pairing,Liu2018Chiral,Kennes2018Strong,Isobe2018Unconventional,Roy2019Unconventional,huang2019antiferromagnetically,Ray2019Wannier,Lian2019Twisted,Stauber2019Kohn,you2019superconductivity,Ceae2107874118,Fernandes2021Charge,khalaf2021charged,Qin2021In,lake2021re} and fractional Chern insulator \cite{Ledwith2020Fractional,Repellin2020Chern,Liu2021Gate,Li2021Spontaneous,Xie2021Fractional}.

TMD heterobilayers are another important class of MSL and are being considered platforms to simulate the Hubbard model. 
Their single-particle physics is modeled by holes with parabolic dispersion subject to a moir\'e potential 
{that yields topologically trivial moir\'e minibands} \cite{Wu_hubbard_2018,Zhang_moire_2020}. In this approach, the massive Dirac structure of TMD is neglected by perturbatively dropping the conduction (remote) band, which is far away from the Fermi energy (of the order of 1eV). This theoretical framework can describe the experimentally observed Mott insulator and Wigner crystal in
the WSe$_2$/WS$_2$ heterobilayer \cite{tang2020simulation,regan2020mott,xu2020correlated,Chu2020Nanoscale,huang2021correlated,li2021imaging}. 

Strikingly, recent experiments report the correlated Chern insulator (CCI) at half filling ($\nu=1$ hole per moir{\'e} unit cell) and quantum valley-spin Hall insulator (QVSHI) at full filling ($\nu=2$ holes per moir{\'e} unit cell) in an AB-stacked MoTe$_2$/WSe$_2$ heterobilayer under 
a vertical electric field \cite{Li_quantum_2021}. 
The experimental observations suggest valley-contrasting Chern bands in the 
TMD heterobilayer that cannot be explained by the existing model \cite{Wu_hubbard_2018,Zhang_moire_2020}. 
This  motivates us to investigate a general problem that whether massive Dirac fermions 
confined in a moir\'e potential can give rise to topological minibands.

In this letter, we study the behavior of massive Dirac fermions in a moir\'e potential. Surprisingly, we show that, {\it no matter how large the Dirac band gap is}, topological flat minibands can emerge when the moir\'e potential has certain symmetries. Our study indicates that the Dirac nature of electrons plays a crucial role in determining the topology of moir\'e minibands. In particular, we find that the Berry curvature induced by Dirac remote bands stabilizes a topological phase, which is absent if remote bands are ignored.
By applying our model to the MoTe$_2$/WSe$_2$ heterobilayer, we demonstrate that the Coulomb interaction can stabilize a CCI at $\nu=1$ and a QVSHI at $\nu=2$ in a vertical electric field, which agrees with the recent experiment \cite{Li_quantum_2021}. 
Furthermore, the potential realizations of our model on the surface of an axion insulator and in a monolayer TMD under 
spatially periodic modulation are also proposed. Therefore, our work unveils a general route towards topological flat minibands in moir\'e systems.

{\it Model.---}The continuum model describing a massive Dirac fermion  in a moir\'e potential reads
\begin{equation}\label{Ht}
    H_\tau = h_{\bm{k},\tau} + V(\bm{r}),\;\;\;\; h_{\bm{k},\tau} = v_F\left(\tau {k}_x\sigma_x+{k}_y\sigma_y\right) + m\sigma_z,
\end{equation}
where $v_F$ is the Fermi velocity, $m$ is the Dirac mass, and $\sigma_{x,y,z}$ are the Pauli matrices acting on pseudospin. $\tau=\pm 1$ determines the chirality of the massive Dirac fermon and is dubbed valley index in TMD \cite{Di_coupled_2012}. The Dirac Hamiltonian $h_{\bm{k},\tau}$ yields a massive Dirac cone $E_{\pm,\bm{k}}=\pm\sqrt{v_F^2\bm{k}^2 + m^2}$ with a direct band gap $\Delta=2m$. Here we consider the moir\'e potential $V(\bm{r})=2V_0\sum_{j=1}^3 \cos(\bm{G}_j\cdot\bm{r}+\phi)$ in TMD heterobilayers \cite{Wu_hubbard_2018,Zhang_moire_2020}, where $\bm{G}_j=\frac{4\pi}{\sqrt{3}a_M}\left(\cos\frac{2\pi j}{3},\sin\frac{2\pi j}{3}\right)$ and $a_M$ is the MSL constant. $H_\tau$ is invariant under the threefold rotation since $\mathcal{C}_3h_{\bm{k},\tau}\mathcal{C}_3^{-1}=h_{R_3\bm{k},\tau}$ and $V(R_3\bm{r})=V(\bm{r})$ where $\mathcal{C}_3=\text{diag}(e^{-\frac{2\pi \tau i}{3}},1)$ \cite{Liu_electronic_2015} and $R_3$ are the threefold rotation operator and matrix.

As far as the energy spectrum is concerned, the massive Dirac fermion described by $h_{\bm{k},\tau}$ can be approximated by a free fermion with effective mass $m^*=\Delta/2v_F^2$ through the second order perturbation theory when the Dirac band gap $\Delta\gg v_F|\bm{k}|$ and $V_0$. Then Eq.~(\ref{Ht}) is reduced to 
\begin{equation}\label{H0}
    H_0 = -\frac{\bm{k}^2}{2m^*} + V(\bm{r}),
\end{equation}
that is widely adopted to describe the moir\'e minibands in TMD heterobilayers \cite{Wu_hubbard_2018,Zhang_moire_2020}. However, as will be shown explicitly below, the topology of minibands can be very different for Eqs. \eqref{Ht} and \eqref{H0} 
because $H_0$ has time-reversal symmetry (TRS) while $H_\tau$ does not. 
The TRS in $H_\tau$ is broken by the massive Dirac femrion, i.e., $\mathcal{T}h_{\bm{k},\tau}\mathcal{T}^{-1}=h_{-\bm{k},-\tau}$ where the TRS operator $\mathcal{T}=\mathcal{K}$ equals the complex conjugate operator $\mathcal{K}$.   
Therefore the topological moir\'e minibands can emerge from Eq.~(\ref{Ht}) but not Eq.~(\ref{H0}).

\begin{figure}[t]
  \begin{center}
  \includegraphics[width=8.5 cm]{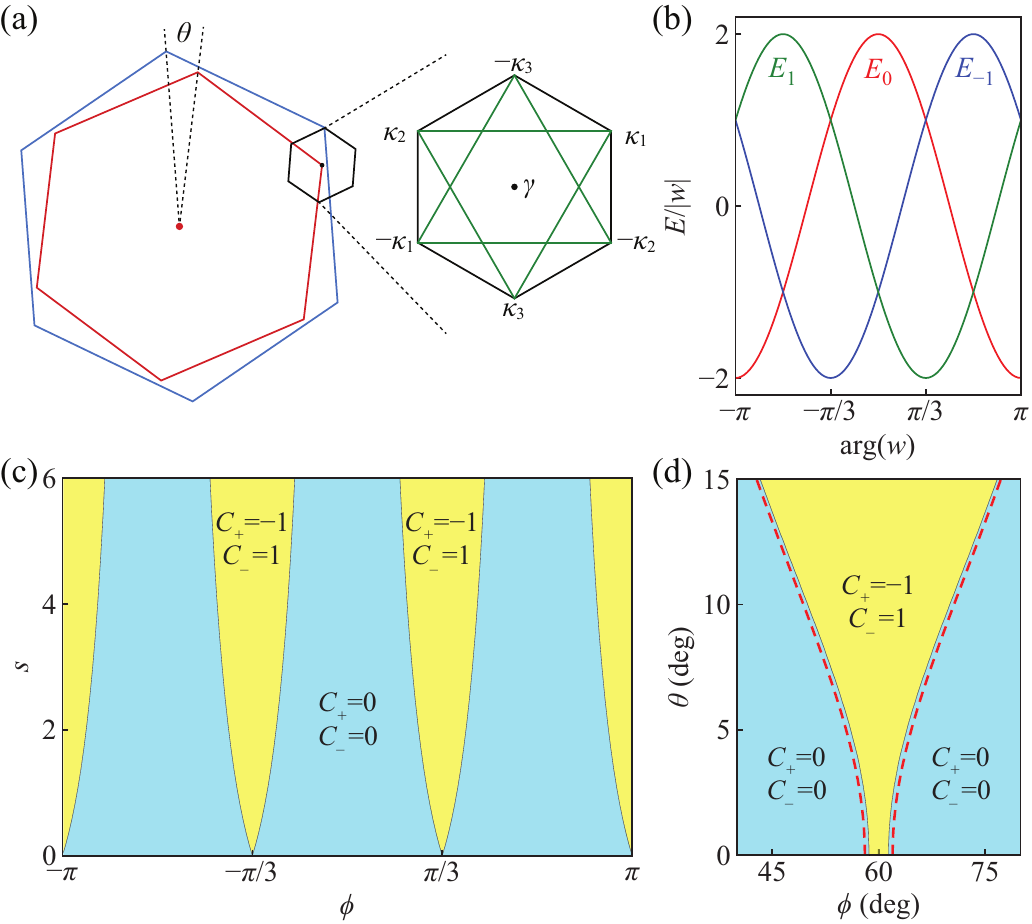}
  \end{center}
\caption{(a) Schematic MBZ in a TMD heterobilayer. The green lines denote the coupling among three degenerate Bloch states at MBZ corners. (b) Eigenvalues of Eq.~(\ref{Vk}) as a function of $\text{arg}(w)$. (c) Topological phase diagram of the continuum model Eq.~(\ref{Ht}) in terms of $\phi$ and $s$. (d) Topological phase diagram of the MoTe$_2$/WSe$_2$ heterobilayer in terms of $\phi$ and $\theta$. In (c) and (d), the yellow regions are the topological phase with valley Chern numbers $C_\pm=\mp1$, while the blue regions are the trivial phase with zero valley Chern numbers. The red dashed lines in (d) are the topological phase boundaries predicted by Eq.~(\ref{sc}).
} 
  \label{fig1}
\end{figure}

{\it Topological phases.---}Due to the $\mathcal{C}_3$ symmetry of $H_\tau$, the Chern number $C_\tau$ of the moir\'e miniband can be determined by its $\mathcal{C}_3$ eigenvalues $\eta_\tau(\bm{k})$ at the $\mathcal{C}_3$-invariant points \cite{Chen_bulk_2012}
\be
e^{\frac{2\pi i }{3}C_\tau}=\eta_\tau(\gamma)\eta_\tau(\kappa)\eta_\tau(-\kappa),
\label{rotchern}
\ee
where $\gamma$ represents the moir\'e Brillouin zone (MBZ) center and $\pm\kappa$ are the MBZ corners. Here we focus on the top valence band. It is easy to show $\eta_\tau(\gamma)=1$, while $\eta_\tau(\pm\kappa)$ can be evaluated to the leading order by the degenerate perturbation theory in which the coupling among three degenerate Bloch states at $\pm\kappa_{1,2,3}$ are considered, as shown in Fig.~\ref{fig1}(a). In the basis of the
Bloch states of the valence band without a moir\'e potential, {i.e.,} 
$\{\ket{u_{\pm\kappa_1,\tau}},\ket{u_{\pm\kappa_2,\tau}},\ket{u_{\pm\kappa_3,\tau}}\}$ with $h_{\bm{k},\tau}\ket{u_{\bm{k},\tau}}=- \sqrt{v_F^2\bm{k}^2+m^2}\ket{u_{\bm{k},\tau}}$, the matrix representation of the moir{\'e} potential operator is
\be\label{Vk}
V_{\pm\kappa,+} = V_{\mp\kappa,-}^* = \left(\begin{array}{ccc}
0 & w(\pm\phi) & w(\pm\phi)^* \\
w(\pm\phi)^* & 0 & w(\pm\phi)  \\
w(\pm\phi) & w(\pm\phi)^* & 0
\end{array}\right),
\ee
whose matrix element
$w(\pm\phi)=\bra{u_{\pm\kappa_1,+}}{V}\ket{u_{\pm\kappa_2,+}}=V_0e^{i(\pm\phi-\frac{\pi}{3})}\left(\frac{1}{2}+\frac{i\sqrt{3}}{2\sqrt{1+s}} \right)$
depends on the dimensionless parameter 
$s={64\pi^2v_F^2}/{9\Delta^2a_M^2}$. Interestingly, $s$ is proportional to the intrinsic Berry curvature $\Omega_I(\bm{k})\approx 2v_F^2/\Delta^2$ (which is valid for $\Delta\gg at|\bm{k}|$ in the MBZ) of the Dirac model in Eq.~(\ref{Ht}) times the MBZ area $A_M=8\pi^2/\sqrt{3}a_M^2$. Namely, $s$ measures the intrinsic Berry curvature 
from the massvie Dirac fermion.

The eigenvalues of Eq.~(\ref{Vk}) are $E_0=2{\rm Re}(w)$ and $E_{\pm1}=-{\rm Re}(w)\pm\sqrt{3}{\rm Im}(w)$, and the corresponding eigenstates have the $\mathcal{C}_3$ eigenvalues $\mathcal{C}_3\ket{E_j}=e^{i\frac{2\pi j}{3}}\ket{E_j}$. In Fig.~\ref{fig1}(b), 
the three eigenvalues are shown as a function of ${\rm arg}(w)$ and the top valence band at $\pm\kappa$ changes among $E_{0}$ and $E_{\pm1}$ through the band crossing at ${\rm arg}(w)=(2n+1)\pi/3$ with $n\in \mathbb{Z}$ where the topological transition can occur. In this way, we can identify $\eta_\tau(\pm\kappa)$ and hence the valley Chern number $C_\tau$ according to Eq.~(\ref{rotchern}). Moreover, the TRS guarantees $\eta_+(\pm\kappa)=\eta_-(\mp\kappa)^*$ and $C_+=-C_-$. A global phase diagram in terms of $\phi$ and $s$ is constructed in Fig.~\ref{fig1}(c). The topological phases with $C_\pm=\mp1$ emerge at $\phi=(2n+1)\pi/3$ and then expand in a wider range of $\phi$ as $s$ increases from zero. This indicates that the intrinsic Berry curvature of massive Dirac fermion measured by $s$ plays a crucial role in determining the topology of moir\'e minibands.  The topological phase boundaries can be obtained analytically by demanding $\text{arg}[w(\pm\phi)]=(2n+1)\pi/3$ that yields 
\be\label{sc}
s = 3\cot^2\left(\phi-\frac{2n\pi}{3}\right) -1,
\ee
with $\phi\in[(2n-1)\pi/3,(2n+1)\pi/3]$.

Significantly, the topological phases at $\phi=(2n+1)\pi/3$ persist to arbitrarily large Dirac band gap since $s\rightarrow 0$ when $\Delta\rightarrow\infty$, as shown in  Fig.~\ref{fig1}(c). To understand the peculiar behavior, it is noticed that the moir\'e potential minima for holes 
form a honeycomb lattice with inversion symmetry $\mathcal{P}$ only at these $\phi$s. Then free fermions coupled to the moir\'e potential, as described by Eq.~(\ref{H0}), give rise to a pair of Dirac cones at MBZ corners that is stabilized by the $\mathcal{PT}$ and $\mathcal{C}_3$ symmetries, same as that in graphene. When the free fermion is replaced by massive Dirac fermion in Eq.~(\ref{Ht}), the $\mathcal{PT}$ symmetry is broken as $\mathcal{PT}h_{\bm{k},\tau}(\mathcal{PT})^{-1}=h_{\bm{k},-\tau}$ that gaps out the Dirac cones and leads to topological minibands. This mechanism to generate topological minibands is protected by the symmetry of moir\'e potential and is independent on the detailed model parameters. The derivation of $\phi$ from $(2n+1)\pi/3$ breaks the $\mathcal{P}$ symmetry and induces a staggered potential on the honeycomb lattice that can drive the topological transition as in the Haldane model \cite{Haldane1988Model}. When $\phi=2n\pi/3$, the moir\'e potential also has $\mathcal{P}$ symmetry but its minima for holes form a triangular lattice that leads to a trivial top valence band \cite{triangular}.

\begin{figure}
\includegraphics[width=8.5 cm]{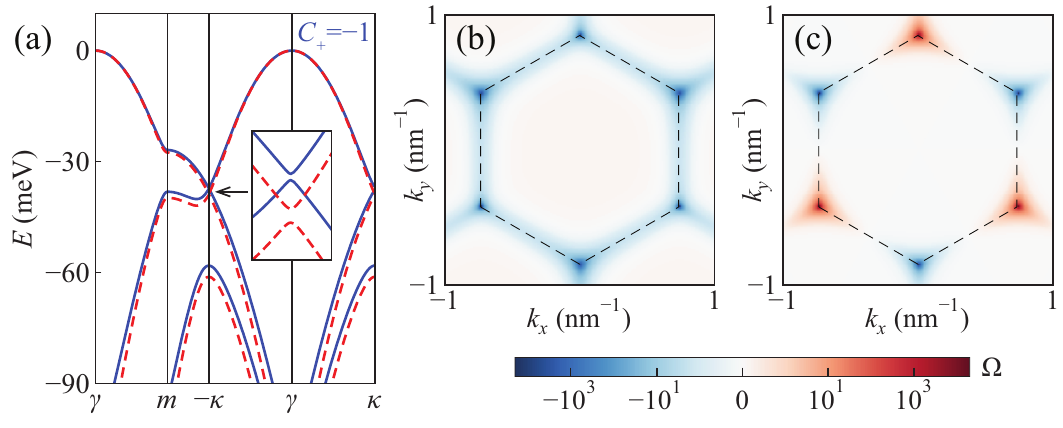}
\caption{(a) Valence bands of the MoTe$_2$/WSe$_2$ heterobilayer with $\theta=1^\circ$, $\phi=59^\circ$, and $V_0=8$ meV. The blue solid and red dashed bands are from the continuum models in Eqs.~(\ref{Ht}) and (\ref{H0}), respectively. (b) and (c) Berry curvatures of the top blue and red bands in (a). The black dashed hexagon encloses the MBZ. 
}
\label{fig2}
\end{figure}

{\it MoTe$_2$/WSe$_2$ heterobilayer.---}To verify the topological phase, we take the MoTe$_2$/WSe$_2$ heterobilayer as an example. MoTe$_2$/WSe$_2$ has the type-I band alignment with a valence band offset about 200$\sim$300 meV. The valence band maximum is from MoTe$_2$ whose Fermi velocity is $v_F=2.526$ eV$\cdot$\r{A} and Dirac band gap is $\Delta=1.017$ eV \cite{Meckbach_ultrafast_2020}. The lattice mismatch is $\delta\sim 7\%$ that results in a MSL with $a_M=a/\sqrt{\delta^2+\theta^2}$ where $\theta$ is twist angle and $a=3.565$ \r{A} is the lattice constant of MoTe$_2$ \cite{Mounet_two_2018}. The direct interlayer tunneling is suppressed by the band offset and by the spin-valley locking in the AB-stacking pattern that requires to flip the electron spin. Therefore, the massive Dirac fermion from MoTe$_2$ coupled to the moir\'e potential provided by WSe$_2$ can be described by Eq.~(\ref{Ht}). By employing the plane wave expansion of the continuum model, we obtain the topological phase diagram in terms of $\phi$ and $\theta$ in Fig.~\ref{fig1}(d). Here the red dashed lines are the topological phase boundaries predicted by Eq.~(\ref{sc}) and are consistent with the direct numerical calculation of Eq. \eqref{Ht}. In  Fig.~\ref{fig1}(d), only the topological phase around $\phi=\pi/3$ is shown and other topological phases can be obtained by shifting $\phi$ by $2n\pi/3$. 

To compare the moir\'e minibands for Eqs. \eqref{Ht} and \eqref{H0}, we choose $\theta=1^\circ$ and $\phi=59^\circ$ in the topological phase and set $V_0=8$ meV. In Fig.~\ref{fig2}(a), the blue and red energy bands are from Eqs.~(\ref{Ht}) and (\ref{H0}), respectively, and show good agreement with each other. Here only the valence bands from the $+$K valley are plotted, and those from the $-$K valley can be obtained by TRS. The Berry curvature of the top blue band is shown in Fig.~\ref{fig2}(b) and yields a valley Chern number $C_+=-1$, while that of the top red band is antisymmetric in  Fig.~\ref{fig2}(c) due to the emergent TRS in Eq.~(\ref{H0}). The Wannier orbitals of the topological and trivial moir\'e minibands from different models are compared in the Supplemental Material \cite{supplement}. On the other hand, deep inside the trivial phase, the Berry curvature derived from Eq.~(\ref{H0}) can be a good approximation to that from Eq.~(\ref{Ht}) \cite{supplement}.

\begin{figure}
\includegraphics[width=8.5 cm]{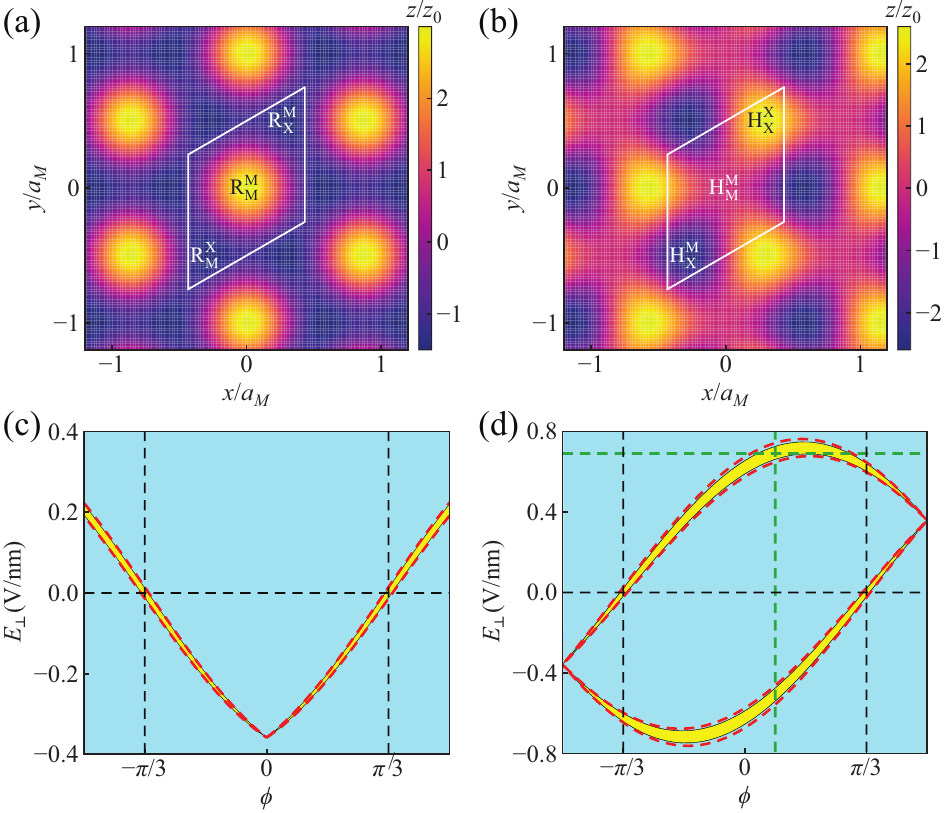}
\caption{(a) and (b) 
{Topography of the corrugated} AA- and AB-stacked TMD heterobilayer. The white parallelogram encloses the moir{\'e} unit cell. (c) and (d) Topological phase diagrams of the  AA- and AB-stacked MoTe$_2$/WSe$_2$ heterobilayer in terms of $\phi$ and $E_\perp$. The yellow regions are the topological phase with valley Chern numbers $C_\pm=\mp1$, while the blue regions are the trivial phase with zero valley Chern numbers. The red dashed lines are the topological phase boundaries predicted by Eq.~(\ref{sc}). The green dashed lines in (d) are for $\phi=\pi/12$ and $E_\perp=0.69$ V/nm. }
\label{fig3}
\end{figure}

{\it Electric-field-driven topological transition.---}{According to the first-principle calculation, the phase of the moir\'e potential in AA- and AB-stacked TMD heterobilayer is unlikely close to} $\phi\sim(2n+1)\pi/3$ \cite{Wu_hubbard_2018,Zhang_moire_2020,Zhang_spin_2021}. Here we show that $\phi$ can be tuned by a vertical electric field. It is noted that the two stacking configurations have different lattice corrugations that have been identified in both the STM measurements \cite{Zhang_interlayer_2017,Shabani_deep_2021} and first-principal calculations \cite{Geng_moire_2020,Zhang_spin_2021}. The electric field couples to the lattice corrugation and modifies the moir\'e potential as
\begin{equation}\label{Hp}
\begin{split}
    H_\tau' & = h_{\bm{k},\tau} + V(\bm{r}) + eE_\perp z(\bm{r}) \\
    & = h_{\bm{k},\tau} + 2V_0' \sum_{j=1}^3 \cos\left(\bm{G}_j\cdot\bm{r}+\frac{\phi+\phi'}{2}+\beta\right),
\end{split}
\end{equation}
where $E_\perp$ is the vertical electric field {and the topography of the corrugated layer is approximated by the lowest harmonics $z(\bm{r}) \approx z_0 \sum_{j=1}^3 \cos(\bm{G}_j\cdot\bm{r}+\phi')$. The role of electric field can be described by a modified moir\'e potential with 
{$V_0'=\sqrt{V_0^2+e^2E_\perp^2z_0^2/4+V_0eE_\perp z_0\cos(\phi-\phi')}$} and $\tan\beta=\frac{2V_0-eE_\perp z_0}{2V_0+eE_\perp z_0}\tan\left(\frac{\phi-\phi'}{2}\right)$.
As $E_\perp$ ramps up, the phase of the moir{\'e} potential in Eq.~(\ref{Hp}) changes continuously from $\phi$ to $\phi'$ when $eE_\perp z_0\gg V_0$, which points to an electric-field-driven topological phase transition.

In AA-stacked TMD heterobilayer, $z(\bm{r})$ is maximal at $\rm{R_M^M}$ 
{and minimal} at $\rm{R_X^M}$ and $\rm{R_M^X}$ 
\cite{Zhang_interlayer_2017,Geng_moire_2020,Shabani_deep_2021}. In AB-stacked TMD heterobilayer, 
$z(\bm{r})$ is maximal (minimal) at $\rm{R_X^X}$ $\left(\rm{R_X^M}\right)$, while $\rm{H_M^M}$ is in  between \cite{Shabani_deep_2021,Zhang_spin_2021}.
{Here M and X refer to the metal and chalcogen, while R and H represent the AA- and AB-stacking. The super- and subscript denote atoms from the top and bottom layer that are aligned locally \cite{supplement}.}
The variation of $z(\bm{r})$ 
in experiments translates into $\phi'\sim 0$ and $-\pi/2$ for the AA- and AB-stacked heterobilayer, as shown in Figs.~\ref{fig3}(a) and \ref{fig3}(b). $\phi$ of the moir{\'e} potential is usually determined by fitting the continuum model to the first-principal energy bands. It has been reported that $\phi\sim\pi/12$ for AB-stacked MoTe$_2$/WSe$_2$ \cite{Zhang_spin_2021} while $\phi$ for AA-stacked MoTe$_2$/WSe$_2$ is still unclear. Nevertheless, most AA-stacked TMD heterobilayers have a $\phi$ of $\pi/6\sim\pi/4$ \cite{Wu_topological_2017,Zhang_moire_2020} and it is natural to expect AA-stacked MoTe$_2$/WSe$_2$ has $\phi$ in the same range. 
The critical $E_\perp$ for the topological transition can be obtained from Eq.~(\ref{sc}) by replacing $\phi$ with the phase of the modified moir{\'e} potential in Eq.~(\ref{Hp}). For $\theta=0^\circ$, $V_0=4.3$ meV, and $z_0=0.024$ nm, the topological phase diagrams in terms of $\phi$ and $E_\perp$ are displayed in Figs.~\ref{fig3}(c) and \ref{fig3} (d) for AA- and AB-stacked MoTe$_2$/WSe$_2$, respectively. The former shows a topological phase for $\phi$ around $\pi/6\sim\pi/4$ in negative $E_\perp$, while the later exhibits two topological phases for $\phi\sim\pi/12$ in both positive and negative $E_\perp$. Note that the positive (negative) $E_\perp$ reduces (enlarges) the valence band offset 
and is applied in the experiment. Therefore, we can focus on $E_\perp>0$ and there is no topological phase for AA-stacked MoTe$_2$/WSe$_2$ as observed in the experiment \cite{Li_continuous_2021}. For AB-stacked MoTe$_2$/WSe$_2$, a topological phase appears for $E_\perp$ within $0.66\sim0.73$ V/nm that agrees well with the experimental result of $0.68\sim0.70$ V/nm \cite{Li_quantum_2021}. 

\begin{figure}[t]
  \begin{center}
  \includegraphics[width=8.5 cm]{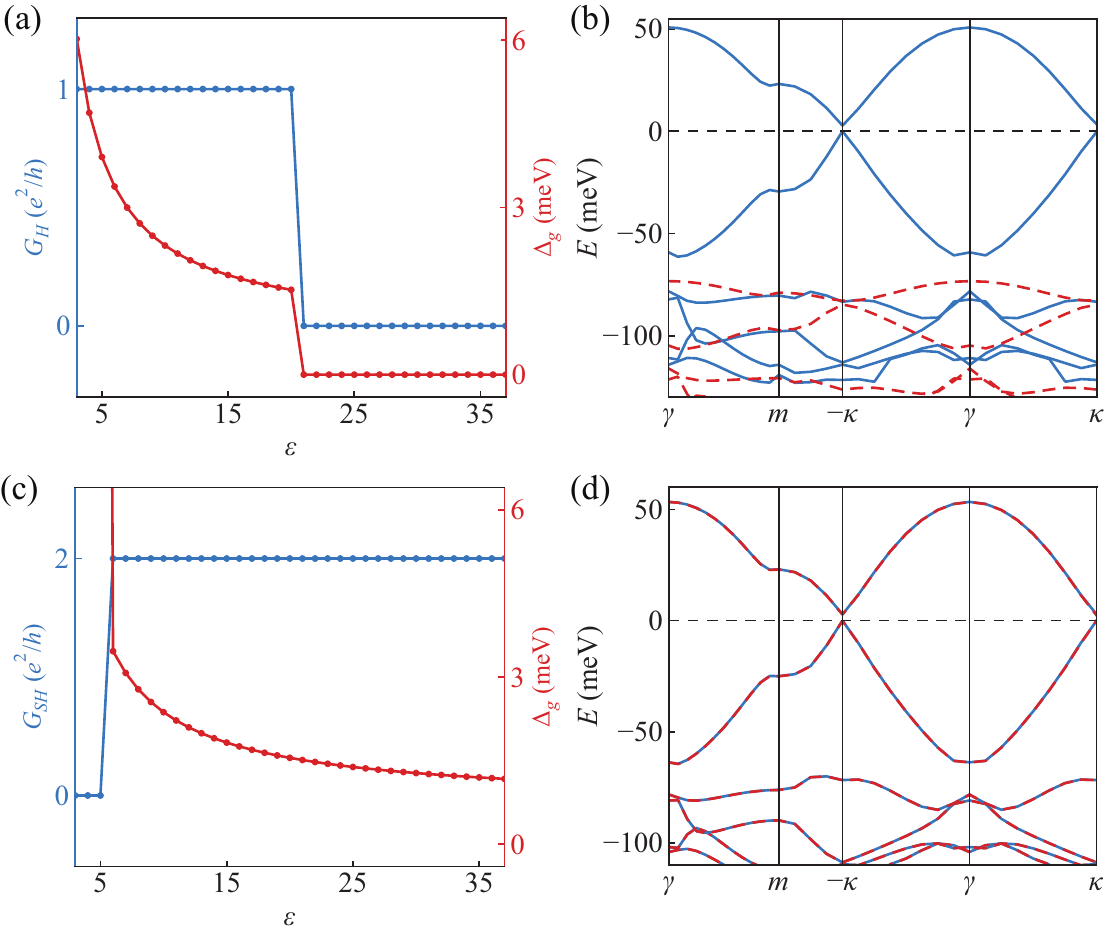}
  \end{center}
\caption{(a) Hall conductance and band gap of the AB-stacked MoTe$_2$/WSe$_2$ heterobilayer at $\nu=1$ and under a vertical electric field $E_\perp=0.69$ V/nm as a function of $\epsilon$. (b) The corresponding energy bands given by the Hartree-Fock approximation for $\epsilon=8$. The blue solid and red dashed bands are originated from the $\pm$K valleys and the chemical potential is set at zero energy. {Besides replacing the Hall conductance by the spin Hall conductance}, (c) and (d) display the same as those in (a) and (b) at $\nu=2$.  
} 
  \label{fig4}
\end{figure}

{\it CCI and QVSHI.---}To stabilize a Chern insulator, it is required to break the TRS, which can be achieved by the Coulomb interaction. 
The Coulomb interaction 
projected onto the moir{\'e} minibands reads
\begin{equation}
\begin{split}
    H&=\sum_{n,\bm{k},\tau}\left(E_{n,\bm{k},\tau}-\mu\right) c_{n,\bm{k},\tau}^{\dagger} c_{n,\bm{k},\tau} + \frac{1}{2A} \sum_{\bm q} \rho(\bm{q}) V_{\bm{q}} \rho(-\bm{q}),
\end{split}\label{H}
\end{equation}
where $c_{n,\bm{k},\tau}$ is the annihilation operator of the eigenstate given by
$H_\tau'\ket{\psi_{n,\bm{k},\tau}}=E_{n,\bm{k},\tau}\ket{\psi_{n,\bm{k},\tau}}$, $A$ is the area of the system, and $\mu$ is the chemical potential. $V_{\bm{q}}=e^2\tanh(qd_\perp)/2\epsilon_0\epsilon q$ is the screened Coulomb interaction in a dual-gated  setup whose gate distant is $d_\perp\sim 10$ nm \cite{Li_quantum_2021}. Here $\epsilon$ is the dielectric constant and $\epsilon_0$ is the vacuum permittivity. The density operator $\rho(\bm{q})=\sum_{n,n'}\sum_{\bm{k},\bm{k}'}\sum_{\tau}\Lambda_{n,n'}^{(\tau)}(\bm{k},\bm{k}',\bm{q}) c_{n,\bm{k},\tau}^\dagger c_{n',\bm{k}',\tau}$ where the form factor $\Lambda_{n,n'}^{(\tau)}(\bm{k},\bm{k}',\bm{q})=\bra{\psi_{n,\bm{k},\tau}}e^{i\bm{q}\cdot \bm{r}}\ket{\psi_{n',\bm{k}',\tau}}$ encodes the correlation between states in different bands and at different momenta.

The interacting Hamiltonian Eq.~(\ref{H}) can be solved self-consistently by using the standard Hartree-Fock approximation \cite{supplement}.
To identify the CCI {at $\nu=1$} and QVSHI {at $\nu=2$} in AB-stacked MoTe$_2$/WSe$_2$, we calculate the Hall conductance $G_H$ and spin Hall conductance $G_{SH}$ as a function of $\epsilon$ under the electric field $E_\perp=0.69$ V/nm at which the CCI was observed in the experiment \cite{Li_quantum_2021}. At $\nu=1$, the Hall conductance drops from $e^2/h$ to 0 at 
{$\epsilon\sim21$}, as shown in Fig.~\ref{fig4}(a). When 
{$\epsilon<21$}, the system becomes a valley-polarized CCI whose energy bands are shown in Fig.~\ref{fig4}(b). Here the blue and red bands are from the $\pm$K valleys, respectively, and the top valence band from the $+$K valley with $C_+=-1$ is empty \cite{vp}. The energy gap $\Delta_g$ decreases with $\epsilon$ and vanishes with $G_H$ at $\epsilon\sim 21$ above which the valley polarization disappears and the system becomes a normal metal. The energy gap {for $\epsilon=8$} in Fig.~\ref{fig4}(b) is 
{$\Delta_g=2.71$} meV that agrees well with the experimental data $\sim$2.5 meV extracted from the capacitance measurement \cite{Li_quantum_2021}. At $\nu=2$, the spin Hall conductance jumps from $0$ to $2e^2/h$ at $\epsilon\sim6$ above which the system becomes a QVSHI, as shown in Figs.~\ref{fig4}(c) and \ref{fig4}(d). In this case, the top valence bands from $\pm$K valleys with opposite Chern numbers $C_\pm=\mp1$ are empty. The energy gap decreases with $\epsilon$. The valley polarization only appears at strong interaction 
{for $\epsilon<6$}, and the top two valence bands from either $+$K or $-$K valley are empty. Because the two bands from the same valley carry opposite Chern numbers, the system becomes a valley-polarized trivial insulator. 

{\it Discussion and summary.---}It is noted that the minimum requirement to realize topological minibands in our study is some nonzero Berry curvature, which appears naturally due to the Dirac structure of TMD. This is different from the other proposals that require the inclusion of interlayer tunneling \cite{tong2017topological,Zhang_spin_2021} or pesudomagnetic field \cite{Xie_theory_2021} in the TMD heterobilayer. In particular, we show that the topological phase, when protected by the symmetry of moir\'e potential, survives to arbitrarily large Dirac band gap that cannot be captured by the existing model Eq.~\eqref{H0}. This mechanism is also verified for a different moir\'e potential with $\mathcal{C}_4$ symmetry that forms a square MSL \cite{supplement}.  
Besides TMD heterobilayers, the massive Dirac fermions on the surface of an axion insulator and in the bulk of a monolayer TMD can also couple to a modulating potential and give rise to topological flat mininbands, as explicitly elucidated in the Supplemental Material \cite{supplement}. In these systems, the modulating potential can be generated by the spatially periodic modulation of magnetic proximity coupling and dielectric screening \cite{xu2021creation}.



A single Dirac cone is generally not allowed to appear in 2D systems with TRS according to the Nielsen–Ninomiya theorem \cite{nielsen1983adler}. In MSL, the emergent valley charge conservation  
allows to assign a valley flavor to fermions. With the Coulomb interaction, one valley flavor is populated while {the other valley remains} empty, which results in the spontaneous breaking of TRS. Therefore, the TMD heterobilayer becomes an exciting platform to study the physics of a single massive Dirac cone, where topological minibands can be realized. A similar situation for massless Dirac cone  can 
be realized on the surface of 3D topological insulators \cite{PhysRevB.103.155157,PhysRevX.11.021024}.

In summary, we spotlight the topological flat minibands that can emerge from the massive Dirac ferminons confined in a moir\'e potential. The topological phase is enabled by the Dirac structure and can be protected by the symmetry of moir\'e potential, which provides a paradigm to study the interplay between electric correlation and nontrivial topology. We take the MoTe$_2$/WSe$_2$ heterobilayer as an example and show that the CCI and QVSHI can be stabilized by the Coulomb interaction. Our work provides a mechanism to the topological states observed in the TMD heterobilayer and points a direction to design topological moir\'e materials.

\begin{acknowledgements}
{\it Acknowledgments.---}The work done at LANL was carried out under the auspices of the U.S. DOE NNSA under contract No. 89233218CNA000001 through the LDRD Program. S. Z. L. was also supported by the U.S. Department of Energy, Office of Science, Basic Energy Sciences, Materials Sciences and Engineering Division, Condensed Matter Theory Program. The work at the University of Texas at Dallas is supported by the Air Force Office of Scientific Research (FA9550-20-1-0220), National Science Foundation (PHY-2110212), and Army Research Office (W911NF-17-1-0128).  H.L. and K.S. acknowledge  support through NSF Grant No. NSF-EFMA-1741618.
\end{acknowledgements}

\bibliography{references}

\begin{thebibliography}{96}%
\makeatletter
\providecommand \@ifxundefined [1]{%
 \@ifx{#1\undefined}
}%
\providecommand \@ifnum [1]{%
 \ifnum #1\expandafter \@firstoftwo
 \else \expandafter \@secondoftwo
 \fi
}%
\providecommand \@ifx [1]{%
 \ifx #1\expandafter \@firstoftwo
 \else \expandafter \@secondoftwo
 \fi
}%
\providecommand \natexlab [1]{#1}%
\providecommand \enquote  [1]{``#1''}%
\providecommand \bibnamefont  [1]{#1}%
\providecommand \bibfnamefont [1]{#1}%
\providecommand \citenamefont [1]{#1}%
\providecommand \href@noop [0]{\@secondoftwo}%
\providecommand \href [0]{\begingroup \@sanitize@url \@href}%
\providecommand \@href[1]{\@@startlink{#1}\@@href}%
\providecommand \@@href[1]{\endgroup#1\@@endlink}%
\providecommand \@sanitize@url [0]{\catcode `\\12\catcode `\$12\catcode
  `\&12\catcode `\#12\catcode `\^12\catcode `\_12\catcode `\%12\relax}%
\providecommand \@@startlink[1]{}%
\providecommand \@@endlink[0]{}%
\providecommand \url  [0]{\begingroup\@sanitize@url \@url }%
\providecommand \@url [1]{\endgroup\@href {#1}{\urlprefix }}%
\providecommand \urlprefix  [0]{URL }%
\providecommand \Eprint [0]{\href }%
\providecommand \doibase [0]{https://doi.org/}%
\providecommand \selectlanguage [0]{\@gobble}%
\providecommand \bibinfo  [0]{\@secondoftwo}%
\providecommand \bibfield  [0]{\@secondoftwo}%
\providecommand \translation [1]{[#1]}%
\providecommand \BibitemOpen [0]{}%
\providecommand \bibitemStop [0]{}%
\providecommand \bibitemNoStop [0]{.\EOS\space}%
\providecommand \EOS [0]{\spacefactor3000\relax}%
\providecommand \BibitemShut  [1]{\csname bibitem#1\endcsname}%
\let\auto@bib@innerbib\@empty
\bibitem [{\citenamefont {Castro~Neto}\ \emph {et~al.}(2009)\citenamefont
  {Castro~Neto}, \citenamefont {Guinea}, \citenamefont {Peres}, \citenamefont
  {Novoselov},\ and\ \citenamefont {Geim}}]{RevModPhys.81.109}%
  \BibitemOpen
  \bibfield  {author} {\bibinfo {author} {\bibfnamefont {A.~H.}\ \bibnamefont
  {Castro~Neto}}, \bibinfo {author} {\bibfnamefont {F.}~\bibnamefont {Guinea}},
  \bibinfo {author} {\bibfnamefont {N.~M.~R.}\ \bibnamefont {Peres}}, \bibinfo
  {author} {\bibfnamefont {K.~S.}\ \bibnamefont {Novoselov}},\ and\ \bibinfo
  {author} {\bibfnamefont {A.~K.}\ \bibnamefont {Geim}},\ }\bibfield  {title}
  {\bibinfo {title} {The electronic properties of graphene},\ }\href
  {https://doi.org/10.1103/RevModPhys.81.109} {\bibfield  {journal} {\bibinfo
  {journal} {Rev. Mod. Phys.}\ }\textbf {\bibinfo {volume} {81}},\ \bibinfo
  {pages} {109} (\bibinfo {year} {2009})}\BibitemShut {NoStop}%
\bibitem [{\citenamefont {Mak}\ \emph {et~al.}(2010)\citenamefont {Mak},
  \citenamefont {Lee}, \citenamefont {Hone}, \citenamefont {Shan},\ and\
  \citenamefont {Heinz}}]{PhysRevLett.105.136805}%
  \BibitemOpen
  \bibfield  {author} {\bibinfo {author} {\bibfnamefont {K.~F.}\ \bibnamefont
  {Mak}}, \bibinfo {author} {\bibfnamefont {C.}~\bibnamefont {Lee}}, \bibinfo
  {author} {\bibfnamefont {J.}~\bibnamefont {Hone}}, \bibinfo {author}
  {\bibfnamefont {J.}~\bibnamefont {Shan}},\ and\ \bibinfo {author}
  {\bibfnamefont {T.~F.}\ \bibnamefont {Heinz}},\ }\bibfield  {title} {\bibinfo
  {title} {Atomically thin ${\mathrm{mos}}_{2}$: A new direct-gap
  semiconductor},\ }\href {https://doi.org/10.1103/PhysRevLett.105.136805}
  {\bibfield  {journal} {\bibinfo  {journal} {Phys. Rev. Lett.}\ }\textbf
  {\bibinfo {volume} {105}},\ \bibinfo {pages} {136805} (\bibinfo {year}
  {2010})}\BibitemShut {NoStop}%
\bibitem [{\citenamefont {Xiao}\ \emph {et~al.}(2012)\citenamefont {Xiao},
  \citenamefont {Liu}, \citenamefont {Feng}, \citenamefont {Xu},\ and\
  \citenamefont {Yao}}]{Di_coupled_2012}%
  \BibitemOpen
  \bibfield  {author} {\bibinfo {author} {\bibfnamefont {D.}~\bibnamefont
  {Xiao}}, \bibinfo {author} {\bibfnamefont {G.-B.}\ \bibnamefont {Liu}},
  \bibinfo {author} {\bibfnamefont {W.}~\bibnamefont {Feng}}, \bibinfo {author}
  {\bibfnamefont {X.}~\bibnamefont {Xu}},\ and\ \bibinfo {author}
  {\bibfnamefont {W.}~\bibnamefont {Yao}},\ }\bibfield  {title} {\bibinfo
  {title} {Coupled spin and valley physics in monolayers of
  ${\mathrm{mos}}_{2}$ and other group-vi dichalcogenides},\ }\href
  {https://doi.org/10.1103/PhysRevLett.108.196802} {\bibfield  {journal}
  {\bibinfo  {journal} {Phys. Rev. Lett.}\ }\textbf {\bibinfo {volume} {108}},\
  \bibinfo {pages} {196802} (\bibinfo {year} {2012})}\BibitemShut {NoStop}%
\bibitem [{\citenamefont {Katsnelson}\ \emph {et~al.}(2006)\citenamefont
  {Katsnelson}, \citenamefont {Novoselov},\ and\ \citenamefont
  {Geim}}]{katsnelson2006chiral}%
  \BibitemOpen
  \bibfield  {author} {\bibinfo {author} {\bibfnamefont {M.}~\bibnamefont
  {Katsnelson}}, \bibinfo {author} {\bibfnamefont {K.}~\bibnamefont
  {Novoselov}},\ and\ \bibinfo {author} {\bibfnamefont {A.}~\bibnamefont
  {Geim}},\ }\bibfield  {title} {\bibinfo {title} {Chiral tunnelling and the
  klein paradox in graphene},\ }\href@noop {} {\bibfield  {journal} {\bibinfo
  {journal} {Nature physics}\ }\textbf {\bibinfo {volume} {2}},\ \bibinfo
  {pages} {620} (\bibinfo {year} {2006})}\BibitemShut {NoStop}%
\bibitem [{\citenamefont {Xiao}\ \emph {et~al.}(2007)\citenamefont {Xiao},
  \citenamefont {Yao},\ and\ \citenamefont {Niu}}]{PhysRevLett.99.236809}%
  \BibitemOpen
  \bibfield  {author} {\bibinfo {author} {\bibfnamefont {D.}~\bibnamefont
  {Xiao}}, \bibinfo {author} {\bibfnamefont {W.}~\bibnamefont {Yao}},\ and\
  \bibinfo {author} {\bibfnamefont {Q.}~\bibnamefont {Niu}},\ }\bibfield
  {title} {\bibinfo {title} {Valley-contrasting physics in graphene: Magnetic
  moment and topological transport},\ }\href
  {https://doi.org/10.1103/PhysRevLett.99.236809} {\bibfield  {journal}
  {\bibinfo  {journal} {Phys. Rev. Lett.}\ }\textbf {\bibinfo {volume} {99}},\
  \bibinfo {pages} {236809} (\bibinfo {year} {2007})}\BibitemShut {NoStop}%
\bibitem [{\citenamefont {Mak}\ \emph {et~al.}(2014)\citenamefont {Mak},
  \citenamefont {McGill}, \citenamefont {Park},\ and\ \citenamefont
  {McEuen}}]{mak2014valley}%
  \BibitemOpen
  \bibfield  {author} {\bibinfo {author} {\bibfnamefont {K.~F.}\ \bibnamefont
  {Mak}}, \bibinfo {author} {\bibfnamefont {K.~L.}\ \bibnamefont {McGill}},
  \bibinfo {author} {\bibfnamefont {J.}~\bibnamefont {Park}},\ and\ \bibinfo
  {author} {\bibfnamefont {P.~L.}\ \bibnamefont {McEuen}},\ }\bibfield  {title}
  {\bibinfo {title} {The valley hall effect in mos2 transistors},\ }\href
  {https://www.science.org/doi/full/10.1126/science.1250140} {\bibfield
  {journal} {\bibinfo  {journal} {Science}\ }\textbf {\bibinfo {volume}
  {344}},\ \bibinfo {pages} {1489} (\bibinfo {year} {2014})}\BibitemShut
  {NoStop}%
\bibitem [{\citenamefont {Yao}\ \emph {et~al.}(2008)\citenamefont {Yao},
  \citenamefont {Xiao},\ and\ \citenamefont {Niu}}]{PhysRevB.77.235406}%
  \BibitemOpen
  \bibfield  {author} {\bibinfo {author} {\bibfnamefont {W.}~\bibnamefont
  {Yao}}, \bibinfo {author} {\bibfnamefont {D.}~\bibnamefont {Xiao}},\ and\
  \bibinfo {author} {\bibfnamefont {Q.}~\bibnamefont {Niu}},\ }\bibfield
  {title} {\bibinfo {title} {Valley-dependent optoelectronics from inversion
  symmetry breaking},\ }\href {https://doi.org/10.1103/PhysRevB.77.235406}
  {\bibfield  {journal} {\bibinfo  {journal} {Phys. Rev. B}\ }\textbf {\bibinfo
  {volume} {77}},\ \bibinfo {pages} {235406} (\bibinfo {year}
  {2008})}\BibitemShut {NoStop}%
\bibitem [{\citenamefont {Cao}\ \emph {et~al.}(2012)\citenamefont {Cao},
  \citenamefont {Wang}, \citenamefont {Han}, \citenamefont {Ye}, \citenamefont
  {Zhu}, \citenamefont {Shi}, \citenamefont {Niu}, \citenamefont {Tan},
  \citenamefont {Wang}, \citenamefont {Liu} \emph {et~al.}}]{cao2012valley}%
  \BibitemOpen
  \bibfield  {author} {\bibinfo {author} {\bibfnamefont {T.}~\bibnamefont
  {Cao}}, \bibinfo {author} {\bibfnamefont {G.}~\bibnamefont {Wang}}, \bibinfo
  {author} {\bibfnamefont {W.}~\bibnamefont {Han}}, \bibinfo {author}
  {\bibfnamefont {H.}~\bibnamefont {Ye}}, \bibinfo {author} {\bibfnamefont
  {C.}~\bibnamefont {Zhu}}, \bibinfo {author} {\bibfnamefont {J.}~\bibnamefont
  {Shi}}, \bibinfo {author} {\bibfnamefont {Q.}~\bibnamefont {Niu}}, \bibinfo
  {author} {\bibfnamefont {P.}~\bibnamefont {Tan}}, \bibinfo {author}
  {\bibfnamefont {E.}~\bibnamefont {Wang}}, \bibinfo {author} {\bibfnamefont
  {B.}~\bibnamefont {Liu}}, \emph {et~al.},\ }\bibfield  {title} {\bibinfo
  {title} {Valley-selective circular dichroism of monolayer molybdenum
  disulphide},\ }\href {https://www.nature.com/articles/ncomms1882} {\bibfield
  {journal} {\bibinfo  {journal} {Nature communications}\ }\textbf {\bibinfo
  {volume} {3}},\ \bibinfo {pages} {1} (\bibinfo {year} {2012})}\BibitemShut
  {NoStop}%
\bibitem [{\citenamefont {Sharpe}\ \emph {et~al.}(2019)\citenamefont {Sharpe},
  \citenamefont {Fox}, \citenamefont {Barnard}, \citenamefont {Finney},
  \citenamefont {Watanabe}, \citenamefont {Taniguchi}, \citenamefont
  {Kastner},\ and\ \citenamefont {Goldhaber-Gordon}}]{Sharpe605}%
  \BibitemOpen
  \bibfield  {author} {\bibinfo {author} {\bibfnamefont {A.~L.}\ \bibnamefont
  {Sharpe}}, \bibinfo {author} {\bibfnamefont {E.~J.}\ \bibnamefont {Fox}},
  \bibinfo {author} {\bibfnamefont {A.~W.}\ \bibnamefont {Barnard}}, \bibinfo
  {author} {\bibfnamefont {J.}~\bibnamefont {Finney}}, \bibinfo {author}
  {\bibfnamefont {K.}~\bibnamefont {Watanabe}}, \bibinfo {author}
  {\bibfnamefont {T.}~\bibnamefont {Taniguchi}}, \bibinfo {author}
  {\bibfnamefont {M.~A.}\ \bibnamefont {Kastner}},\ and\ \bibinfo {author}
  {\bibfnamefont {D.}~\bibnamefont {Goldhaber-Gordon}},\ }\bibfield  {title}
  {\bibinfo {title} {Emergent ferromagnetism near three-quarters filling in
  twisted bilayer graphene},\ }\href {https://doi.org/10.1126/science.aaw3780}
  {\bibfield  {journal} {\bibinfo  {journal} {Science}\ }\textbf {\bibinfo
  {volume} {365}},\ \bibinfo {pages} {605} (\bibinfo {year}
  {2019})}\BibitemShut {NoStop}%
\bibitem [{\citenamefont {Zhang}\ \emph
  {et~al.}(2019{\natexlab{a}})\citenamefont {Zhang}, \citenamefont {Mao},
  \citenamefont {Cao}, \citenamefont {Jarillo-Herrero},\ and\ \citenamefont
  {Senthil}}]{zhang2019nearly}%
  \BibitemOpen
  \bibfield  {author} {\bibinfo {author} {\bibfnamefont {Y.-H.}\ \bibnamefont
  {Zhang}}, \bibinfo {author} {\bibfnamefont {D.}~\bibnamefont {Mao}}, \bibinfo
  {author} {\bibfnamefont {Y.}~\bibnamefont {Cao}}, \bibinfo {author}
  {\bibfnamefont {P.}~\bibnamefont {Jarillo-Herrero}},\ and\ \bibinfo {author}
  {\bibfnamefont {T.}~\bibnamefont {Senthil}},\ }\bibfield  {title} {\bibinfo
  {title} {Nearly flat chern bands in moir\'e superlattices},\ }\href
  {https://doi.org/10.1103/PhysRevB.99.075127} {\bibfield  {journal} {\bibinfo
  {journal} {Phys. Rev. B}\ }\textbf {\bibinfo {volume} {99}},\ \bibinfo
  {pages} {075127} (\bibinfo {year} {2019}{\natexlab{a}})}\BibitemShut
  {NoStop}%
\bibitem [{\citenamefont {Song}\ \emph {et~al.}(2019)\citenamefont {Song},
  \citenamefont {Wang}, \citenamefont {Shi}, \citenamefont {Li}, \citenamefont
  {Fang},\ and\ \citenamefont {Bernevig}}]{Song2019All}%
  \BibitemOpen
  \bibfield  {author} {\bibinfo {author} {\bibfnamefont {Z.}~\bibnamefont
  {Song}}, \bibinfo {author} {\bibfnamefont {Z.}~\bibnamefont {Wang}}, \bibinfo
  {author} {\bibfnamefont {W.}~\bibnamefont {Shi}}, \bibinfo {author}
  {\bibfnamefont {G.}~\bibnamefont {Li}}, \bibinfo {author} {\bibfnamefont
  {C.}~\bibnamefont {Fang}},\ and\ \bibinfo {author} {\bibfnamefont {B.~A.}\
  \bibnamefont {Bernevig}},\ }\bibfield  {title} {\bibinfo {title} {All magic
  angles in twisted bilayer graphene are topological},\ }\href
  {https://doi.org/10.1103/PhysRevLett.123.036401} {\bibfield  {journal}
  {\bibinfo  {journal} {Phys. Rev. Lett.}\ }\textbf {\bibinfo {volume} {123}},\
  \bibinfo {pages} {036401} (\bibinfo {year} {2019})}\BibitemShut {NoStop}%
\bibitem [{\citenamefont {Zhang}\ \emph
  {et~al.}(2019{\natexlab{b}})\citenamefont {Zhang}, \citenamefont {Mao},\ and\
  \citenamefont {Senthil}}]{Zhang2019Twisted}%
  \BibitemOpen
  \bibfield  {author} {\bibinfo {author} {\bibfnamefont {Y.-H.}\ \bibnamefont
  {Zhang}}, \bibinfo {author} {\bibfnamefont {D.}~\bibnamefont {Mao}},\ and\
  \bibinfo {author} {\bibfnamefont {T.}~\bibnamefont {Senthil}},\ }\bibfield
  {title} {\bibinfo {title} {Twisted bilayer graphene aligned with hexagonal
  boron nitride: Anomalous hall effect and a lattice model},\ }\href
  {https://doi.org/10.1103/PhysRevResearch.1.033126} {\bibfield  {journal}
  {\bibinfo  {journal} {Phys. Rev. Research}\ }\textbf {\bibinfo {volume}
  {1}},\ \bibinfo {pages} {033126} (\bibinfo {year}
  {2019}{\natexlab{b}})}\BibitemShut {NoStop}%
\bibitem [{\citenamefont {Lee}\ \emph {et~al.}(2019)\citenamefont {Lee},
  \citenamefont {Khalaf}, \citenamefont {Liu}, \citenamefont {Liu},
  \citenamefont {Hao}, \citenamefont {Kim},\ and\ \citenamefont
  {Vishwanath}}]{lee2019theory}%
  \BibitemOpen
  \bibfield  {author} {\bibinfo {author} {\bibfnamefont {J.~Y.}\ \bibnamefont
  {Lee}}, \bibinfo {author} {\bibfnamefont {E.}~\bibnamefont {Khalaf}},
  \bibinfo {author} {\bibfnamefont {S.}~\bibnamefont {Liu}}, \bibinfo {author}
  {\bibfnamefont {X.}~\bibnamefont {Liu}}, \bibinfo {author} {\bibfnamefont
  {Z.}~\bibnamefont {Hao}}, \bibinfo {author} {\bibfnamefont {P.}~\bibnamefont
  {Kim}},\ and\ \bibinfo {author} {\bibfnamefont {A.}~\bibnamefont
  {Vishwanath}},\ }\bibfield  {title} {\bibinfo {title} {Theory of correlated
  insulating behaviour and spin-triplet superconductivity in twisted double
  bilayer graphene},\ }\href
  {https://www.nature.com/articles/s41467-019-12981-1} {\bibfield  {journal}
  {\bibinfo  {journal} {Nature communications}\ }\textbf {\bibinfo {volume}
  {10}},\ \bibinfo {pages} {1} (\bibinfo {year} {2019})}\BibitemShut {NoStop}%
\bibitem [{\citenamefont {Serlin}\ \emph {et~al.}(2020)\citenamefont {Serlin},
  \citenamefont {Tschirhart}, \citenamefont {Polshyn}, \citenamefont {Zhang},
  \citenamefont {Zhu}, \citenamefont {Watanabe}, \citenamefont {Taniguchi},
  \citenamefont {Balents},\ and\ \citenamefont {Young}}]{serlin2020intrinsic}%
  \BibitemOpen
  \bibfield  {author} {\bibinfo {author} {\bibfnamefont {M.}~\bibnamefont
  {Serlin}}, \bibinfo {author} {\bibfnamefont {C.}~\bibnamefont {Tschirhart}},
  \bibinfo {author} {\bibfnamefont {H.}~\bibnamefont {Polshyn}}, \bibinfo
  {author} {\bibfnamefont {Y.}~\bibnamefont {Zhang}}, \bibinfo {author}
  {\bibfnamefont {J.}~\bibnamefont {Zhu}}, \bibinfo {author} {\bibfnamefont
  {K.}~\bibnamefont {Watanabe}}, \bibinfo {author} {\bibfnamefont
  {T.}~\bibnamefont {Taniguchi}}, \bibinfo {author} {\bibfnamefont
  {L.}~\bibnamefont {Balents}},\ and\ \bibinfo {author} {\bibfnamefont
  {A.}~\bibnamefont {Young}},\ }\bibfield  {title} {\bibinfo {title} {Intrinsic
  quantized anomalous hall effect in a moir{\'e} heterostructure},\ }\href
  {https://doi.org/10.1126/science.aay5533} {\bibfield  {journal} {\bibinfo
  {journal} {Science}\ }\textbf {\bibinfo {volume} {367}},\ \bibinfo {pages}
  {900} (\bibinfo {year} {2020})}\BibitemShut {NoStop}%
\bibitem [{\citenamefont {Stepanov}\ \emph
  {et~al.}(2020{\natexlab{a}})\citenamefont {Stepanov}, \citenamefont {Xie},
  \citenamefont {Taniguchi}, \citenamefont {Watanabe}, \citenamefont {Lu},
  \citenamefont {MacDonald}, \citenamefont {Bernevig},\ and\ \citenamefont
  {Efetov}}]{stepanov2020competing}%
  \BibitemOpen
  \bibfield  {author} {\bibinfo {author} {\bibfnamefont {P.}~\bibnamefont
  {Stepanov}}, \bibinfo {author} {\bibfnamefont {M.}~\bibnamefont {Xie}},
  \bibinfo {author} {\bibfnamefont {T.}~\bibnamefont {Taniguchi}}, \bibinfo
  {author} {\bibfnamefont {K.}~\bibnamefont {Watanabe}}, \bibinfo {author}
  {\bibfnamefont {X.}~\bibnamefont {Lu}}, \bibinfo {author} {\bibfnamefont
  {A.~H.}\ \bibnamefont {MacDonald}}, \bibinfo {author} {\bibfnamefont {B.~A.}\
  \bibnamefont {Bernevig}},\ and\ \bibinfo {author} {\bibfnamefont {D.~K.}\
  \bibnamefont {Efetov}},\ }\href@noop {} {\bibinfo {title} {Competing
  zero-field chern insulators in superconducting twisted bilayer graphene}}
  (\bibinfo {year} {2020}{\natexlab{a}}),\ \Eprint
  {https://arxiv.org/abs/2012.15126} {arXiv:2012.15126 [cond-mat.mes-hall]}
  \BibitemShut {NoStop}%
\bibitem [{\citenamefont {Xie}\ and\ \citenamefont
  {MacDonald}(2020)}]{Xie2020Nature}%
  \BibitemOpen
  \bibfield  {author} {\bibinfo {author} {\bibfnamefont {M.}~\bibnamefont
  {Xie}}\ and\ \bibinfo {author} {\bibfnamefont {A.~H.}\ \bibnamefont
  {MacDonald}},\ }\bibfield  {title} {\bibinfo {title} {Nature of the
  correlated insulator states in twisted bilayer graphene},\ }\href
  {https://doi.org/10.1103/PhysRevLett.124.097601} {\bibfield  {journal}
  {\bibinfo  {journal} {Phys. Rev. Lett.}\ }\textbf {\bibinfo {volume} {124}},\
  \bibinfo {pages} {097601} (\bibinfo {year} {2020})}\BibitemShut {NoStop}%
\bibitem [{\citenamefont {Bultinck}\ \emph {et~al.}(2020)\citenamefont
  {Bultinck}, \citenamefont {Chatterjee},\ and\ \citenamefont
  {Zaletel}}]{Bultinck2020Mechanism}%
  \BibitemOpen
  \bibfield  {author} {\bibinfo {author} {\bibfnamefont {N.}~\bibnamefont
  {Bultinck}}, \bibinfo {author} {\bibfnamefont {S.}~\bibnamefont
  {Chatterjee}},\ and\ \bibinfo {author} {\bibfnamefont {M.~P.}\ \bibnamefont
  {Zaletel}},\ }\bibfield  {title} {\bibinfo {title} {Mechanism for anomalous
  hall ferromagnetism in twisted bilayer graphene},\ }\href
  {https://doi.org/10.1103/PhysRevLett.124.166601} {\bibfield  {journal}
  {\bibinfo  {journal} {Phys. Rev. Lett.}\ }\textbf {\bibinfo {volume} {124}},\
  \bibinfo {pages} {166601} (\bibinfo {year} {2020})}\BibitemShut {NoStop}%
\bibitem [{\citenamefont {Wu}\ and\ \citenamefont
  {Das~Sarma}(2020)}]{Wu2020Collective}%
  \BibitemOpen
  \bibfield  {author} {\bibinfo {author} {\bibfnamefont {F.}~\bibnamefont
  {Wu}}\ and\ \bibinfo {author} {\bibfnamefont {S.}~\bibnamefont {Das~Sarma}},\
  }\bibfield  {title} {\bibinfo {title} {Collective excitations of quantum
  anomalous hall ferromagnets in twisted bilayer graphene},\ }\href
  {https://doi.org/10.1103/PhysRevLett.124.046403} {\bibfield  {journal}
  {\bibinfo  {journal} {Phys. Rev. Lett.}\ }\textbf {\bibinfo {volume} {124}},\
  \bibinfo {pages} {046403} (\bibinfo {year} {2020})}\BibitemShut {NoStop}%
\bibitem [{\citenamefont {Su}\ and\ \citenamefont {Lin}(2020)}]{Su2020Current}%
  \BibitemOpen
  \bibfield  {author} {\bibinfo {author} {\bibfnamefont {Y.}~\bibnamefont
  {Su}}\ and\ \bibinfo {author} {\bibfnamefont {S.-Z.}\ \bibnamefont {Lin}},\
  }\bibfield  {title} {\bibinfo {title} {Current-induced reversal of anomalous
  hall conductance in twisted bilayer graphene},\ }\href
  {https://doi.org/10.1103/PhysRevLett.125.226401} {\bibfield  {journal}
  {\bibinfo  {journal} {Phys. Rev. Lett.}\ }\textbf {\bibinfo {volume} {125}},\
  \bibinfo {pages} {226401} (\bibinfo {year} {2020})}\BibitemShut {NoStop}%
\bibitem [{\citenamefont {Nuckolls}\ \emph {et~al.}(2020)\citenamefont
  {Nuckolls}, \citenamefont {Oh}, \citenamefont {Wong}, \citenamefont {Lian},
  \citenamefont {Watanabe}, \citenamefont {Taniguchi}, \citenamefont
  {Bernevig},\ and\ \citenamefont {Yazdani}}]{nuckolls2020strongly}%
  \BibitemOpen
  \bibfield  {author} {\bibinfo {author} {\bibfnamefont {K.~P.}\ \bibnamefont
  {Nuckolls}}, \bibinfo {author} {\bibfnamefont {M.}~\bibnamefont {Oh}},
  \bibinfo {author} {\bibfnamefont {D.}~\bibnamefont {Wong}}, \bibinfo {author}
  {\bibfnamefont {B.}~\bibnamefont {Lian}}, \bibinfo {author} {\bibfnamefont
  {K.}~\bibnamefont {Watanabe}}, \bibinfo {author} {\bibfnamefont
  {T.}~\bibnamefont {Taniguchi}}, \bibinfo {author} {\bibfnamefont {B.~A.}\
  \bibnamefont {Bernevig}},\ and\ \bibinfo {author} {\bibfnamefont
  {A.}~\bibnamefont {Yazdani}},\ }\bibfield  {title} {\bibinfo {title}
  {Strongly correlated chern insulators in magic-angle twisted bilayer
  graphene},\ }\href {https://doi.org/10.1038/s41586-020-3028-8} {\bibfield
  {journal} {\bibinfo  {journal} {Nature}\ }\textbf {\bibinfo {volume} {588}},\
  \bibinfo {pages} {610} (\bibinfo {year} {2020})}\BibitemShut {NoStop}%
\bibitem [{\citenamefont {Choi}\ \emph {et~al.}(2021)\citenamefont {Choi},
  \citenamefont {Kim}, \citenamefont {Peng}, \citenamefont {Thomson},
  \citenamefont {Lewandowski}, \citenamefont {Polski}, \citenamefont {Zhang},
  \citenamefont {Arora}, \citenamefont {Watanabe}, \citenamefont {Taniguchi}
  \emph {et~al.}}]{choi2021correlation}%
  \BibitemOpen
  \bibfield  {author} {\bibinfo {author} {\bibfnamefont {Y.}~\bibnamefont
  {Choi}}, \bibinfo {author} {\bibfnamefont {H.}~\bibnamefont {Kim}}, \bibinfo
  {author} {\bibfnamefont {Y.}~\bibnamefont {Peng}}, \bibinfo {author}
  {\bibfnamefont {A.}~\bibnamefont {Thomson}}, \bibinfo {author} {\bibfnamefont
  {C.}~\bibnamefont {Lewandowski}}, \bibinfo {author} {\bibfnamefont
  {R.}~\bibnamefont {Polski}}, \bibinfo {author} {\bibfnamefont
  {Y.}~\bibnamefont {Zhang}}, \bibinfo {author} {\bibfnamefont {H.~S.}\
  \bibnamefont {Arora}}, \bibinfo {author} {\bibfnamefont {K.}~\bibnamefont
  {Watanabe}}, \bibinfo {author} {\bibfnamefont {T.}~\bibnamefont {Taniguchi}},
  \emph {et~al.},\ }\bibfield  {title} {\bibinfo {title} {Correlation-driven
  topological phases in magic-angle twisted bilayer graphene},\ }\href
  {https://www.nature.com/articles/s41586-020-03159-7} {\bibfield  {journal}
  {\bibinfo  {journal} {Nature}\ }\textbf {\bibinfo {volume} {589}},\ \bibinfo
  {pages} {536} (\bibinfo {year} {2021})}\BibitemShut {NoStop}%
\bibitem [{\citenamefont {Wu}\ \emph {et~al.}(2021)\citenamefont {Wu},
  \citenamefont {Zhang}, \citenamefont {Watanabe}, \citenamefont {Taniguchi},\
  and\ \citenamefont {Andrei}}]{wu2021chern}%
  \BibitemOpen
  \bibfield  {author} {\bibinfo {author} {\bibfnamefont {S.}~\bibnamefont
  {Wu}}, \bibinfo {author} {\bibfnamefont {Z.}~\bibnamefont {Zhang}}, \bibinfo
  {author} {\bibfnamefont {K.}~\bibnamefont {Watanabe}}, \bibinfo {author}
  {\bibfnamefont {T.}~\bibnamefont {Taniguchi}},\ and\ \bibinfo {author}
  {\bibfnamefont {E.~Y.}\ \bibnamefont {Andrei}},\ }\bibfield  {title}
  {\bibinfo {title} {Chern insulators, van hove singularities and topological
  flat bands in magic-angle twisted bilayer graphene},\ }\href
  {https://www.nature.com/articles/s41563-020-00911-2} {\bibfield  {journal}
  {\bibinfo  {journal} {Nature materials}\ }\textbf {\bibinfo {volume} {20}},\
  \bibinfo {pages} {488} (\bibinfo {year} {2021})}\BibitemShut {NoStop}%
\bibitem [{\citenamefont {Das}\ \emph {et~al.}(2021)\citenamefont {Das},
  \citenamefont {Lu}, \citenamefont {Herzog-Arbeitman}, \citenamefont {Song},
  \citenamefont {Watanabe}, \citenamefont {Taniguchi}, \citenamefont
  {Bernevig},\ and\ \citenamefont {Efetov}}]{das2021symmetry}%
  \BibitemOpen
  \bibfield  {author} {\bibinfo {author} {\bibfnamefont {I.}~\bibnamefont
  {Das}}, \bibinfo {author} {\bibfnamefont {X.}~\bibnamefont {Lu}}, \bibinfo
  {author} {\bibfnamefont {J.}~\bibnamefont {Herzog-Arbeitman}}, \bibinfo
  {author} {\bibfnamefont {Z.-D.}\ \bibnamefont {Song}}, \bibinfo {author}
  {\bibfnamefont {K.}~\bibnamefont {Watanabe}}, \bibinfo {author}
  {\bibfnamefont {T.}~\bibnamefont {Taniguchi}}, \bibinfo {author}
  {\bibfnamefont {B.~A.}\ \bibnamefont {Bernevig}},\ and\ \bibinfo {author}
  {\bibfnamefont {D.~K.}\ \bibnamefont {Efetov}},\ }\bibfield  {title}
  {\bibinfo {title} {Symmetry-broken chern insulators and rashba-like
  landau-level crossings in magic-angle bilayer graphene},\ }\href
  {https://doi.org/10.1038/s41567-021-01186-3} {\bibfield  {journal} {\bibinfo
  {journal} {Nature Physics}\ }\textbf {\bibinfo {volume} {17}},\ \bibinfo
  {pages} {710} (\bibinfo {year} {2021})}\BibitemShut {NoStop}%
\bibitem [{\citenamefont {Park}\ \emph
  {et~al.}(2021{\natexlab{a}})\citenamefont {Park}, \citenamefont {Cao},
  \citenamefont {Watanabe}, \citenamefont {Taniguchi},\ and\ \citenamefont
  {Jarillo-Herrero}}]{park2021flavour}%
  \BibitemOpen
  \bibfield  {author} {\bibinfo {author} {\bibfnamefont {J.~M.}\ \bibnamefont
  {Park}}, \bibinfo {author} {\bibfnamefont {Y.}~\bibnamefont {Cao}}, \bibinfo
  {author} {\bibfnamefont {K.}~\bibnamefont {Watanabe}}, \bibinfo {author}
  {\bibfnamefont {T.}~\bibnamefont {Taniguchi}},\ and\ \bibinfo {author}
  {\bibfnamefont {P.}~\bibnamefont {Jarillo-Herrero}},\ }\bibfield  {title}
  {\bibinfo {title} {Flavour hund’s coupling, chern gaps and charge
  diffusivity in moir{\'e} graphene},\ }\href
  {https://www.nature.com/articles/s41586-021-03366-w} {\bibfield  {journal}
  {\bibinfo  {journal} {Nature}\ }\textbf {\bibinfo {volume} {592}},\ \bibinfo
  {pages} {43} (\bibinfo {year} {2021}{\natexlab{a}})}\BibitemShut {NoStop}%
\bibitem [{\citenamefont {Wang}\ \emph
  {et~al.}(2021{\natexlab{a}})\citenamefont {Wang}, \citenamefont
  {Herzog-Arbeitman}, \citenamefont {Burg}, \citenamefont {Zhu}, \citenamefont
  {Watanabe}, \citenamefont {Taniguchi}, \citenamefont {MacDonald},
  \citenamefont {Bernevig},\ and\ \citenamefont {Tutuc}}]{wang2021topological}%
  \BibitemOpen
  \bibfield  {author} {\bibinfo {author} {\bibfnamefont {Y.}~\bibnamefont
  {Wang}}, \bibinfo {author} {\bibfnamefont {J.}~\bibnamefont
  {Herzog-Arbeitman}}, \bibinfo {author} {\bibfnamefont {G.~W.}\ \bibnamefont
  {Burg}}, \bibinfo {author} {\bibfnamefont {J.}~\bibnamefont {Zhu}}, \bibinfo
  {author} {\bibfnamefont {K.}~\bibnamefont {Watanabe}}, \bibinfo {author}
  {\bibfnamefont {T.}~\bibnamefont {Taniguchi}}, \bibinfo {author}
  {\bibfnamefont {A.~H.}\ \bibnamefont {MacDonald}}, \bibinfo {author}
  {\bibfnamefont {B.~A.}\ \bibnamefont {Bernevig}},\ and\ \bibinfo {author}
  {\bibfnamefont {E.}~\bibnamefont {Tutuc}},\ }\href@noop {} {\bibinfo {title}
  {Topological edge transport in twisted double-bilayer graphene}} (\bibinfo
  {year} {2021}{\natexlab{a}}),\ \Eprint {https://arxiv.org/abs/2101.03621}
  {arXiv:2101.03621 [cond-mat.mes-hall]} \BibitemShut {NoStop}%
\bibitem [{\citenamefont {Pierce}\ \emph {et~al.}(2021)\citenamefont {Pierce},
  \citenamefont {Xie}, \citenamefont {Park}, \citenamefont {Khalaf},
  \citenamefont {Lee}, \citenamefont {Cao}, \citenamefont {Parker},
  \citenamefont {Forrester}, \citenamefont {Chen}, \citenamefont {Watanabe}
  \emph {et~al.}}]{pierce2021unconventional}%
  \BibitemOpen
  \bibfield  {author} {\bibinfo {author} {\bibfnamefont {A.~T.}\ \bibnamefont
  {Pierce}}, \bibinfo {author} {\bibfnamefont {Y.}~\bibnamefont {Xie}},
  \bibinfo {author} {\bibfnamefont {J.~M.}\ \bibnamefont {Park}}, \bibinfo
  {author} {\bibfnamefont {E.}~\bibnamefont {Khalaf}}, \bibinfo {author}
  {\bibfnamefont {S.~H.}\ \bibnamefont {Lee}}, \bibinfo {author} {\bibfnamefont
  {Y.}~\bibnamefont {Cao}}, \bibinfo {author} {\bibfnamefont {D.~E.}\
  \bibnamefont {Parker}}, \bibinfo {author} {\bibfnamefont {P.~R.}\
  \bibnamefont {Forrester}}, \bibinfo {author} {\bibfnamefont {S.}~\bibnamefont
  {Chen}}, \bibinfo {author} {\bibfnamefont {K.}~\bibnamefont {Watanabe}},
  \emph {et~al.},\ }\bibfield  {title} {\bibinfo {title} {Unconventional
  sequence of correlated chern insulators in magic-angle twisted bilayer
  graphene},\ }\href {https://www.nature.com/articles/s41567-021-01347-4}
  {\bibfield  {journal} {\bibinfo  {journal} {Nature Physics}\ } (\bibinfo
  {year} {2021})}\BibitemShut {NoStop}%
\bibitem [{\citenamefont {He}\ \emph {et~al.}(2021)\citenamefont {He},
  \citenamefont {Cai}, \citenamefont {Zhang}, \citenamefont {Liu},
  \citenamefont {Li}, \citenamefont {Taniguchi}, \citenamefont {Watanabe},
  \citenamefont {Cobden}, \citenamefont {Yankowitz},\ and\ \citenamefont
  {Xu}}]{he2021chirality}%
  \BibitemOpen
  \bibfield  {author} {\bibinfo {author} {\bibfnamefont {M.}~\bibnamefont
  {He}}, \bibinfo {author} {\bibfnamefont {J.}~\bibnamefont {Cai}}, \bibinfo
  {author} {\bibfnamefont {Y.-H.}\ \bibnamefont {Zhang}}, \bibinfo {author}
  {\bibfnamefont {Y.}~\bibnamefont {Liu}}, \bibinfo {author} {\bibfnamefont
  {Y.}~\bibnamefont {Li}}, \bibinfo {author} {\bibfnamefont {T.}~\bibnamefont
  {Taniguchi}}, \bibinfo {author} {\bibfnamefont {K.}~\bibnamefont {Watanabe}},
  \bibinfo {author} {\bibfnamefont {D.~H.}\ \bibnamefont {Cobden}}, \bibinfo
  {author} {\bibfnamefont {M.}~\bibnamefont {Yankowitz}},\ and\ \bibinfo
  {author} {\bibfnamefont {X.}~\bibnamefont {Xu}},\ }\href@noop {} {\bibinfo
  {title} {Chirality-dependent topological states in twisted double bilayer
  graphene}} (\bibinfo {year} {2021}),\ \Eprint
  {https://arxiv.org/abs/2109.08255} {arXiv:2109.08255 [cond-mat.mes-hall]}
  \BibitemShut {NoStop}%
\bibitem [{\citenamefont {Chittari}\ \emph {et~al.}(2019)\citenamefont
  {Chittari}, \citenamefont {Chen}, \citenamefont {Zhang}, \citenamefont
  {Wang},\ and\ \citenamefont {Jung}}]{Chittari2019Gate}%
  \BibitemOpen
  \bibfield  {author} {\bibinfo {author} {\bibfnamefont {B.~L.}\ \bibnamefont
  {Chittari}}, \bibinfo {author} {\bibfnamefont {G.}~\bibnamefont {Chen}},
  \bibinfo {author} {\bibfnamefont {Y.}~\bibnamefont {Zhang}}, \bibinfo
  {author} {\bibfnamefont {F.}~\bibnamefont {Wang}},\ and\ \bibinfo {author}
  {\bibfnamefont {J.}~\bibnamefont {Jung}},\ }\bibfield  {title} {\bibinfo
  {title} {Gate-tunable topological flat bands in trilayer graphene
  boron-nitride moir\'e superlattices},\ }\href
  {https://doi.org/10.1103/PhysRevLett.122.016401} {\bibfield  {journal}
  {\bibinfo  {journal} {Phys. Rev. Lett.}\ }\textbf {\bibinfo {volume} {122}},\
  \bibinfo {pages} {016401} (\bibinfo {year} {2019})}\BibitemShut {NoStop}%
\bibitem [{\citenamefont {Zhang}\ and\ \citenamefont
  {Senthil}(2019)}]{Zhang2019Bridging}%
  \BibitemOpen
  \bibfield  {author} {\bibinfo {author} {\bibfnamefont {Y.-H.}\ \bibnamefont
  {Zhang}}\ and\ \bibinfo {author} {\bibfnamefont {T.}~\bibnamefont
  {Senthil}},\ }\bibfield  {title} {\bibinfo {title} {Bridging hubbard model
  physics and quantum hall physics in trilayer
  $\text{graphene}/h\ensuremath{-}\mathrm{BN}$ moir\'e superlattice},\ }\href
  {https://doi.org/10.1103/PhysRevB.99.205150} {\bibfield  {journal} {\bibinfo
  {journal} {Phys. Rev. B}\ }\textbf {\bibinfo {volume} {99}},\ \bibinfo
  {pages} {205150} (\bibinfo {year} {2019})}\BibitemShut {NoStop}%
\bibitem [{\citenamefont {Chen}\ \emph {et~al.}(2020)\citenamefont {Chen},
  \citenamefont {Sharpe}, \citenamefont {Fox}, \citenamefont {Zhang},
  \citenamefont {Wang}, \citenamefont {Jiang}, \citenamefont {Lyu},
  \citenamefont {Li}, \citenamefont {Watanabe}, \citenamefont {Taniguchi} \emph
  {et~al.}}]{chen2020tunable}%
  \BibitemOpen
  \bibfield  {author} {\bibinfo {author} {\bibfnamefont {G.}~\bibnamefont
  {Chen}}, \bibinfo {author} {\bibfnamefont {A.~L.}\ \bibnamefont {Sharpe}},
  \bibinfo {author} {\bibfnamefont {E.~J.}\ \bibnamefont {Fox}}, \bibinfo
  {author} {\bibfnamefont {Y.-H.}\ \bibnamefont {Zhang}}, \bibinfo {author}
  {\bibfnamefont {S.}~\bibnamefont {Wang}}, \bibinfo {author} {\bibfnamefont
  {L.}~\bibnamefont {Jiang}}, \bibinfo {author} {\bibfnamefont
  {B.}~\bibnamefont {Lyu}}, \bibinfo {author} {\bibfnamefont {H.}~\bibnamefont
  {Li}}, \bibinfo {author} {\bibfnamefont {K.}~\bibnamefont {Watanabe}},
  \bibinfo {author} {\bibfnamefont {T.}~\bibnamefont {Taniguchi}}, \emph
  {et~al.},\ }\bibfield  {title} {\bibinfo {title} {Tunable correlated chern
  insulator and ferromagnetism in a moir{\'e} superlattice},\ }\href
  {https://doi.org/10.1038/s41586-020-2049-7} {\bibfield  {journal} {\bibinfo
  {journal} {Nature}\ }\textbf {\bibinfo {volume} {579}},\ \bibinfo {pages}
  {56} (\bibinfo {year} {2020})}\BibitemShut {NoStop}%
\bibitem [{\citenamefont {Wu}\ \emph {et~al.}(2019)\citenamefont {Wu},
  \citenamefont {Lovorn}, \citenamefont {Tutuc}, \citenamefont {Martin},\ and\
  \citenamefont {MacDonald}}]{Wu2019Topological}%
  \BibitemOpen
  \bibfield  {author} {\bibinfo {author} {\bibfnamefont {F.}~\bibnamefont
  {Wu}}, \bibinfo {author} {\bibfnamefont {T.}~\bibnamefont {Lovorn}}, \bibinfo
  {author} {\bibfnamefont {E.}~\bibnamefont {Tutuc}}, \bibinfo {author}
  {\bibfnamefont {I.}~\bibnamefont {Martin}},\ and\ \bibinfo {author}
  {\bibfnamefont {A.~H.}\ \bibnamefont {MacDonald}},\ }\bibfield  {title}
  {\bibinfo {title} {Topological insulators in twisted transition metal
  dichalcogenide homobilayers},\ }\href
  {https://doi.org/10.1103/PhysRevLett.122.086402} {\bibfield  {journal}
  {\bibinfo  {journal} {Phys. Rev. Lett.}\ }\textbf {\bibinfo {volume} {122}},\
  \bibinfo {pages} {086402} (\bibinfo {year} {2019})}\BibitemShut {NoStop}%
\bibitem [{\citenamefont {Cao}\ \emph {et~al.}(2018)\citenamefont {Cao},
  \citenamefont {Fatemi}, \citenamefont {Fang}, \citenamefont {Watanabe},
  \citenamefont {Taniguchi}, \citenamefont {Kaxiras},\ and\ \citenamefont
  {Jarillo-Herrero}}]{cao_unconventional_2018}%
  \BibitemOpen
  \bibfield  {author} {\bibinfo {author} {\bibfnamefont {Y.}~\bibnamefont
  {Cao}}, \bibinfo {author} {\bibfnamefont {V.}~\bibnamefont {Fatemi}},
  \bibinfo {author} {\bibfnamefont {S.}~\bibnamefont {Fang}}, \bibinfo {author}
  {\bibfnamefont {K.}~\bibnamefont {Watanabe}}, \bibinfo {author}
  {\bibfnamefont {T.}~\bibnamefont {Taniguchi}}, \bibinfo {author}
  {\bibfnamefont {E.}~\bibnamefont {Kaxiras}},\ and\ \bibinfo {author}
  {\bibfnamefont {P.}~\bibnamefont {Jarillo-Herrero}},\ }\bibfield  {title}
  {\bibinfo {title} {Unconventional superconductivity in magic-angle graphene
  superlattices},\ }\href {https://doi.org/10.1038/nature26160} {\bibfield
  {journal} {\bibinfo  {journal} {Nature}\ }\textbf {\bibinfo {volume} {556}},\
  \bibinfo {pages} {43} (\bibinfo {year} {2018})}\BibitemShut {NoStop}%
\bibitem [{\citenamefont {Yankowitz}\ \emph {et~al.}(2019)\citenamefont
  {Yankowitz}, \citenamefont {Chen}, \citenamefont {Polshyn}, \citenamefont
  {Zhang}, \citenamefont {Watanabe}, \citenamefont {Taniguchi}, \citenamefont
  {Graf}, \citenamefont {Young},\ and\ \citenamefont {Dean}}]{Yankowitz1059}%
  \BibitemOpen
  \bibfield  {author} {\bibinfo {author} {\bibfnamefont {M.}~\bibnamefont
  {Yankowitz}}, \bibinfo {author} {\bibfnamefont {S.}~\bibnamefont {Chen}},
  \bibinfo {author} {\bibfnamefont {H.}~\bibnamefont {Polshyn}}, \bibinfo
  {author} {\bibfnamefont {Y.}~\bibnamefont {Zhang}}, \bibinfo {author}
  {\bibfnamefont {K.}~\bibnamefont {Watanabe}}, \bibinfo {author}
  {\bibfnamefont {T.}~\bibnamefont {Taniguchi}}, \bibinfo {author}
  {\bibfnamefont {D.}~\bibnamefont {Graf}}, \bibinfo {author} {\bibfnamefont
  {A.~F.}\ \bibnamefont {Young}},\ and\ \bibinfo {author} {\bibfnamefont
  {C.~R.}\ \bibnamefont {Dean}},\ }\bibfield  {title} {\bibinfo {title} {Tuning
  superconductivity in twisted bilayer graphene},\ }\href
  {https://doi.org/10.1126/science.aav1910} {\bibfield  {journal} {\bibinfo
  {journal} {Science}\ }\textbf {\bibinfo {volume} {363}},\ \bibinfo {pages}
  {1059} (\bibinfo {year} {2019})}\BibitemShut {NoStop}%
\bibitem [{\citenamefont {Chen}\ \emph {et~al.}(2019)\citenamefont {Chen},
  \citenamefont {Sharpe}, \citenamefont {Gallagher}, \citenamefont {Rosen},
  \citenamefont {Fox}, \citenamefont {Jiang}, \citenamefont {Lyu},
  \citenamefont {Li}, \citenamefont {Watanabe}, \citenamefont {Taniguchi} \emph
  {et~al.}}]{chen2019signatures}%
  \BibitemOpen
  \bibfield  {author} {\bibinfo {author} {\bibfnamefont {G.}~\bibnamefont
  {Chen}}, \bibinfo {author} {\bibfnamefont {A.~L.}\ \bibnamefont {Sharpe}},
  \bibinfo {author} {\bibfnamefont {P.}~\bibnamefont {Gallagher}}, \bibinfo
  {author} {\bibfnamefont {I.~T.}\ \bibnamefont {Rosen}}, \bibinfo {author}
  {\bibfnamefont {E.~J.}\ \bibnamefont {Fox}}, \bibinfo {author} {\bibfnamefont
  {L.}~\bibnamefont {Jiang}}, \bibinfo {author} {\bibfnamefont
  {B.}~\bibnamefont {Lyu}}, \bibinfo {author} {\bibfnamefont {H.}~\bibnamefont
  {Li}}, \bibinfo {author} {\bibfnamefont {K.}~\bibnamefont {Watanabe}},
  \bibinfo {author} {\bibfnamefont {T.}~\bibnamefont {Taniguchi}}, \emph
  {et~al.},\ }\bibfield  {title} {\bibinfo {title} {Signatures of tunable
  superconductivity in a trilayer graphene moir{\'e} superlattice},\ }\href
  {https://doi.org/10.1038/s41586-019-1393-y} {\bibfield  {journal} {\bibinfo
  {journal} {Nature}\ }\textbf {\bibinfo {volume} {572}},\ \bibinfo {pages}
  {215} (\bibinfo {year} {2019})}\BibitemShut {NoStop}%
\bibitem [{\citenamefont {Lu}\ \emph {et~al.}(2019)\citenamefont {Lu},
  \citenamefont {Stepanov}, \citenamefont {Yang}, \citenamefont {Xie},
  \citenamefont {Aamir}, \citenamefont {Das}, \citenamefont {Urgell},
  \citenamefont {Watanabe}, \citenamefont {Taniguchi}, \citenamefont {Zhang}
  \emph {et~al.}}]{lu2019superconductors}%
  \BibitemOpen
  \bibfield  {author} {\bibinfo {author} {\bibfnamefont {X.}~\bibnamefont
  {Lu}}, \bibinfo {author} {\bibfnamefont {P.}~\bibnamefont {Stepanov}},
  \bibinfo {author} {\bibfnamefont {W.}~\bibnamefont {Yang}}, \bibinfo {author}
  {\bibfnamefont {M.}~\bibnamefont {Xie}}, \bibinfo {author} {\bibfnamefont
  {M.~A.}\ \bibnamefont {Aamir}}, \bibinfo {author} {\bibfnamefont
  {I.}~\bibnamefont {Das}}, \bibinfo {author} {\bibfnamefont {C.}~\bibnamefont
  {Urgell}}, \bibinfo {author} {\bibfnamefont {K.}~\bibnamefont {Watanabe}},
  \bibinfo {author} {\bibfnamefont {T.}~\bibnamefont {Taniguchi}}, \bibinfo
  {author} {\bibfnamefont {G.}~\bibnamefont {Zhang}}, \emph {et~al.},\
  }\bibfield  {title} {\bibinfo {title} {Superconductors, orbital magnets and
  correlated states in magic-angle bilayer graphene},\ }\href
  {https://doi.org/10.1038/s41586-019-1695-0} {\bibfield  {journal} {\bibinfo
  {journal} {Nature}\ }\textbf {\bibinfo {volume} {574}},\ \bibinfo {pages}
  {653} (\bibinfo {year} {2019})}\BibitemShut {NoStop}%
\bibitem [{\citenamefont {Stepanov}\ \emph
  {et~al.}(2020{\natexlab{b}})\citenamefont {Stepanov}, \citenamefont {Das},
  \citenamefont {Lu}, \citenamefont {Fahimniya}, \citenamefont {Watanabe},
  \citenamefont {Taniguchi}, \citenamefont {Koppens}, \citenamefont {Lischner},
  \citenamefont {Levitov},\ and\ \citenamefont {Efetov}}]{stepanov2020untying}%
  \BibitemOpen
  \bibfield  {author} {\bibinfo {author} {\bibfnamefont {P.}~\bibnamefont
  {Stepanov}}, \bibinfo {author} {\bibfnamefont {I.}~\bibnamefont {Das}},
  \bibinfo {author} {\bibfnamefont {X.}~\bibnamefont {Lu}}, \bibinfo {author}
  {\bibfnamefont {A.}~\bibnamefont {Fahimniya}}, \bibinfo {author}
  {\bibfnamefont {K.}~\bibnamefont {Watanabe}}, \bibinfo {author}
  {\bibfnamefont {T.}~\bibnamefont {Taniguchi}}, \bibinfo {author}
  {\bibfnamefont {F.~H.}\ \bibnamefont {Koppens}}, \bibinfo {author}
  {\bibfnamefont {J.}~\bibnamefont {Lischner}}, \bibinfo {author}
  {\bibfnamefont {L.}~\bibnamefont {Levitov}},\ and\ \bibinfo {author}
  {\bibfnamefont {D.~K.}\ \bibnamefont {Efetov}},\ }\bibfield  {title}
  {\bibinfo {title} {Untying the insulating and superconducting orders in
  magic-angle graphene},\ }\href
  {https://www.nature.com/articles/s41586-020-2459-6} {\bibfield  {journal}
  {\bibinfo  {journal} {Nature}\ }\textbf {\bibinfo {volume} {583}},\ \bibinfo
  {pages} {375} (\bibinfo {year} {2020}{\natexlab{b}})}\BibitemShut {NoStop}%
\bibitem [{\citenamefont {Arora}\ \emph {et~al.}(2020)\citenamefont {Arora},
  \citenamefont {Polski}, \citenamefont {Zhang}, \citenamefont {Thomson},
  \citenamefont {Choi}, \citenamefont {Kim}, \citenamefont {Lin}, \citenamefont
  {Wilson}, \citenamefont {Xu}, \citenamefont {Chu} \emph
  {et~al.}}]{arora2020superconductivity}%
  \BibitemOpen
  \bibfield  {author} {\bibinfo {author} {\bibfnamefont {H.~S.}\ \bibnamefont
  {Arora}}, \bibinfo {author} {\bibfnamefont {R.}~\bibnamefont {Polski}},
  \bibinfo {author} {\bibfnamefont {Y.}~\bibnamefont {Zhang}}, \bibinfo
  {author} {\bibfnamefont {A.}~\bibnamefont {Thomson}}, \bibinfo {author}
  {\bibfnamefont {Y.}~\bibnamefont {Choi}}, \bibinfo {author} {\bibfnamefont
  {H.}~\bibnamefont {Kim}}, \bibinfo {author} {\bibfnamefont {Z.}~\bibnamefont
  {Lin}}, \bibinfo {author} {\bibfnamefont {I.~Z.}\ \bibnamefont {Wilson}},
  \bibinfo {author} {\bibfnamefont {X.}~\bibnamefont {Xu}}, \bibinfo {author}
  {\bibfnamefont {J.-H.}\ \bibnamefont {Chu}}, \emph {et~al.},\ }\bibfield
  {title} {\bibinfo {title} {Superconductivity in metallic twisted bilayer
  graphene stabilized by wse 2},\ }\href
  {https://www.nature.com/articles/s41586-020-2473-8} {\bibfield  {journal}
  {\bibinfo  {journal} {Nature}\ }\textbf {\bibinfo {volume} {583}},\ \bibinfo
  {pages} {379} (\bibinfo {year} {2020})}\BibitemShut {NoStop}%
\bibitem [{\citenamefont {Saito}\ \emph {et~al.}(2020)\citenamefont {Saito},
  \citenamefont {Ge}, \citenamefont {Watanabe}, \citenamefont {Taniguchi},\
  and\ \citenamefont {Young}}]{saito2020independent}%
  \BibitemOpen
  \bibfield  {author} {\bibinfo {author} {\bibfnamefont {Y.}~\bibnamefont
  {Saito}}, \bibinfo {author} {\bibfnamefont {J.}~\bibnamefont {Ge}}, \bibinfo
  {author} {\bibfnamefont {K.}~\bibnamefont {Watanabe}}, \bibinfo {author}
  {\bibfnamefont {T.}~\bibnamefont {Taniguchi}},\ and\ \bibinfo {author}
  {\bibfnamefont {A.~F.}\ \bibnamefont {Young}},\ }\bibfield  {title} {\bibinfo
  {title} {Independent superconductors and correlated insulators in twisted
  bilayer graphene},\ }\href
  {https://www.nature.com/articles/s41567-020-0928-3} {\bibfield  {journal}
  {\bibinfo  {journal} {Nature Physics}\ }\textbf {\bibinfo {volume} {16}},\
  \bibinfo {pages} {926} (\bibinfo {year} {2020})}\BibitemShut {NoStop}%
\bibitem [{\citenamefont {Cao}\ \emph {et~al.}(2021{\natexlab{a}})\citenamefont
  {Cao}, \citenamefont {Rodan-Legrain}, \citenamefont {Park}, \citenamefont
  {Yuan}, \citenamefont {Watanabe}, \citenamefont {Taniguchi}, \citenamefont
  {Fernandes}, \citenamefont {Fu},\ and\ \citenamefont
  {Jarillo-Herrero}}]{cao2021nematicity}%
  \BibitemOpen
  \bibfield  {author} {\bibinfo {author} {\bibfnamefont {Y.}~\bibnamefont
  {Cao}}, \bibinfo {author} {\bibfnamefont {D.}~\bibnamefont {Rodan-Legrain}},
  \bibinfo {author} {\bibfnamefont {J.~M.}\ \bibnamefont {Park}}, \bibinfo
  {author} {\bibfnamefont {N.~F.}\ \bibnamefont {Yuan}}, \bibinfo {author}
  {\bibfnamefont {K.}~\bibnamefont {Watanabe}}, \bibinfo {author}
  {\bibfnamefont {T.}~\bibnamefont {Taniguchi}}, \bibinfo {author}
  {\bibfnamefont {R.~M.}\ \bibnamefont {Fernandes}}, \bibinfo {author}
  {\bibfnamefont {L.}~\bibnamefont {Fu}},\ and\ \bibinfo {author}
  {\bibfnamefont {P.}~\bibnamefont {Jarillo-Herrero}},\ }\bibfield  {title}
  {\bibinfo {title} {Nematicity and competing orders in superconducting
  magic-angle graphene},\ }\href
  {https://www.science.org/doi/full/10.1126/science.abc2836} {\bibfield
  {journal} {\bibinfo  {journal} {science}\ }\textbf {\bibinfo {volume}
  {372}},\ \bibinfo {pages} {264} (\bibinfo {year}
  {2021}{\natexlab{a}})}\BibitemShut {NoStop}%
\bibitem [{\citenamefont {Park}\ \emph
  {et~al.}(2021{\natexlab{b}})\citenamefont {Park}, \citenamefont {Cao},
  \citenamefont {Watanabe}, \citenamefont {Taniguchi},\ and\ \citenamefont
  {Jarillo-Herrero}}]{park2021tunable}%
  \BibitemOpen
  \bibfield  {author} {\bibinfo {author} {\bibfnamefont {J.~M.}\ \bibnamefont
  {Park}}, \bibinfo {author} {\bibfnamefont {Y.}~\bibnamefont {Cao}}, \bibinfo
  {author} {\bibfnamefont {K.}~\bibnamefont {Watanabe}}, \bibinfo {author}
  {\bibfnamefont {T.}~\bibnamefont {Taniguchi}},\ and\ \bibinfo {author}
  {\bibfnamefont {P.}~\bibnamefont {Jarillo-Herrero}},\ }\bibfield  {title}
  {\bibinfo {title} {Tunable strongly coupled superconductivity in magic-angle
  twisted trilayer graphene},\ }\href
  {https://www.nature.com/articles/s41586-021-03192-0} {\bibfield  {journal}
  {\bibinfo  {journal} {Nature}\ }\textbf {\bibinfo {volume} {590}},\ \bibinfo
  {pages} {249} (\bibinfo {year} {2021}{\natexlab{b}})}\BibitemShut {NoStop}%
\bibitem [{\citenamefont {Hao}\ \emph {et~al.}(2021)\citenamefont {Hao},
  \citenamefont {Zimmerman}, \citenamefont {Ledwith}, \citenamefont {Khalaf},
  \citenamefont {Najafabadi}, \citenamefont {Watanabe}, \citenamefont
  {Taniguchi}, \citenamefont {Vishwanath},\ and\ \citenamefont
  {Kim}}]{hao2021electric}%
  \BibitemOpen
  \bibfield  {author} {\bibinfo {author} {\bibfnamefont {Z.}~\bibnamefont
  {Hao}}, \bibinfo {author} {\bibfnamefont {A.}~\bibnamefont {Zimmerman}},
  \bibinfo {author} {\bibfnamefont {P.}~\bibnamefont {Ledwith}}, \bibinfo
  {author} {\bibfnamefont {E.}~\bibnamefont {Khalaf}}, \bibinfo {author}
  {\bibfnamefont {D.~H.}\ \bibnamefont {Najafabadi}}, \bibinfo {author}
  {\bibfnamefont {K.}~\bibnamefont {Watanabe}}, \bibinfo {author}
  {\bibfnamefont {T.}~\bibnamefont {Taniguchi}}, \bibinfo {author}
  {\bibfnamefont {A.}~\bibnamefont {Vishwanath}},\ and\ \bibinfo {author}
  {\bibfnamefont {P.}~\bibnamefont {Kim}},\ }\bibfield  {title} {\bibinfo
  {title} {Electric field--tunable superconductivity in alternating-twist
  magic-angle trilayer graphene},\ }\href
  {https://www.science.org/doi/full/10.1126/science.abg0399} {\bibfield
  {journal} {\bibinfo  {journal} {Science}\ }\textbf {\bibinfo {volume}
  {371}},\ \bibinfo {pages} {1133} (\bibinfo {year} {2021})}\BibitemShut
  {NoStop}%
\bibitem [{\citenamefont {Cao}\ \emph {et~al.}(2021{\natexlab{b}})\citenamefont
  {Cao}, \citenamefont {Park}, \citenamefont {Watanabe}, \citenamefont
  {Taniguchi},\ and\ \citenamefont {Jarillo-Herrero}}]{cao2021pauli}%
  \BibitemOpen
  \bibfield  {author} {\bibinfo {author} {\bibfnamefont {Y.}~\bibnamefont
  {Cao}}, \bibinfo {author} {\bibfnamefont {J.~M.}\ \bibnamefont {Park}},
  \bibinfo {author} {\bibfnamefont {K.}~\bibnamefont {Watanabe}}, \bibinfo
  {author} {\bibfnamefont {T.}~\bibnamefont {Taniguchi}},\ and\ \bibinfo
  {author} {\bibfnamefont {P.}~\bibnamefont {Jarillo-Herrero}},\ }\bibfield
  {title} {\bibinfo {title} {Pauli-limit violation and re-entrant
  superconductivity in moir{\'e} graphene},\ }\href
  {https://www.nature.com/articles/s41586-021-03685-y} {\bibfield  {journal}
  {\bibinfo  {journal} {Nature}\ }\textbf {\bibinfo {volume} {595}},\ \bibinfo
  {pages} {526} (\bibinfo {year} {2021}{\natexlab{b}})}\BibitemShut {NoStop}%
\bibitem [{\citenamefont {Kim}\ \emph {et~al.}(2021)\citenamefont {Kim},
  \citenamefont {Choi}, \citenamefont {Lewandowski}, \citenamefont {Thomson},
  \citenamefont {Zhang}, \citenamefont {Polski}, \citenamefont {Watanabe},
  \citenamefont {Taniguchi}, \citenamefont {Alicea},\ and\ \citenamefont
  {Nadj-Perge}}]{kim2021spectroscopic}%
  \BibitemOpen
  \bibfield  {author} {\bibinfo {author} {\bibfnamefont {H.}~\bibnamefont
  {Kim}}, \bibinfo {author} {\bibfnamefont {Y.}~\bibnamefont {Choi}}, \bibinfo
  {author} {\bibfnamefont {C.}~\bibnamefont {Lewandowski}}, \bibinfo {author}
  {\bibfnamefont {A.}~\bibnamefont {Thomson}}, \bibinfo {author} {\bibfnamefont
  {Y.}~\bibnamefont {Zhang}}, \bibinfo {author} {\bibfnamefont
  {R.}~\bibnamefont {Polski}}, \bibinfo {author} {\bibfnamefont
  {K.}~\bibnamefont {Watanabe}}, \bibinfo {author} {\bibfnamefont
  {T.}~\bibnamefont {Taniguchi}}, \bibinfo {author} {\bibfnamefont
  {J.}~\bibnamefont {Alicea}},\ and\ \bibinfo {author} {\bibfnamefont
  {S.}~\bibnamefont {Nadj-Perge}},\ }\href@noop {} {\bibinfo {title}
  {Spectroscopic signatures of strong correlations and unconventional
  superconductivity in twisted trilayer graphene}} (\bibinfo {year} {2021}),\
  \Eprint {https://arxiv.org/abs/2109.12127} {arXiv:2109.12127
  [cond-mat.mes-hall]} \BibitemShut {NoStop}%
\bibitem [{\citenamefont {Xu}\ and\ \citenamefont
  {Balents}(2018)}]{Xu2018topological}%
  \BibitemOpen
  \bibfield  {author} {\bibinfo {author} {\bibfnamefont {C.}~\bibnamefont
  {Xu}}\ and\ \bibinfo {author} {\bibfnamefont {L.}~\bibnamefont {Balents}},\
  }\bibfield  {title} {\bibinfo {title} {Topological superconductivity in
  twisted multilayer graphene},\ }\href
  {https://doi.org/10.1103/PhysRevLett.121.087001} {\bibfield  {journal}
  {\bibinfo  {journal} {Phys. Rev. Lett.}\ }\textbf {\bibinfo {volume} {121}},\
  \bibinfo {pages} {087001} (\bibinfo {year} {2018})}\BibitemShut {NoStop}%
\bibitem [{\citenamefont {Guo}\ \emph {et~al.}(2018)\citenamefont {Guo},
  \citenamefont {Zhu}, \citenamefont {Feng},\ and\ \citenamefont
  {Scalettar}}]{Guo2018Pairing}%
  \BibitemOpen
  \bibfield  {author} {\bibinfo {author} {\bibfnamefont {H.}~\bibnamefont
  {Guo}}, \bibinfo {author} {\bibfnamefont {X.}~\bibnamefont {Zhu}}, \bibinfo
  {author} {\bibfnamefont {S.}~\bibnamefont {Feng}},\ and\ \bibinfo {author}
  {\bibfnamefont {R.~T.}\ \bibnamefont {Scalettar}},\ }\bibfield  {title}
  {\bibinfo {title} {Pairing symmetry of interacting fermions on a twisted
  bilayer graphene superlattice},\ }\href
  {https://doi.org/10.1103/PhysRevB.97.235453} {\bibfield  {journal} {\bibinfo
  {journal} {Phys. Rev. B}\ }\textbf {\bibinfo {volume} {97}},\ \bibinfo
  {pages} {235453} (\bibinfo {year} {2018})}\BibitemShut {NoStop}%
\bibitem [{\citenamefont {Wu}\ \emph {et~al.}(2018{\natexlab{a}})\citenamefont
  {Wu}, \citenamefont {MacDonald},\ and\ \citenamefont
  {Martin}}]{Wu2018Theory}%
  \BibitemOpen
  \bibfield  {author} {\bibinfo {author} {\bibfnamefont {F.}~\bibnamefont
  {Wu}}, \bibinfo {author} {\bibfnamefont {A.~H.}\ \bibnamefont {MacDonald}},\
  and\ \bibinfo {author} {\bibfnamefont {I.}~\bibnamefont {Martin}},\
  }\bibfield  {title} {\bibinfo {title} {Theory of phonon-mediated
  superconductivity in twisted bilayer graphene},\ }\href
  {https://doi.org/10.1103/PhysRevLett.121.257001} {\bibfield  {journal}
  {\bibinfo  {journal} {Phys. Rev. Lett.}\ }\textbf {\bibinfo {volume} {121}},\
  \bibinfo {pages} {257001} (\bibinfo {year} {2018}{\natexlab{a}})}\BibitemShut
  {NoStop}%
\bibitem [{\citenamefont {Fidrysiak}\ \emph {et~al.}(2018)\citenamefont
  {Fidrysiak}, \citenamefont {Zegrodnik},\ and\ \citenamefont
  {Spa\l{}ek}}]{Fidrysiak2018Unconventional}%
  \BibitemOpen
  \bibfield  {author} {\bibinfo {author} {\bibfnamefont {M.}~\bibnamefont
  {Fidrysiak}}, \bibinfo {author} {\bibfnamefont {M.}~\bibnamefont
  {Zegrodnik}},\ and\ \bibinfo {author} {\bibfnamefont {J.}~\bibnamefont
  {Spa\l{}ek}},\ }\bibfield  {title} {\bibinfo {title} {Unconventional
  topological superconductivity and phase diagram for an effective two-orbital
  model as applied to twisted bilayer graphene},\ }\href
  {https://doi.org/10.1103/PhysRevB.98.085436} {\bibfield  {journal} {\bibinfo
  {journal} {Phys. Rev. B}\ }\textbf {\bibinfo {volume} {98}},\ \bibinfo
  {pages} {085436} (\bibinfo {year} {2018})}\BibitemShut {NoStop}%
\bibitem [{\citenamefont {Su}\ and\ \citenamefont {Lin}(2018)}]{Su2018Pairing}%
  \BibitemOpen
  \bibfield  {author} {\bibinfo {author} {\bibfnamefont {Y.}~\bibnamefont
  {Su}}\ and\ \bibinfo {author} {\bibfnamefont {S.-Z.}\ \bibnamefont {Lin}},\
  }\bibfield  {title} {\bibinfo {title} {Pairing symmetry and spontaneous
  vortex-antivortex lattice in superconducting twisted-bilayer graphene:
  Bogoliubov-de gennes approach},\ }\href
  {https://doi.org/10.1103/PhysRevB.98.195101} {\bibfield  {journal} {\bibinfo
  {journal} {Phys. Rev. B}\ }\textbf {\bibinfo {volume} {98}},\ \bibinfo
  {pages} {195101} (\bibinfo {year} {2018})}\BibitemShut {NoStop}%
\bibitem [{\citenamefont {Liu}\ \emph {et~al.}(2018)\citenamefont {Liu},
  \citenamefont {Zhang}, \citenamefont {Chen},\ and\ \citenamefont
  {Yang}}]{Liu2018Chiral}%
  \BibitemOpen
  \bibfield  {author} {\bibinfo {author} {\bibfnamefont {C.-C.}\ \bibnamefont
  {Liu}}, \bibinfo {author} {\bibfnamefont {L.-D.}\ \bibnamefont {Zhang}},
  \bibinfo {author} {\bibfnamefont {W.-Q.}\ \bibnamefont {Chen}},\ and\
  \bibinfo {author} {\bibfnamefont {F.}~\bibnamefont {Yang}},\ }\bibfield
  {title} {\bibinfo {title} {Chiral spin density wave and $d+id$
  superconductivity in the magic-angle-twisted bilayer graphene},\ }\href
  {https://doi.org/10.1103/PhysRevLett.121.217001} {\bibfield  {journal}
  {\bibinfo  {journal} {Phys. Rev. Lett.}\ }\textbf {\bibinfo {volume} {121}},\
  \bibinfo {pages} {217001} (\bibinfo {year} {2018})}\BibitemShut {NoStop}%
\bibitem [{\citenamefont {Kennes}\ \emph {et~al.}(2018)\citenamefont {Kennes},
  \citenamefont {Lischner},\ and\ \citenamefont {Karrasch}}]{Kennes2018Strong}%
  \BibitemOpen
  \bibfield  {author} {\bibinfo {author} {\bibfnamefont {D.~M.}\ \bibnamefont
  {Kennes}}, \bibinfo {author} {\bibfnamefont {J.}~\bibnamefont {Lischner}},\
  and\ \bibinfo {author} {\bibfnamefont {C.}~\bibnamefont {Karrasch}},\
  }\bibfield  {title} {\bibinfo {title} {Strong correlations and
  $d+\mathit{id}$ superconductivity in twisted bilayer graphene},\ }\href
  {https://doi.org/10.1103/PhysRevB.98.241407} {\bibfield  {journal} {\bibinfo
  {journal} {Phys. Rev. B}\ }\textbf {\bibinfo {volume} {98}},\ \bibinfo
  {pages} {241407} (\bibinfo {year} {2018})}\BibitemShut {NoStop}%
\bibitem [{\citenamefont {Isobe}\ \emph {et~al.}(2018)\citenamefont {Isobe},
  \citenamefont {Yuan},\ and\ \citenamefont {Fu}}]{Isobe2018Unconventional}%
  \BibitemOpen
  \bibfield  {author} {\bibinfo {author} {\bibfnamefont {H.}~\bibnamefont
  {Isobe}}, \bibinfo {author} {\bibfnamefont {N.~F.~Q.}\ \bibnamefont {Yuan}},\
  and\ \bibinfo {author} {\bibfnamefont {L.}~\bibnamefont {Fu}},\ }\bibfield
  {title} {\bibinfo {title} {Unconventional superconductivity and density waves
  in twisted bilayer graphene},\ }\href
  {https://doi.org/10.1103/PhysRevX.8.041041} {\bibfield  {journal} {\bibinfo
  {journal} {Phys. Rev. X}\ }\textbf {\bibinfo {volume} {8}},\ \bibinfo {pages}
  {041041} (\bibinfo {year} {2018})}\BibitemShut {NoStop}%
\bibitem [{\citenamefont {Roy}\ and\ \citenamefont {Juri\ifmmode \check{c}\else
  \v{c}\fi{}i\ifmmode~\acute{c}\else
  \'{c}\fi{}}(2019)}]{Roy2019Unconventional}%
  \BibitemOpen
  \bibfield  {author} {\bibinfo {author} {\bibfnamefont {B.}~\bibnamefont
  {Roy}}\ and\ \bibinfo {author} {\bibfnamefont {V.}~\bibnamefont {Juri\ifmmode
  \check{c}\else \v{c}\fi{}i\ifmmode~\acute{c}\else \'{c}\fi{}}},\ }\bibfield
  {title} {\bibinfo {title} {Unconventional superconductivity in nearly flat
  bands in twisted bilayer graphene},\ }\href
  {https://doi.org/10.1103/PhysRevB.99.121407} {\bibfield  {journal} {\bibinfo
  {journal} {Phys. Rev. B}\ }\textbf {\bibinfo {volume} {99}},\ \bibinfo
  {pages} {121407} (\bibinfo {year} {2019})}\BibitemShut {NoStop}%
\bibitem [{\citenamefont {Huang}\ \emph {et~al.}(2019)\citenamefont {Huang},
  \citenamefont {Zhang},\ and\ \citenamefont
  {Ma}}]{huang2019antiferromagnetically}%
  \BibitemOpen
  \bibfield  {author} {\bibinfo {author} {\bibfnamefont {T.}~\bibnamefont
  {Huang}}, \bibinfo {author} {\bibfnamefont {L.}~\bibnamefont {Zhang}},\ and\
  \bibinfo {author} {\bibfnamefont {T.}~\bibnamefont {Ma}},\ }\bibfield
  {title} {\bibinfo {title} {Antiferromagnetically ordered mott insulator and
  d+ id superconductivity in twisted bilayer graphene: A quantum monte carlo
  study},\ }\href
  {https://www.sciencedirect.com/science/article/abs/pii/S2095927319300805?via\%3Dihub}
  {\bibfield  {journal} {\bibinfo  {journal} {Science Bulletin}\ }\textbf
  {\bibinfo {volume} {64}},\ \bibinfo {pages} {310} (\bibinfo {year}
  {2019})}\BibitemShut {NoStop}%
\bibitem [{\citenamefont {Ray}\ \emph {et~al.}(2019)\citenamefont {Ray},
  \citenamefont {Jung},\ and\ \citenamefont {Das}}]{Ray2019Wannier}%
  \BibitemOpen
  \bibfield  {author} {\bibinfo {author} {\bibfnamefont {S.}~\bibnamefont
  {Ray}}, \bibinfo {author} {\bibfnamefont {J.}~\bibnamefont {Jung}},\ and\
  \bibinfo {author} {\bibfnamefont {T.}~\bibnamefont {Das}},\ }\bibfield
  {title} {\bibinfo {title} {Wannier pairs in superconducting twisted bilayer
  graphene and related systems},\ }\href
  {https://doi.org/10.1103/PhysRevB.99.134515} {\bibfield  {journal} {\bibinfo
  {journal} {Phys. Rev. B}\ }\textbf {\bibinfo {volume} {99}},\ \bibinfo
  {pages} {134515} (\bibinfo {year} {2019})}\BibitemShut {NoStop}%
\bibitem [{\citenamefont {Lian}\ \emph {et~al.}(2019)\citenamefont {Lian},
  \citenamefont {Wang},\ and\ \citenamefont {Bernevig}}]{Lian2019Twisted}%
  \BibitemOpen
  \bibfield  {author} {\bibinfo {author} {\bibfnamefont {B.}~\bibnamefont
  {Lian}}, \bibinfo {author} {\bibfnamefont {Z.}~\bibnamefont {Wang}},\ and\
  \bibinfo {author} {\bibfnamefont {B.~A.}\ \bibnamefont {Bernevig}},\
  }\bibfield  {title} {\bibinfo {title} {Twisted bilayer graphene: A
  phonon-driven superconductor},\ }\href
  {https://doi.org/10.1103/PhysRevLett.122.257002} {\bibfield  {journal}
  {\bibinfo  {journal} {Phys. Rev. Lett.}\ }\textbf {\bibinfo {volume} {122}},\
  \bibinfo {pages} {257002} (\bibinfo {year} {2019})}\BibitemShut {NoStop}%
\bibitem [{\citenamefont {Gonz\'alez}\ and\ \citenamefont
  {Stauber}(2019)}]{Stauber2019Kohn}%
  \BibitemOpen
  \bibfield  {author} {\bibinfo {author} {\bibfnamefont {J.}~\bibnamefont
  {Gonz\'alez}}\ and\ \bibinfo {author} {\bibfnamefont {T.}~\bibnamefont
  {Stauber}},\ }\bibfield  {title} {\bibinfo {title} {Kohn-luttinger
  superconductivity in twisted bilayer graphene},\ }\href
  {https://doi.org/10.1103/PhysRevLett.122.026801} {\bibfield  {journal}
  {\bibinfo  {journal} {Phys. Rev. Lett.}\ }\textbf {\bibinfo {volume} {122}},\
  \bibinfo {pages} {026801} (\bibinfo {year} {2019})}\BibitemShut {NoStop}%
\bibitem [{\citenamefont {You}\ and\ \citenamefont
  {Vishwanath}(2019)}]{you2019superconductivity}%
  \BibitemOpen
  \bibfield  {author} {\bibinfo {author} {\bibfnamefont {Y.-Z.}\ \bibnamefont
  {You}}\ and\ \bibinfo {author} {\bibfnamefont {A.}~\bibnamefont
  {Vishwanath}},\ }\bibfield  {title} {\bibinfo {title} {Superconductivity from
  valley fluctuations and approximate so (4) symmetry in a weak coupling theory
  of twisted bilayer graphene},\ }\href
  {https://www.nature.com/articles/s41535-019-0153-4} {\bibfield  {journal}
  {\bibinfo  {journal} {npj Quantum Materials}\ }\textbf {\bibinfo {volume}
  {4}},\ \bibinfo {pages} {1} (\bibinfo {year} {2019})}\BibitemShut {NoStop}%
\bibitem [{\citenamefont {Cea}\ and\ \citenamefont
  {Guinea}(2021)}]{Ceae2107874118}%
  \BibitemOpen
  \bibfield  {author} {\bibinfo {author} {\bibfnamefont {T.}~\bibnamefont
  {Cea}}\ and\ \bibinfo {author} {\bibfnamefont {F.}~\bibnamefont {Guinea}},\
  }\bibfield  {title} {\bibinfo {title} {Coulomb interaction, phonons, and
  superconductivity in twisted bilayer graphene},\ }\bibfield  {journal}
  {\bibinfo  {journal} {Proceedings of the National Academy of Sciences}\
  }\textbf {\bibinfo {volume} {118}},\ \href
  {https://doi.org/10.1073/pnas.2107874118} {10.1073/pnas.2107874118} (\bibinfo
  {year} {2021})\BibitemShut {NoStop}%
\bibitem [{\citenamefont {Fernandes}\ and\ \citenamefont
  {Fu}(2021)}]{Fernandes2021Charge}%
  \BibitemOpen
  \bibfield  {author} {\bibinfo {author} {\bibfnamefont {R.~M.}\ \bibnamefont
  {Fernandes}}\ and\ \bibinfo {author} {\bibfnamefont {L.}~\bibnamefont {Fu}},\
  }\bibfield  {title} {\bibinfo {title} {Charge-$4e$ superconductivity from
  multicomponent nematic pairing: Application to twisted bilayer graphene},\
  }\href {https://doi.org/10.1103/PhysRevLett.127.047001} {\bibfield  {journal}
  {\bibinfo  {journal} {Phys. Rev. Lett.}\ }\textbf {\bibinfo {volume} {127}},\
  \bibinfo {pages} {047001} (\bibinfo {year} {2021})}\BibitemShut {NoStop}%
\bibitem [{\citenamefont {Khalaf}\ \emph {et~al.}(2021)\citenamefont {Khalaf},
  \citenamefont {Chatterjee}, \citenamefont {Bultinck}, \citenamefont
  {Zaletel},\ and\ \citenamefont {Vishwanath}}]{khalaf2021charged}%
  \BibitemOpen
  \bibfield  {author} {\bibinfo {author} {\bibfnamefont {E.}~\bibnamefont
  {Khalaf}}, \bibinfo {author} {\bibfnamefont {S.}~\bibnamefont {Chatterjee}},
  \bibinfo {author} {\bibfnamefont {N.}~\bibnamefont {Bultinck}}, \bibinfo
  {author} {\bibfnamefont {M.~P.}\ \bibnamefont {Zaletel}},\ and\ \bibinfo
  {author} {\bibfnamefont {A.}~\bibnamefont {Vishwanath}},\ }\bibfield  {title}
  {\bibinfo {title} {Charged skyrmions and topological origin of
  superconductivity in magic-angle graphene},\ }\href
  {https://www.science.org/doi/10.1126/sciadv.abf5299} {\bibfield  {journal}
  {\bibinfo  {journal} {Science advances}\ }\textbf {\bibinfo {volume} {7}},\
  \bibinfo {pages} {eabf5299} (\bibinfo {year} {2021})}\BibitemShut {NoStop}%
\bibitem [{\citenamefont {Qin}\ and\ \citenamefont
  {MacDonald}(2021)}]{Qin2021In}%
  \BibitemOpen
  \bibfield  {author} {\bibinfo {author} {\bibfnamefont {W.}~\bibnamefont
  {Qin}}\ and\ \bibinfo {author} {\bibfnamefont {A.~H.}\ \bibnamefont
  {MacDonald}},\ }\bibfield  {title} {\bibinfo {title} {In-plane critical
  magnetic fields in magic-angle twisted trilayer graphene},\ }\href
  {https://doi.org/10.1103/PhysRevLett.127.097001} {\bibfield  {journal}
  {\bibinfo  {journal} {Phys. Rev. Lett.}\ }\textbf {\bibinfo {volume} {127}},\
  \bibinfo {pages} {097001} (\bibinfo {year} {2021})}\BibitemShut {NoStop}%
\bibitem [{\citenamefont {Lake}\ and\ \citenamefont
  {Senthil}(2021)}]{lake2021re}%
  \BibitemOpen
  \bibfield  {author} {\bibinfo {author} {\bibfnamefont {E.}~\bibnamefont
  {Lake}}\ and\ \bibinfo {author} {\bibfnamefont {T.}~\bibnamefont {Senthil}},\
  }\href@noop {} {\bibinfo {title} {Re-entrant superconductivity through a
  quantum lifshitz transition in twisted trilayer graphene}} (\bibinfo {year}
  {2021}),\ \Eprint {https://arxiv.org/abs/2104.13920} {arXiv:2104.13920
  [cond-mat.mes-hall]} \BibitemShut {NoStop}%
\bibitem [{\citenamefont {Ledwith}\ \emph {et~al.}(2020)\citenamefont
  {Ledwith}, \citenamefont {Tarnopolsky}, \citenamefont {Khalaf},\ and\
  \citenamefont {Vishwanath}}]{Ledwith2020Fractional}%
  \BibitemOpen
  \bibfield  {author} {\bibinfo {author} {\bibfnamefont {P.~J.}\ \bibnamefont
  {Ledwith}}, \bibinfo {author} {\bibfnamefont {G.}~\bibnamefont
  {Tarnopolsky}}, \bibinfo {author} {\bibfnamefont {E.}~\bibnamefont
  {Khalaf}},\ and\ \bibinfo {author} {\bibfnamefont {A.}~\bibnamefont
  {Vishwanath}},\ }\bibfield  {title} {\bibinfo {title} {Fractional chern
  insulator states in twisted bilayer graphene: An analytical approach},\
  }\href {https://doi.org/10.1103/PhysRevResearch.2.023237} {\bibfield
  {journal} {\bibinfo  {journal} {Phys. Rev. Research}\ }\textbf {\bibinfo
  {volume} {2}},\ \bibinfo {pages} {023237} (\bibinfo {year}
  {2020})}\BibitemShut {NoStop}%
\bibitem [{\citenamefont {Repellin}\ and\ \citenamefont
  {Senthil}(2020)}]{Repellin2020Chern}%
  \BibitemOpen
  \bibfield  {author} {\bibinfo {author} {\bibfnamefont {C.}~\bibnamefont
  {Repellin}}\ and\ \bibinfo {author} {\bibfnamefont {T.}~\bibnamefont
  {Senthil}},\ }\bibfield  {title} {\bibinfo {title} {Chern bands of twisted
  bilayer graphene: Fractional chern insulators and spin phase transition},\
  }\href {https://doi.org/10.1103/PhysRevResearch.2.023238} {\bibfield
  {journal} {\bibinfo  {journal} {Phys. Rev. Research}\ }\textbf {\bibinfo
  {volume} {2}},\ \bibinfo {pages} {023238} (\bibinfo {year}
  {2020})}\BibitemShut {NoStop}%
\bibitem [{\citenamefont {Liu}\ \emph {et~al.}(2021)\citenamefont {Liu},
  \citenamefont {Abouelkomsan},\ and\ \citenamefont {Bergholtz}}]{Liu2021Gate}%
  \BibitemOpen
  \bibfield  {author} {\bibinfo {author} {\bibfnamefont {Z.}~\bibnamefont
  {Liu}}, \bibinfo {author} {\bibfnamefont {A.}~\bibnamefont {Abouelkomsan}},\
  and\ \bibinfo {author} {\bibfnamefont {E.~J.}\ \bibnamefont {Bergholtz}},\
  }\bibfield  {title} {\bibinfo {title} {Gate-tunable fractional chern
  insulators in twisted double bilayer graphene},\ }\href
  {https://doi.org/10.1103/PhysRevLett.126.026801} {\bibfield  {journal}
  {\bibinfo  {journal} {Phys. Rev. Lett.}\ }\textbf {\bibinfo {volume} {126}},\
  \bibinfo {pages} {026801} (\bibinfo {year} {2021})}\BibitemShut {NoStop}%
\bibitem [{\citenamefont {Li}\ \emph {et~al.}(2021{\natexlab{a}})\citenamefont
  {Li}, \citenamefont {Kumar}, \citenamefont {Sun},\ and\ \citenamefont
  {Lin}}]{Li2021Spontaneous}%
  \BibitemOpen
  \bibfield  {author} {\bibinfo {author} {\bibfnamefont {H.}~\bibnamefont
  {Li}}, \bibinfo {author} {\bibfnamefont {U.}~\bibnamefont {Kumar}}, \bibinfo
  {author} {\bibfnamefont {K.}~\bibnamefont {Sun}},\ and\ \bibinfo {author}
  {\bibfnamefont {S.-Z.}\ \bibnamefont {Lin}},\ }\bibfield  {title} {\bibinfo
  {title} {Spontaneous fractional chern insulators in transition metal
  dichalcogenide moir\'e superlattices},\ }\href
  {https://doi.org/10.1103/PhysRevResearch.3.L032070} {\bibfield  {journal}
  {\bibinfo  {journal} {Phys. Rev. Research}\ }\textbf {\bibinfo {volume}
  {3}},\ \bibinfo {pages} {L032070} (\bibinfo {year}
  {2021}{\natexlab{a}})}\BibitemShut {NoStop}%
\bibitem [{\citenamefont {Xie}\ \emph {et~al.}(2021)\citenamefont {Xie},
  \citenamefont {Pierce}, \citenamefont {Park}, \citenamefont {Parker},
  \citenamefont {Khalaf}, \citenamefont {Ledwith}, \citenamefont {Cao},
  \citenamefont {Lee}, \citenamefont {Chen}, \citenamefont {Forrester} \emph
  {et~al.}}]{Xie2021Fractional}%
  \BibitemOpen
  \bibfield  {author} {\bibinfo {author} {\bibfnamefont {Y.}~\bibnamefont
  {Xie}}, \bibinfo {author} {\bibfnamefont {A.~T.}\ \bibnamefont {Pierce}},
  \bibinfo {author} {\bibfnamefont {J.~M.}\ \bibnamefont {Park}}, \bibinfo
  {author} {\bibfnamefont {D.~E.}\ \bibnamefont {Parker}}, \bibinfo {author}
  {\bibfnamefont {E.}~\bibnamefont {Khalaf}}, \bibinfo {author} {\bibfnamefont
  {P.}~\bibnamefont {Ledwith}}, \bibinfo {author} {\bibfnamefont
  {Y.}~\bibnamefont {Cao}}, \bibinfo {author} {\bibfnamefont {S.~H.}\
  \bibnamefont {Lee}}, \bibinfo {author} {\bibfnamefont {S.}~\bibnamefont
  {Chen}}, \bibinfo {author} {\bibfnamefont {P.~R.}\ \bibnamefont {Forrester}},
  \emph {et~al.},\ }\href@noop {} {\bibinfo {title} {Fractional chern
  insulators in magic-angle twisted bilayer graphene}} (\bibinfo {year}
  {2021}),\ \Eprint {https://arxiv.org/abs/2107.10854} {arXiv:2107.10854
  [cond-mat.mes-hall]} \BibitemShut {NoStop}%
\bibitem [{\citenamefont {Wu}\ \emph {et~al.}(2018{\natexlab{b}})\citenamefont
  {Wu}, \citenamefont {Lovorn}, \citenamefont {Tutuc},\ and\ \citenamefont
  {MacDonald}}]{Wu_hubbard_2018}%
  \BibitemOpen
  \bibfield  {author} {\bibinfo {author} {\bibfnamefont {F.}~\bibnamefont
  {Wu}}, \bibinfo {author} {\bibfnamefont {T.}~\bibnamefont {Lovorn}}, \bibinfo
  {author} {\bibfnamefont {E.}~\bibnamefont {Tutuc}},\ and\ \bibinfo {author}
  {\bibfnamefont {A.~H.}\ \bibnamefont {MacDonald}},\ }\bibfield  {title}
  {\bibinfo {title} {Hubbard model physics in transition metal dichalcogenide
  moir\'e bands},\ }\href {https://doi.org/10.1103/PhysRevLett.121.026402}
  {\bibfield  {journal} {\bibinfo  {journal} {Phys. Rev. Lett.}\ }\textbf
  {\bibinfo {volume} {121}},\ \bibinfo {pages} {026402} (\bibinfo {year}
  {2018}{\natexlab{b}})}\BibitemShut {NoStop}%
\bibitem [{\citenamefont {Zhang}\ \emph {et~al.}(2020)\citenamefont {Zhang},
  \citenamefont {Yuan},\ and\ \citenamefont {Fu}}]{Zhang_moire_2020}%
  \BibitemOpen
  \bibfield  {author} {\bibinfo {author} {\bibfnamefont {Y.}~\bibnamefont
  {Zhang}}, \bibinfo {author} {\bibfnamefont {N.~F.~Q.}\ \bibnamefont {Yuan}},\
  and\ \bibinfo {author} {\bibfnamefont {L.}~\bibnamefont {Fu}},\ }\bibfield
  {title} {\bibinfo {title} {Moir\'e quantum chemistry: Charge transfer in
  transition metal dichalcogenide superlattices},\ }\href
  {https://doi.org/10.1103/PhysRevB.102.201115} {\bibfield  {journal} {\bibinfo
   {journal} {Phys. Rev. B}\ }\textbf {\bibinfo {volume} {102}},\ \bibinfo
  {pages} {201115} (\bibinfo {year} {2020})}\BibitemShut {NoStop}%
\bibitem [{\citenamefont {Tang}\ \emph {et~al.}(2020)\citenamefont {Tang},
  \citenamefont {Li}, \citenamefont {Li}, \citenamefont {Xu}, \citenamefont
  {Liu}, \citenamefont {Barmak}, \citenamefont {Watanabe}, \citenamefont
  {Taniguchi}, \citenamefont {MacDonald}, \citenamefont {Shan},\ and\
  \citenamefont {Mak}}]{tang2020simulation}%
  \BibitemOpen
  \bibfield  {author} {\bibinfo {author} {\bibfnamefont {Y.}~\bibnamefont
  {Tang}}, \bibinfo {author} {\bibfnamefont {L.}~\bibnamefont {Li}}, \bibinfo
  {author} {\bibfnamefont {T.}~\bibnamefont {Li}}, \bibinfo {author}
  {\bibfnamefont {Y.}~\bibnamefont {Xu}}, \bibinfo {author} {\bibfnamefont
  {S.}~\bibnamefont {Liu}}, \bibinfo {author} {\bibfnamefont {K.}~\bibnamefont
  {Barmak}}, \bibinfo {author} {\bibfnamefont {K.}~\bibnamefont {Watanabe}},
  \bibinfo {author} {\bibfnamefont {T.}~\bibnamefont {Taniguchi}}, \bibinfo
  {author} {\bibfnamefont {A.~H.}\ \bibnamefont {MacDonald}}, \bibinfo {author}
  {\bibfnamefont {J.}~\bibnamefont {Shan}},\ and\ \bibinfo {author}
  {\bibfnamefont {K.~F.}\ \bibnamefont {Mak}},\ }\bibfield  {title} {\bibinfo
  {title} {Simulation of hubbard model physics in wse$_2$/ws$_2$ moir\'e
  superlattices},\ }\href {https://doi.org/10.1038/s41586-020-2085-3}
  {\bibfield  {journal} {\bibinfo  {journal} {Nature}\ }\textbf {\bibinfo
  {volume} {579}},\ \bibinfo {pages} {353} (\bibinfo {year}
  {2020})}\BibitemShut {NoStop}%
\bibitem [{\citenamefont {Regan}\ \emph {et~al.}(2020)\citenamefont {Regan},
  \citenamefont {Wang}, \citenamefont {Jin}, \citenamefont {Utama},
  \citenamefont {Gao}, \citenamefont {Wei}, \citenamefont {Zhao}, \citenamefont
  {Zhao}, \citenamefont {Zhang}, \citenamefont {Yumigeta} \emph
  {et~al.}}]{regan2020mott}%
  \BibitemOpen
  \bibfield  {author} {\bibinfo {author} {\bibfnamefont {E.~C.}\ \bibnamefont
  {Regan}}, \bibinfo {author} {\bibfnamefont {D.}~\bibnamefont {Wang}},
  \bibinfo {author} {\bibfnamefont {C.}~\bibnamefont {Jin}}, \bibinfo {author}
  {\bibfnamefont {M.~I.~B.}\ \bibnamefont {Utama}}, \bibinfo {author}
  {\bibfnamefont {B.}~\bibnamefont {Gao}}, \bibinfo {author} {\bibfnamefont
  {X.}~\bibnamefont {Wei}}, \bibinfo {author} {\bibfnamefont {S.}~\bibnamefont
  {Zhao}}, \bibinfo {author} {\bibfnamefont {W.}~\bibnamefont {Zhao}}, \bibinfo
  {author} {\bibfnamefont {Z.}~\bibnamefont {Zhang}}, \bibinfo {author}
  {\bibfnamefont {K.}~\bibnamefont {Yumigeta}}, \emph {et~al.},\ }\bibfield
  {title} {\bibinfo {title} {Mott and generalized wigner crystal states in
  wse$_2$/ws$_2$ moir\'e superlattices},\ }\href
  {https://doi.org/10.1038/s41586-020-2092-4} {\bibfield  {journal} {\bibinfo
  {journal} {Nature}\ }\textbf {\bibinfo {volume} {579}},\ \bibinfo {pages}
  {359} (\bibinfo {year} {2020})}\BibitemShut {NoStop}%
\bibitem [{\citenamefont {Xu}\ \emph {et~al.}(2020)\citenamefont {Xu},
  \citenamefont {Liu}, \citenamefont {Rhodes}, \citenamefont {Watanabe},
  \citenamefont {Taniguchi}, \citenamefont {Hone}, \citenamefont {Elser},
  \citenamefont {Mak},\ and\ \citenamefont {Shan}}]{xu2020correlated}%
  \BibitemOpen
  \bibfield  {author} {\bibinfo {author} {\bibfnamefont {Y.}~\bibnamefont
  {Xu}}, \bibinfo {author} {\bibfnamefont {S.}~\bibnamefont {Liu}}, \bibinfo
  {author} {\bibfnamefont {D.~A.}\ \bibnamefont {Rhodes}}, \bibinfo {author}
  {\bibfnamefont {K.}~\bibnamefont {Watanabe}}, \bibinfo {author}
  {\bibfnamefont {T.}~\bibnamefont {Taniguchi}}, \bibinfo {author}
  {\bibfnamefont {J.}~\bibnamefont {Hone}}, \bibinfo {author} {\bibfnamefont
  {V.}~\bibnamefont {Elser}}, \bibinfo {author} {\bibfnamefont {K.~F.}\
  \bibnamefont {Mak}},\ and\ \bibinfo {author} {\bibfnamefont {J.}~\bibnamefont
  {Shan}},\ }\bibfield  {title} {\bibinfo {title} {Correlated insulating states
  at fractional fillings of moir\'e superlattices},\ }\href
  {https://doi.org/10.1038/s41586-020-2868-6} {\bibfield  {journal} {\bibinfo
  {journal} {Nature}\ }\textbf {\bibinfo {volume} {587}},\ \bibinfo {pages}
  {214} (\bibinfo {year} {2020})}\BibitemShut {NoStop}%
\bibitem [{\citenamefont {Chu}\ \emph {et~al.}(2020)\citenamefont {Chu},
  \citenamefont {Regan}, \citenamefont {Ma}, \citenamefont {Wang},
  \citenamefont {Xu}, \citenamefont {Utama}, \citenamefont {Yumigeta},
  \citenamefont {Blei}, \citenamefont {Watanabe}, \citenamefont {Taniguchi},
  \citenamefont {Tongay}, \citenamefont {Wang},\ and\ \citenamefont
  {Lai}}]{Chu2020Nanoscale}%
  \BibitemOpen
  \bibfield  {author} {\bibinfo {author} {\bibfnamefont {Z.}~\bibnamefont
  {Chu}}, \bibinfo {author} {\bibfnamefont {E.~C.}\ \bibnamefont {Regan}},
  \bibinfo {author} {\bibfnamefont {X.}~\bibnamefont {Ma}}, \bibinfo {author}
  {\bibfnamefont {D.}~\bibnamefont {Wang}}, \bibinfo {author} {\bibfnamefont
  {Z.}~\bibnamefont {Xu}}, \bibinfo {author} {\bibfnamefont {M.~I.~B.}\
  \bibnamefont {Utama}}, \bibinfo {author} {\bibfnamefont {K.}~\bibnamefont
  {Yumigeta}}, \bibinfo {author} {\bibfnamefont {M.}~\bibnamefont {Blei}},
  \bibinfo {author} {\bibfnamefont {K.}~\bibnamefont {Watanabe}}, \bibinfo
  {author} {\bibfnamefont {T.}~\bibnamefont {Taniguchi}}, \bibinfo {author}
  {\bibfnamefont {S.}~\bibnamefont {Tongay}}, \bibinfo {author} {\bibfnamefont
  {F.}~\bibnamefont {Wang}},\ and\ \bibinfo {author} {\bibfnamefont
  {K.}~\bibnamefont {Lai}},\ }\bibfield  {title} {\bibinfo {title} {Nanoscale
  conductivity imaging of correlated electronic states in
  ${\mathrm{wse}}_{2}/{\mathrm{ws}}_{2}$ moir\'e superlattices},\ }\href
  {https://doi.org/10.1103/PhysRevLett.125.186803} {\bibfield  {journal}
  {\bibinfo  {journal} {Phys. Rev. Lett.}\ }\textbf {\bibinfo {volume} {125}},\
  \bibinfo {pages} {186803} (\bibinfo {year} {2020})}\BibitemShut {NoStop}%
\bibitem [{\citenamefont {Huang}\ \emph {et~al.}(2021)\citenamefont {Huang},
  \citenamefont {Wang}, \citenamefont {Miao}, \citenamefont {Wang},
  \citenamefont {Li}, \citenamefont {Lian}, \citenamefont {Taniguchi},
  \citenamefont {Watanabe}, \citenamefont {Okamoto}, \citenamefont {Xiao} \emph
  {et~al.}}]{huang2021correlated}%
  \BibitemOpen
  \bibfield  {author} {\bibinfo {author} {\bibfnamefont {X.}~\bibnamefont
  {Huang}}, \bibinfo {author} {\bibfnamefont {T.}~\bibnamefont {Wang}},
  \bibinfo {author} {\bibfnamefont {S.}~\bibnamefont {Miao}}, \bibinfo {author}
  {\bibfnamefont {C.}~\bibnamefont {Wang}}, \bibinfo {author} {\bibfnamefont
  {Z.}~\bibnamefont {Li}}, \bibinfo {author} {\bibfnamefont {Z.}~\bibnamefont
  {Lian}}, \bibinfo {author} {\bibfnamefont {T.}~\bibnamefont {Taniguchi}},
  \bibinfo {author} {\bibfnamefont {K.}~\bibnamefont {Watanabe}}, \bibinfo
  {author} {\bibfnamefont {S.}~\bibnamefont {Okamoto}}, \bibinfo {author}
  {\bibfnamefont {D.}~\bibnamefont {Xiao}}, \emph {et~al.},\ }\bibfield
  {title} {\bibinfo {title} {Correlated insulating states at fractional
  fillings of the ws$_2$/wse$_2$ moir{\'e} lattice},\ }\href
  {https://doi.org/10.1038/s41567-021-01171-w} {\bibfield  {journal} {\bibinfo
  {journal} {Nature Physics}\ }\textbf {\bibinfo {volume} {17}},\ \bibinfo
  {pages} {715} (\bibinfo {year} {2021})}\BibitemShut {NoStop}%
\bibitem [{\citenamefont {Li}\ \emph {et~al.}(2021{\natexlab{b}})\citenamefont
  {Li}, \citenamefont {Li}, \citenamefont {Regan}, \citenamefont {Wang},
  \citenamefont {Zhao}, \citenamefont {Kahn}, \citenamefont {Yumigeta},
  \citenamefont {Blei}, \citenamefont {Taniguchi}, \citenamefont {Watanabe}
  \emph {et~al.}}]{li2021imaging}%
  \BibitemOpen
  \bibfield  {author} {\bibinfo {author} {\bibfnamefont {H.}~\bibnamefont
  {Li}}, \bibinfo {author} {\bibfnamefont {S.}~\bibnamefont {Li}}, \bibinfo
  {author} {\bibfnamefont {E.~C.}\ \bibnamefont {Regan}}, \bibinfo {author}
  {\bibfnamefont {D.}~\bibnamefont {Wang}}, \bibinfo {author} {\bibfnamefont
  {W.}~\bibnamefont {Zhao}}, \bibinfo {author} {\bibfnamefont {S.}~\bibnamefont
  {Kahn}}, \bibinfo {author} {\bibfnamefont {K.}~\bibnamefont {Yumigeta}},
  \bibinfo {author} {\bibfnamefont {M.}~\bibnamefont {Blei}}, \bibinfo {author}
  {\bibfnamefont {T.}~\bibnamefont {Taniguchi}}, \bibinfo {author}
  {\bibfnamefont {K.}~\bibnamefont {Watanabe}}, \emph {et~al.},\ }\bibfield
  {title} {\bibinfo {title} {Imaging two-dimensional generalized wigner
  crystals},\ }\href {https://www.nature.com/articles/s41586-021-03874-9}
  {\bibfield  {journal} {\bibinfo  {journal} {Nature}\ }\textbf {\bibinfo
  {volume} {597}},\ \bibinfo {pages} {650} (\bibinfo {year}
  {2021}{\natexlab{b}})}\BibitemShut {NoStop}%
\bibitem [{\citenamefont {Li}\ \emph {et~al.}(2021{\natexlab{c}})\citenamefont
  {Li}, \citenamefont {Jiang}, \citenamefont {Shen}, \citenamefont {Zhang},
  \citenamefont {Li}, \citenamefont {Tao}, \citenamefont {Devakul},
  \citenamefont {Watanabe}, \citenamefont {Taniguchi}, \citenamefont {Fu} \emph
  {et~al.}}]{Li_quantum_2021}%
  \BibitemOpen
  \bibfield  {author} {\bibinfo {author} {\bibfnamefont {T.}~\bibnamefont
  {Li}}, \bibinfo {author} {\bibfnamefont {S.}~\bibnamefont {Jiang}}, \bibinfo
  {author} {\bibfnamefont {B.}~\bibnamefont {Shen}}, \bibinfo {author}
  {\bibfnamefont {Y.}~\bibnamefont {Zhang}}, \bibinfo {author} {\bibfnamefont
  {L.}~\bibnamefont {Li}}, \bibinfo {author} {\bibfnamefont {Z.}~\bibnamefont
  {Tao}}, \bibinfo {author} {\bibfnamefont {T.}~\bibnamefont {Devakul}},
  \bibinfo {author} {\bibfnamefont {K.}~\bibnamefont {Watanabe}}, \bibinfo
  {author} {\bibfnamefont {T.}~\bibnamefont {Taniguchi}}, \bibinfo {author}
  {\bibfnamefont {L.}~\bibnamefont {Fu}}, \emph {et~al.},\ }\bibfield  {title}
  {\bibinfo {title} {Quantum anomalous hall effect from intertwined moir{\'e}
  bands},\ }\href {https://www.nature.com/articles/s41586-021-04171-1}
  {\bibfield  {journal} {\bibinfo  {journal} {Nature}\ }\textbf {\bibinfo
  {volume} {600}},\ \bibinfo {pages} {641} (\bibinfo {year}
  {2021}{\natexlab{c}})}\BibitemShut {NoStop}%
\bibitem [{\citenamefont {Liu}\ \emph {et~al.}(2015)\citenamefont {Liu},
  \citenamefont {Xiao}, \citenamefont {Yao}, \citenamefont {Xu},\ and\
  \citenamefont {Yao}}]{Liu_electronic_2015}%
  \BibitemOpen
  \bibfield  {author} {\bibinfo {author} {\bibfnamefont {G.-B.}\ \bibnamefont
  {Liu}}, \bibinfo {author} {\bibfnamefont {D.}~\bibnamefont {Xiao}}, \bibinfo
  {author} {\bibfnamefont {Y.}~\bibnamefont {Yao}}, \bibinfo {author}
  {\bibfnamefont {X.}~\bibnamefont {Xu}},\ and\ \bibinfo {author}
  {\bibfnamefont {W.}~\bibnamefont {Yao}},\ }\bibfield  {title} {\bibinfo
  {title} {Electronic structures and theoretical modelling of two-dimensional
  group-vib transition metal dichalcogenides},\ }\href
  {https://doi.org/10.1039/C4CS00301B} {\bibfield  {journal} {\bibinfo
  {journal} {Chem. Soc. Rev.}\ }\textbf {\bibinfo {volume} {44}},\ \bibinfo
  {pages} {2643} (\bibinfo {year} {2015})}\BibitemShut {NoStop}%
\bibitem [{\citenamefont {Fang}\ \emph {et~al.}(2012)\citenamefont {Fang},
  \citenamefont {Gilbert},\ and\ \citenamefont {Bernevig}}]{Chen_bulk_2012}%
  \BibitemOpen
  \bibfield  {author} {\bibinfo {author} {\bibfnamefont {C.}~\bibnamefont
  {Fang}}, \bibinfo {author} {\bibfnamefont {M.~J.}\ \bibnamefont {Gilbert}},\
  and\ \bibinfo {author} {\bibfnamefont {B.~A.}\ \bibnamefont {Bernevig}},\
  }\bibfield  {title} {\bibinfo {title} {Bulk topological invariants in
  noninteracting point group symmetric insulators},\ }\href
  {https://doi.org/10.1103/PhysRevB.86.115112} {\bibfield  {journal} {\bibinfo
  {journal} {Phys. Rev. B}\ }\textbf {\bibinfo {volume} {86}},\ \bibinfo
  {pages} {115112} (\bibinfo {year} {2012})}\BibitemShut {NoStop}%
\bibitem [{\citenamefont {Haldane}(1988)}]{Haldane1988Model}%
  \BibitemOpen
  \bibfield  {author} {\bibinfo {author} {\bibfnamefont {F.~D.~M.}\
  \bibnamefont {Haldane}},\ }\bibfield  {title} {\bibinfo {title} {Model for a
  quantum hall effect without landau levels: Condensed-matter realization of
  the "parity anomaly"},\ }\href {https://doi.org/10.1103/PhysRevLett.61.2015}
  {\bibfield  {journal} {\bibinfo  {journal} {Phys. Rev. Lett.}\ }\textbf
  {\bibinfo {volume} {61}},\ \bibinfo {pages} {2015} (\bibinfo {year}
  {1988})}\BibitemShut {NoStop}%
\bibitem [{tri()}]{triangular}%
  \BibitemOpen
  \href@noop {} {}\bibinfo {note} {In triangular lattice, there is one site per
  unit cell and the $\mathcal{PT}$ and $\mathcal{C}_3$ symmetries can not
  stabilize any Dirac cone because there is only one single band.}\BibitemShut
  {Stop}%
\bibitem [{\citenamefont {Meckbach}\ \emph {et~al.}(2020)\citenamefont
  {Meckbach}, \citenamefont {Hader}, \citenamefont {Huttner}, \citenamefont
  {Neuhaus}, \citenamefont {Steiner}, \citenamefont {Stroucken}, \citenamefont
  {Moloney},\ and\ \citenamefont {Koch}}]{Meckbach_ultrafast_2020}%
  \BibitemOpen
  \bibfield  {author} {\bibinfo {author} {\bibfnamefont {L.}~\bibnamefont
  {Meckbach}}, \bibinfo {author} {\bibfnamefont {J.}~\bibnamefont {Hader}},
  \bibinfo {author} {\bibfnamefont {U.}~\bibnamefont {Huttner}}, \bibinfo
  {author} {\bibfnamefont {J.}~\bibnamefont {Neuhaus}}, \bibinfo {author}
  {\bibfnamefont {J.~T.}\ \bibnamefont {Steiner}}, \bibinfo {author}
  {\bibfnamefont {T.}~\bibnamefont {Stroucken}}, \bibinfo {author}
  {\bibfnamefont {J.~V.}\ \bibnamefont {Moloney}},\ and\ \bibinfo {author}
  {\bibfnamefont {S.~W.}\ \bibnamefont {Koch}},\ }\bibfield  {title} {\bibinfo
  {title} {Ultrafast band-gap renormalization and build-up of optical gain in
  monolayer ${\mathrm{mote}}_{2}$},\ }\href
  {https://doi.org/10.1103/PhysRevB.101.075401} {\bibfield  {journal} {\bibinfo
   {journal} {Phys. Rev. B}\ }\textbf {\bibinfo {volume} {101}},\ \bibinfo
  {pages} {075401} (\bibinfo {year} {2020})}\BibitemShut {NoStop}%
\bibitem [{\citenamefont {Mounet}\ \emph {et~al.}(2018)\citenamefont {Mounet},
  \citenamefont {Gibertini}, \citenamefont {Schwaller}, \citenamefont {Campi},
  \citenamefont {Merkys}, \citenamefont {Marrazzo}, \citenamefont {Sohier},
  \citenamefont {Castelli}, \citenamefont {Cepellotti}, \citenamefont {Pizzi},\
  and\ \citenamefont {Marzari}}]{Mounet_two_2018}%
  \BibitemOpen
  \bibfield  {author} {\bibinfo {author} {\bibfnamefont {N.}~\bibnamefont
  {Mounet}}, \bibinfo {author} {\bibfnamefont {M.}~\bibnamefont {Gibertini}},
  \bibinfo {author} {\bibfnamefont {P.}~\bibnamefont {Schwaller}}, \bibinfo
  {author} {\bibfnamefont {D.}~\bibnamefont {Campi}}, \bibinfo {author}
  {\bibfnamefont {A.}~\bibnamefont {Merkys}}, \bibinfo {author} {\bibfnamefont
  {A.}~\bibnamefont {Marrazzo}}, \bibinfo {author} {\bibfnamefont
  {T.}~\bibnamefont {Sohier}}, \bibinfo {author} {\bibfnamefont {I.~E.}\
  \bibnamefont {Castelli}}, \bibinfo {author} {\bibfnamefont {A.}~\bibnamefont
  {Cepellotti}}, \bibinfo {author} {\bibfnamefont {G.}~\bibnamefont {Pizzi}},\
  and\ \bibinfo {author} {\bibfnamefont {N.}~\bibnamefont {Marzari}},\
  }\bibfield  {title} {\bibinfo {title} {Two-dimensional materials from
  high-throughput computational exfoliation of experimentally known
  compounds},\ }\href {https://doi.org/10.1038/s41565-017-0035-5} {\bibfield
  {journal} {\bibinfo  {journal} {Nature nanotechnology}\ }\textbf {\bibinfo
  {volume} {13}},\ \bibinfo {pages} {246} (\bibinfo {year} {2018})}\BibitemShut
  {NoStop}%
\bibitem [{sup()}]{supplement}%
  \BibitemOpen
  \href@noop {} {}\bibinfo {note} {See Supplemental Materials for (i) Wannier
  orbitals of moir\'e minibands, (ii) comparison between the two continuum
  models in the trivial phase, (iii) local stacking configurations in AA- and
  AB-stacked TMD heterobilayers, (IV) details of Hartree-Fock calculations, (V)
  topological phase protected by the $\mathcal{C}_4$ symmetry of moir\'e
  potential, and (VI) potential realizations of massive Dirac fermions confined
  in a moir\'e potential.}\BibitemShut {Stop}%
\bibitem [{\citenamefont {Zhang}\ \emph {et~al.}(2021)\citenamefont {Zhang},
  \citenamefont {Devakul},\ and\ \citenamefont {Fu}}]{Zhang_spin_2021}%
  \BibitemOpen
  \bibfield  {author} {\bibinfo {author} {\bibfnamefont {Y.}~\bibnamefont
  {Zhang}}, \bibinfo {author} {\bibfnamefont {T.}~\bibnamefont {Devakul}},\
  and\ \bibinfo {author} {\bibfnamefont {L.}~\bibnamefont {Fu}},\ }\bibfield
  {title} {\bibinfo {title} {Spin-textured chern bands in ab-stacked transition
  metal dichalcogenide bilayers},\ }\bibfield  {journal} {\bibinfo  {journal}
  {Proceedings of the National Academy of Sciences}\ }\textbf {\bibinfo
  {volume} {118}},\ \href {https://doi.org/10.1073/pnas.2112673118}
  {10.1073/pnas.2112673118} (\bibinfo {year} {2021}),\ \Eprint
  {https://arxiv.org/abs/https://www.pnas.org/content/118/36/e2112673118.full.pdf}
  {https://www.pnas.org/content/118/36/e2112673118.full.pdf} \BibitemShut
  {NoStop}%
\bibitem [{\citenamefont {Zhang}\ \emph {et~al.}(2017)\citenamefont {Zhang},
  \citenamefont {Chuu}, \citenamefont {Ren}, \citenamefont {Li}, \citenamefont
  {Li}, \citenamefont {Jin}, \citenamefont {Chou},\ and\ \citenamefont
  {Shih}}]{Zhang_interlayer_2017}%
  \BibitemOpen
  \bibfield  {author} {\bibinfo {author} {\bibfnamefont {C.}~\bibnamefont
  {Zhang}}, \bibinfo {author} {\bibfnamefont {C.-P.}\ \bibnamefont {Chuu}},
  \bibinfo {author} {\bibfnamefont {X.}~\bibnamefont {Ren}}, \bibinfo {author}
  {\bibfnamefont {M.-Y.}\ \bibnamefont {Li}}, \bibinfo {author} {\bibfnamefont
  {L.-J.}\ \bibnamefont {Li}}, \bibinfo {author} {\bibfnamefont
  {C.}~\bibnamefont {Jin}}, \bibinfo {author} {\bibfnamefont {M.-Y.}\
  \bibnamefont {Chou}},\ and\ \bibinfo {author} {\bibfnamefont {C.-K.}\
  \bibnamefont {Shih}},\ }\bibfield  {title} {\bibinfo {title} {Interlayer
  couplings, moir{\'e} patterns, and 2d electronic superlattices in
  mos$_2$/wse$_2$ hetero-bilayers},\ }\href
  {https://doi.org/10.1126/sciadv.1601459} {\bibfield  {journal} {\bibinfo
  {journal} {Science advances}\ }\textbf {\bibinfo {volume} {3}},\ \bibinfo
  {pages} {e1601459} (\bibinfo {year} {2017})}\BibitemShut {NoStop}%
\bibitem [{\citenamefont {Shabani}\ \emph {et~al.}(2021)\citenamefont
  {Shabani}, \citenamefont {Halbertal}, \citenamefont {Wu}, \citenamefont
  {Chen}, \citenamefont {Liu}, \citenamefont {Hone}, \citenamefont {Yao},
  \citenamefont {Basov}, \citenamefont {Zhu},\ and\ \citenamefont
  {Pasupathy}}]{Shabani_deep_2021}%
  \BibitemOpen
  \bibfield  {author} {\bibinfo {author} {\bibfnamefont {S.}~\bibnamefont
  {Shabani}}, \bibinfo {author} {\bibfnamefont {D.}~\bibnamefont {Halbertal}},
  \bibinfo {author} {\bibfnamefont {W.}~\bibnamefont {Wu}}, \bibinfo {author}
  {\bibfnamefont {M.}~\bibnamefont {Chen}}, \bibinfo {author} {\bibfnamefont
  {S.}~\bibnamefont {Liu}}, \bibinfo {author} {\bibfnamefont {J.}~\bibnamefont
  {Hone}}, \bibinfo {author} {\bibfnamefont {W.}~\bibnamefont {Yao}}, \bibinfo
  {author} {\bibfnamefont {D.~N.}\ \bibnamefont {Basov}}, \bibinfo {author}
  {\bibfnamefont {X.}~\bibnamefont {Zhu}},\ and\ \bibinfo {author}
  {\bibfnamefont {A.~N.}\ \bibnamefont {Pasupathy}},\ }\bibfield  {title}
  {\bibinfo {title} {Deep moir{\'e} potentials in twisted transition metal
  dichalcogenide bilayers},\ }\href
  {https://doi.org/10.1038/s41567-021-01174-7} {\bibfield  {journal} {\bibinfo
  {journal} {Nature Physics}\ }\textbf {\bibinfo {volume} {17}},\ \bibinfo
  {pages} {720} (\bibinfo {year} {2021})}\BibitemShut {NoStop}%
\bibitem [{\citenamefont {Geng}\ \emph {et~al.}(2020)\citenamefont {Geng},
  \citenamefont {Wang}, \citenamefont {Liu}, \citenamefont {Ohno},\ and\
  \citenamefont {Nara}}]{Geng_moire_2020}%
  \BibitemOpen
  \bibfield  {author} {\bibinfo {author} {\bibfnamefont {W.~T.}\ \bibnamefont
  {Geng}}, \bibinfo {author} {\bibfnamefont {V.}~\bibnamefont {Wang}}, \bibinfo
  {author} {\bibfnamefont {Y.~C.}\ \bibnamefont {Liu}}, \bibinfo {author}
  {\bibfnamefont {T.}~\bibnamefont {Ohno}},\ and\ \bibinfo {author}
  {\bibfnamefont {J.}~\bibnamefont {Nara}},\ }\bibfield  {title} {\bibinfo
  {title} {Moiré potential, lattice corrugation, and band gap spatial
  variation in a twist-free mote2/mos2 heterobilayer},\ }\href
  {https://doi.org/10.1021/acs.jpclett.0c00605} {\bibfield  {journal} {\bibinfo
   {journal} {The Journal of Physical Chemistry Letters}\ }\textbf {\bibinfo
  {volume} {11}},\ \bibinfo {pages} {2637} (\bibinfo {year} {2020})},\ \Eprint
  {https://arxiv.org/abs/https://doi.org/10.1021/acs.jpclett.0c00605}
  {https://doi.org/10.1021/acs.jpclett.0c00605} \BibitemShut {NoStop}%
\bibitem [{\citenamefont {Wu}\ \emph {et~al.}(2017)\citenamefont {Wu},
  \citenamefont {Lovorn},\ and\ \citenamefont
  {MacDonald}}]{Wu_topological_2017}%
  \BibitemOpen
  \bibfield  {author} {\bibinfo {author} {\bibfnamefont {F.}~\bibnamefont
  {Wu}}, \bibinfo {author} {\bibfnamefont {T.}~\bibnamefont {Lovorn}},\ and\
  \bibinfo {author} {\bibfnamefont {A.~H.}\ \bibnamefont {MacDonald}},\
  }\bibfield  {title} {\bibinfo {title} {Topological exciton bands in moir\'e
  heterojunctions},\ }\href {https://doi.org/10.1103/PhysRevLett.118.147401}
  {\bibfield  {journal} {\bibinfo  {journal} {Phys. Rev. Lett.}\ }\textbf
  {\bibinfo {volume} {118}},\ \bibinfo {pages} {147401} (\bibinfo {year}
  {2017})}\BibitemShut {NoStop}%
\bibitem [{\citenamefont {Li}\ \emph {et~al.}(2021{\natexlab{d}})\citenamefont
  {Li}, \citenamefont {Jiang}, \citenamefont {Li}, \citenamefont {Zhang},
  \citenamefont {Kang}, \citenamefont {Zhu}, \citenamefont {Watanabe},
  \citenamefont {Taniguchi}, \citenamefont {Chowdhury}, \citenamefont {Fu}, ,
  \citenamefont {Shan},\ and\ \citenamefont {Mak}}]{Li_continuous_2021}%
  \BibitemOpen
  \bibfield  {author} {\bibinfo {author} {\bibfnamefont {T.}~\bibnamefont
  {Li}}, \bibinfo {author} {\bibfnamefont {S.}~\bibnamefont {Jiang}}, \bibinfo
  {author} {\bibfnamefont {L.}~\bibnamefont {Li}}, \bibinfo {author}
  {\bibfnamefont {Y.}~\bibnamefont {Zhang}}, \bibinfo {author} {\bibfnamefont
  {K.}~\bibnamefont {Kang}}, \bibinfo {author} {\bibfnamefont {J.}~\bibnamefont
  {Zhu}}, \bibinfo {author} {\bibfnamefont {K.}~\bibnamefont {Watanabe}},
  \bibinfo {author} {\bibfnamefont {T.}~\bibnamefont {Taniguchi}}, \bibinfo
  {author} {\bibfnamefont {D.}~\bibnamefont {Chowdhury}}, \bibinfo {author}
  {\bibfnamefont {L.}~\bibnamefont {Fu}}, , \bibinfo {author} {\bibfnamefont
  {J.}~\bibnamefont {Shan}},\ and\ \bibinfo {author} {\bibfnamefont {K.~F.}\
  \bibnamefont {Mak}},\ }\bibfield  {title} {\bibinfo {title} {Continuous mott
  transition in semiconductor moiré superlattices},\ }\href
  {https://www.nature.com/articles/s41586-021-03853-0} {\bibfield  {journal}
  {\bibinfo  {journal} {Nature}\ }\textbf {\bibinfo {volume} {597}},\ \bibinfo
  {pages} {350} (\bibinfo {year} {2021}{\natexlab{d}})}\BibitemShut {NoStop}%
\bibitem [{vp()}]{vp}%
  \BibitemOpen
  \href@noop {} {}\bibinfo {note} {There are two degenerate ground states with
  the top valence band from either the $+$K or $-$K valley is empty at
  $\nu=1$.}\BibitemShut {Stop}%
\bibitem [{\citenamefont {Tong}\ \emph {et~al.}(2017)\citenamefont {Tong},
  \citenamefont {Yu}, \citenamefont {Zhu}, \citenamefont {Wang}, \citenamefont
  {Xu},\ and\ \citenamefont {Yao}}]{tong2017topological}%
  \BibitemOpen
  \bibfield  {author} {\bibinfo {author} {\bibfnamefont {Q.}~\bibnamefont
  {Tong}}, \bibinfo {author} {\bibfnamefont {H.}~\bibnamefont {Yu}}, \bibinfo
  {author} {\bibfnamefont {Q.}~\bibnamefont {Zhu}}, \bibinfo {author}
  {\bibfnamefont {Y.}~\bibnamefont {Wang}}, \bibinfo {author} {\bibfnamefont
  {X.}~\bibnamefont {Xu}},\ and\ \bibinfo {author} {\bibfnamefont
  {W.}~\bibnamefont {Yao}},\ }\bibfield  {title} {\bibinfo {title} {Topological
  mosaics in moir{\'e} superlattices of van der waals heterobilayers},\ }\href
  {https://www.nature.com/articles/nphys3968} {\bibfield  {journal} {\bibinfo
  {journal} {Nature Physics}\ }\textbf {\bibinfo {volume} {13}},\ \bibinfo
  {pages} {356} (\bibinfo {year} {2017})}\BibitemShut {NoStop}%
\bibitem [{\citenamefont {Xie}\ \emph {et~al.}(2022)\citenamefont {Xie},
  \citenamefont {Zhang}, \citenamefont {Hu}, \citenamefont {Mak},\ and\
  \citenamefont {Law}}]{Xie_theory_2021}%
  \BibitemOpen
  \bibfield  {author} {\bibinfo {author} {\bibfnamefont {Y.-M.}\ \bibnamefont
  {Xie}}, \bibinfo {author} {\bibfnamefont {C.-P.}\ \bibnamefont {Zhang}},
  \bibinfo {author} {\bibfnamefont {J.-X.}\ \bibnamefont {Hu}}, \bibinfo
  {author} {\bibfnamefont {K.~F.}\ \bibnamefont {Mak}},\ and\ \bibinfo {author}
  {\bibfnamefont {K.~T.}\ \bibnamefont {Law}},\ }\bibfield  {title} {\bibinfo
  {title} {Valley-polarized quantum anomalous hall state in moir\'e
  ${\mathrm{mote}}_{2}/{\mathrm{wse}}_{2}$ heterobilayers},\ }\href
  {https://doi.org/10.1103/PhysRevLett.128.026402} {\bibfield  {journal}
  {\bibinfo  {journal} {Phys. Rev. Lett.}\ }\textbf {\bibinfo {volume} {128}},\
  \bibinfo {pages} {026402} (\bibinfo {year} {2022})}\BibitemShut {NoStop}%
\bibitem [{\citenamefont {Xu}\ \emph {et~al.}(2021)\citenamefont {Xu},
  \citenamefont {Horn}, \citenamefont {Zhu}, \citenamefont {Tang},
  \citenamefont {Ma}, \citenamefont {Li}, \citenamefont {Liu}, \citenamefont
  {Watanabe}, \citenamefont {Taniguchi}, \citenamefont {Hone} \emph
  {et~al.}}]{xu2021creation}%
  \BibitemOpen
  \bibfield  {author} {\bibinfo {author} {\bibfnamefont {Y.}~\bibnamefont
  {Xu}}, \bibinfo {author} {\bibfnamefont {C.}~\bibnamefont {Horn}}, \bibinfo
  {author} {\bibfnamefont {J.}~\bibnamefont {Zhu}}, \bibinfo {author}
  {\bibfnamefont {Y.}~\bibnamefont {Tang}}, \bibinfo {author} {\bibfnamefont
  {L.}~\bibnamefont {Ma}}, \bibinfo {author} {\bibfnamefont {L.}~\bibnamefont
  {Li}}, \bibinfo {author} {\bibfnamefont {S.}~\bibnamefont {Liu}}, \bibinfo
  {author} {\bibfnamefont {K.}~\bibnamefont {Watanabe}}, \bibinfo {author}
  {\bibfnamefont {T.}~\bibnamefont {Taniguchi}}, \bibinfo {author}
  {\bibfnamefont {J.~C.}\ \bibnamefont {Hone}}, \emph {et~al.},\ }\bibfield
  {title} {\bibinfo {title} {Creation of moir{\'e} bands in a monolayer
  semiconductor by spatially periodic dielectric screening},\ }\href
  {https://www.nature.com/articles/s41563-020-00888-y} {\bibfield  {journal}
  {\bibinfo  {journal} {Nature Materials}\ }\textbf {\bibinfo {volume} {20}},\
  \bibinfo {pages} {645} (\bibinfo {year} {2021})}\BibitemShut {NoStop}%
\bibitem [{\citenamefont {Nielsen}\ and\ \citenamefont
  {Ninomiya}(1981)}]{nielsen1983adler}%
  \BibitemOpen
  \bibfield  {author} {\bibinfo {author} {\bibfnamefont {H.~B.}\ \bibnamefont
  {Nielsen}}\ and\ \bibinfo {author} {\bibfnamefont {M.}~\bibnamefont
  {Ninomiya}},\ }\bibfield  {title} {\bibinfo {title} {A no-go theorem for
  regularizing chiral fermions},\ }\href
  {https://www.sciencedirect.com/science/article/pii/0370269381910261?via\%3Dihub}
  {\bibfield  {journal} {\bibinfo  {journal} {Physics Letters B}\ }\textbf
  {\bibinfo {volume} {105}},\ \bibinfo {pages} {219} (\bibinfo {year}
  {1981})}\BibitemShut {NoStop}%
\bibitem [{\citenamefont {Cano}\ \emph {et~al.}(2021)\citenamefont {Cano},
  \citenamefont {Fang}, \citenamefont {Pixley},\ and\ \citenamefont
  {Wilson}}]{PhysRevB.103.155157}%
  \BibitemOpen
  \bibfield  {author} {\bibinfo {author} {\bibfnamefont {J.}~\bibnamefont
  {Cano}}, \bibinfo {author} {\bibfnamefont {S.}~\bibnamefont {Fang}}, \bibinfo
  {author} {\bibfnamefont {J.~H.}\ \bibnamefont {Pixley}},\ and\ \bibinfo
  {author} {\bibfnamefont {J.~H.}\ \bibnamefont {Wilson}},\ }\bibfield  {title}
  {\bibinfo {title} {Moir\'e superlattice on the surface of a topological
  insulator},\ }\href {https://doi.org/10.1103/PhysRevB.103.155157} {\bibfield
  {journal} {\bibinfo  {journal} {Phys. Rev. B}\ }\textbf {\bibinfo {volume}
  {103}},\ \bibinfo {pages} {155157} (\bibinfo {year} {2021})}\BibitemShut
  {NoStop}%
\bibitem [{\citenamefont {Wang}\ \emph
  {et~al.}(2021{\natexlab{b}})\citenamefont {Wang}, \citenamefont {Yuan},\ and\
  \citenamefont {Fu}}]{PhysRevX.11.021024}%
  \BibitemOpen
  \bibfield  {author} {\bibinfo {author} {\bibfnamefont {T.}~\bibnamefont
  {Wang}}, \bibinfo {author} {\bibfnamefont {N.~F.~Q.}\ \bibnamefont {Yuan}},\
  and\ \bibinfo {author} {\bibfnamefont {L.}~\bibnamefont {Fu}},\ }\bibfield
  {title} {\bibinfo {title} {Moir\'e surface states and enhanced
  superconductivity in topological insulators},\ }\href
  {https://doi.org/10.1103/PhysRevX.11.021024} {\bibfield  {journal} {\bibinfo
  {journal} {Phys. Rev. X}\ }\textbf {\bibinfo {volume} {11}},\ \bibinfo
  {pages} {021024} (\bibinfo {year} {2021}{\natexlab{b}})}\BibitemShut
  {NoStop}%
\end{thebibliography}%


\begin{thebibliography}{11}
\bibitem{Koshino_Maximally_2018s} M. Koshino, N. F. Q. Yuan, T. Koretsune, M. Ochi, K. Kuroki, and L. Fu, Maximally localized wannier orbitals and the extended Hubbard model for twisted bilayer graphene, Phys. Rev. X {\bf 8}, 031087 (2018).

\bibitem{Kang_Symmetry_2018s} J. Kang and O. Vafek, Symmetry, maximally localized Wannier states, and a low-energy model for twisted bilayer graphene narrow bands, Phys. Rev. X {\bf 8}, 031088 (2018).

\bibitem{Wu2019Topologicals} F. Wu, T. Lovorn, E. Tutuc, I. Martin, and A. H. MacDonald, Topological insulators in twisted transition metal dichalcogenide homobilayers, Phys. Rev. Lett. {\bf 122}, 086402 (2019).

\bibitem{sun2012topologicals} K. Sun, W. V. Liu, A. Hemmerich, and S. D. Sarma, Topological semimetal in a fermionic optical lattice, Nature Physics {\bf 8},
67 (2012).

\bibitem{Qi2008Topologicals} X.-L. Qi, T. L. Hughes, and S.-C. Zhang, Topological field theory of time-reversal invariant insulators, Phys. Rev. B 78, 195424 (2008).

\bibitem{mogi2017magnetics} M. Mogi, M. Kawamura, R. Yoshimi, A. Tsukazaki, Y. Kozuka, N. Shirakawa, K. Takahashi, M. Kawasaki, and Y. Tokura, A magnetic heterostructure of topological insulators as a candidate for an axion insulator, Nature materials 16, 516 (2017).

\bibitem{Xiao2018Realizations} D. Xiao, J. Jiang, J.-H. Shin, W. Wang, F. Wang, Y.-F. Zhao, C. Liu, W. Wu, M. H. W. Chan, N. Samarth, and C.-Z. Chang, Realization of the axion insulator state in quantum anomalous hall sandwich heterostructures, Phys. Rev. Lett. 120, 056801 (2018).

\bibitem{xu2021creations} Y. Xu, C. Horn, J. Zhu, Y. Tang, L. Ma, L. Li, S. Liu, K.Watanabe, T. Taniguchi, J. C. Hone, et al., Creation of moir´e bands in a monolayer semiconductor by spatially periodic dielectric screening, Nature Materials 20, 645 (2021).

\end{thebibliography}

\widetext
\clearpage
\begin{center}
\textbf{\large Supplemental Material: Massive Dirac fermions in moir\'e superlattices: a route toward correlated Chern insulators}
\end{center}
\setcounter{equation}{0}
\setcounter{figure}{0}
\setcounter{table}{0}
\setcounter{page}{1}
\makeatletter
\renewcommand{\theequation}{S\arabic{equation}}
\renewcommand{\thefigure}{S\arabic{figure}}
\renewcommand{\bibnumfmt}[1]{[S#1]}
\renewcommand{\citenumfont}[1]{S#1}

\twocolumngrid

\section{I. Wannier orbitals of moir{\'e} minibands}

We compare the Wannier orbitals of moir{\'e} minibands from the two different continuum models in Eq. (1) and (2) of the main text. In Fig \ref{figs3}, we extract the top four minibands from Fig. 2(a) of the main text. The blue solid bands are from the continuum model Eq. (1) and have valley Chern numbers $C_+=\mp1$, while the red dashed bands are from the continuum model Eq. (2) and have zero valley Chern numbers.

For the top trivial miniband given by the continuum model Eq. (2), the Wannier orbital can be obtained from the Fourier transform of the Bloch wave as
\be \label{W1}
   W_{n,\tau}(\bm{r}-\bm{R})= \frac{1}{\sqrt{N}} \sum_{\bm{k}} e^{-i\bm{k}\cdot\bm{R}} \psi_{n,\bm{k},\tau}(\bm{r}),
\ee
where $\bm{R}$ is the moir\'e superlattice vector and $N$ is number of moir\'e unit cells. In Fig. {\ref{figs4}}, we show the Wannier orbital of the top trivial band for $\bm{R}=0$. Here we choose a gauge in which the Bloch wave is real at $\rm{H_M^M}$ at the origin. The Wannier center at $\rm{H_M^M}$ forms a triangular lattice, as shown in Fig. \ref{figs4}. Therefore, the Coulomb interaction in the trivial moir\'e miniband can simulate the Hubbard model on a triangular lattice.

\begin{figure}[b]
\includegraphics[width=7 cm]{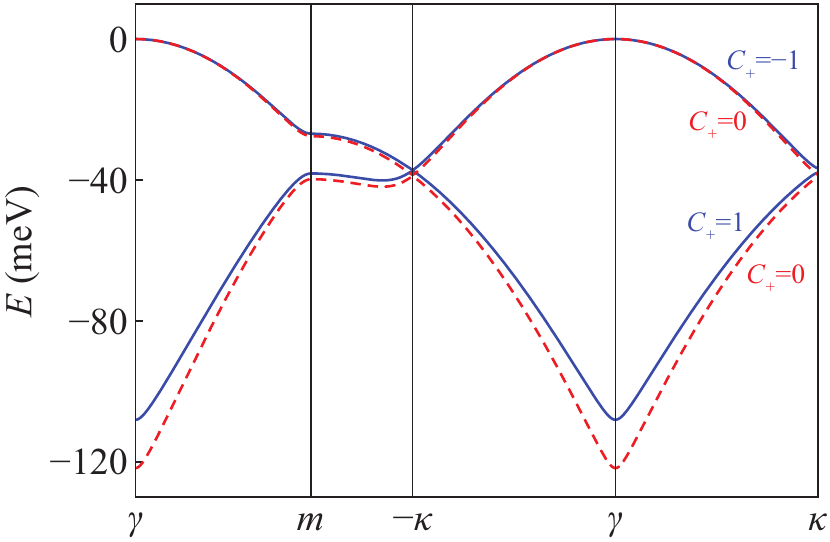}
\caption{The top two valence bands extracted from Fig. 2(a) of the main text. The blue solid and red dashed bands are from the continuum models in Eq. (1) and (2) of the main text, respectively. The valley Chern numbers of the valence bands are marked in the figure.}
\label{figs3}
\end{figure}

\begin{figure}
\includegraphics[width=7 cm]{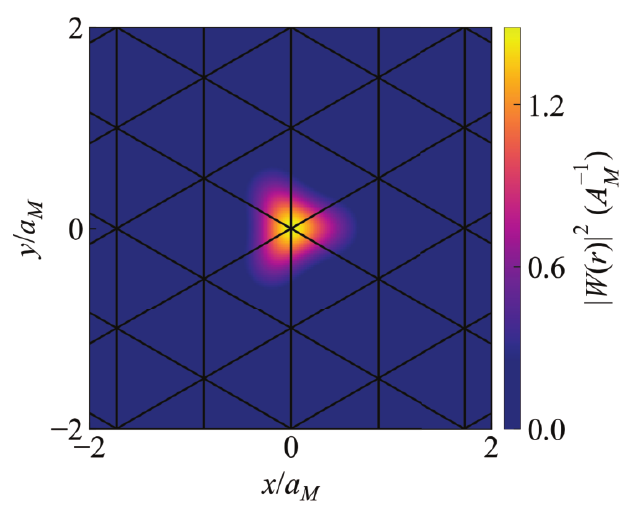}
\caption{The Wannier orbital of the top valence band (red dashed line) in Fig. \ref{figs3} whose valley Chern number is $C_+=0$. The Wannier center at $\rm{H_M^M}$ is connected by the black triangular lattice.}
\label{figs4}
\end{figure}

On the other hand, the moir\'e minibands are topologically nontrivial according to the continuum model  Eq. (1), and Eq. (\ref{W1}) is inapplicable to the Chern bands due to the Wannier obstruction. In this case, we need to consider the two topological minibands with opposite Chern numbers in Fig. \ref{figs3}. For the two-band system, the Wannier orbitals can be prepared as
\be
\begin{split}
&W_{1,\tau}(\bm{r}-\bm{R})= \frac{1}{\sqrt{2N}} \sum_{\bm{k}} e^{-i\bm{k}\cdot\bm{R}} \left[\psi_{1,\bm{k},\tau}(\bm{r}) + \psi_{2,\bm{k},\tau}(\bm{r})\right], \\
&W_{2,\tau}(\bm{r}-\bm{R})= \frac{1}{\sqrt{2N}} \sum_{\bm{k}} e^{-i\bm{k}\cdot\bm{R}}e^{i\theta_{1,\bm{k},\tau}} \left[\psi_{1,\bm{k},\tau}(\bm{r}) - \psi_{2,\bm{k},\tau}(\bm{r})\right],
\end{split}
\ee
where $\psi_{1,\bm{k},\tau}(\bm{r})$ and $\psi_{2,\bm{k},\tau}(\bm{r})$ are the Bloch waves of the first and second valence band. 
For the Dirac model in Eq. (1), the Bloch wave is a two-component spinor, i.e., $\psi_{n,\bm{k},\tau}=(\psi_{n,\bm{k},\tau}^c,\psi_{n,\bm{k},\tau}^v)^\top$. We fix a gauge such that $\psi_{n,\bm{k},\tau}^c(\bm{r})$ is real at $\rm{H_M^M}$ at the origin. Then we perform another gauge transformation $e^{i\theta_{n,\bm{k},\tau}}\psi_{n,\bm{k},\tau}(\bm{r})$ such that $e^{i\theta_{n,\bm{k},\tau}}\psi_{n,\bm{k},\tau}^v(\bm{r})$ is real at $\rm{H_X^M}$ at $\bm{r}=(a_M/\sqrt{3},0)$. 
Note that $\rm{H_M^M}$ and $\rm{H_X^M}$ are the moir{\'e} potential minima for holes when $\phi\sim (2n+1)\pi/3$. For $\bm{R}=0$, the {squared amplitudes of the} two Wannier orbitals are shown in Figs. \ref{figs5} (a) and \ref{figs5}(b). It is easy to show that the two Wannier orbitals are orthogonal. The Wannier centers at $\rm{H_M^M}$ and $\rm{H_X^M}$ form a honeycomb lattice and each Wannier center is surrounded by three peaks of wavefunction amplitude maxima. The peculiar three-peak structure of the Wannier orbitals is similar to that in twisted bilayer graphene \cite{Koshino_Maximally_2018s,Kang_Symmetry_2018s}. The two moir\'e minibands with opposite Chern numbers together can realize the Haldane model as that proposed in TMD homobilayers \cite{Wu2019Topologicals}. Further considering the time-reversal counterparts from two distinct valleys as well as the Coulomb interaction, the system can simulate the Kane-Mele-Hubbard model.

\begin{figure}
\includegraphics[width=7 cm]{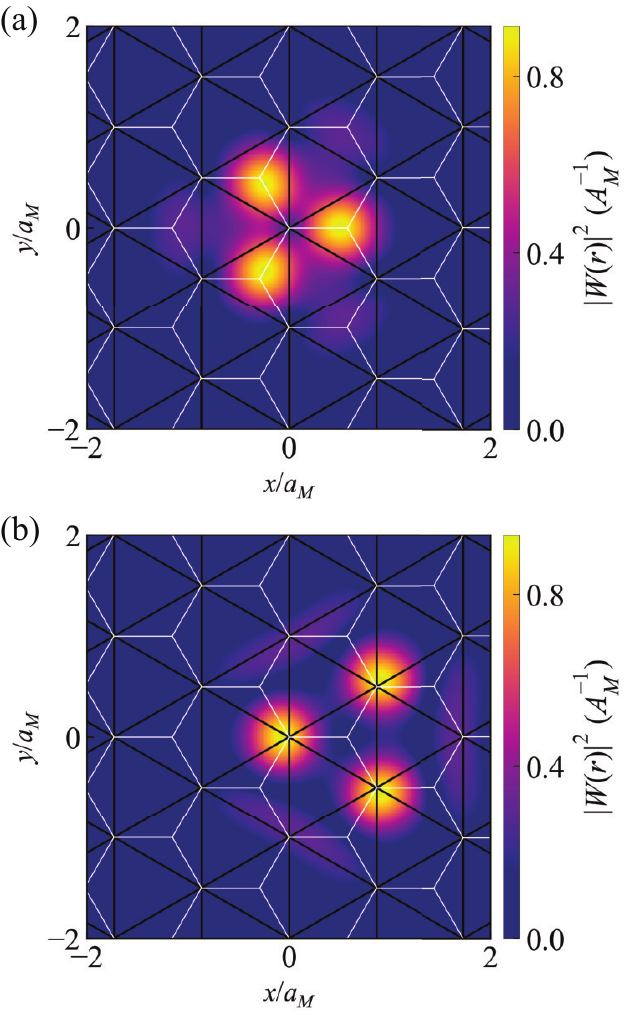}
\caption{(a) and (b) The Wannier orbitals for the top two valence bands (blue solid lines) in Fig. \ref{figs3} whose valley Chern numbers are $C_+=\mp1$. The Wannier centers at $\rm{H_M^M}$ in (a) and at $\rm{H_M^X}$ in (b) correspond to the A and B sublattices of the white honeycomb lattice.}
\label{figs5}
\end{figure}

\section{II. Comparison between the two continuum models in the trivial phase}

In this section, we focus on the topologically trivial phase and compare the energy bands and Berry curvatures from the two continuum models. Here the blue solid and red dashed bands are from Eqs. (1) and (2) of the main text. We choose $\theta=1^\circ$ and $\phi=40^\circ$ in the trivial phase in Fig. 1(d), and set $V_0=8$ meV. The other model parameters are same as those specified in the main text. The moir\'e minibands from the two different models exhibit good agreement with each other, as shown in Fig. \ref{figs2}(a). The Berry curvatures of the top valence bands from Eqs. (1) and (2) are shown in Figs. \ref{figs2}(b) and \ref{figs2}(c), respectively. The two Berry curvature profiles look similar to each other. Nevertheless, the Berry curvature in \ref{figs2}(c) is strictly antisymmetric, i.e., $\Omega(\bm{k})=-\Omega(-\bm{k})$, due to the emergent TRS of Eq. (2) in the main text, while that in \ref{figs2}(b) is just approximately antisymmetric. Our results show that, deep inside the trivial phase, the continuum model Eq. (2) can be a good approximation to that in Eq. (1) in both energy band and Berry curvature.

\begin{figure}
\includegraphics[width=8.5 cm]{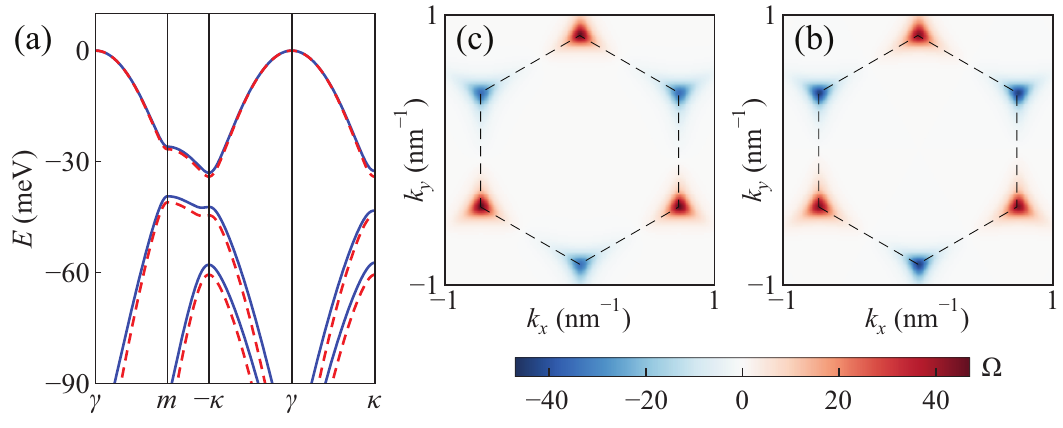}
\caption{(a) Valence bands of the MoTe$_2$/WSe$_2$ heterobilayer with $\theta=1^\circ$, $\phi=40^\circ$, and $V_0=8$ meV. The blue solid and red dashed bands are from the continuum models in Eqs. (1) and (2) of the main text. (b) and (c) Berry curvatures of the top blue and red bands in (a). The black dashed hexagon encloses the MBZ.}
\label{figs2}
\end{figure}

\section{III. Local stacking configurations in AA- and AB-stacked TMD heterobilayers}

The lattice structures of AA- and AB-stacked TMD heterobilayers are shown in Figs. \ref{figs1}(a) and \ref{figs1}(b), respectively. In the AA stacking (R stacking), the two layers have the same lattice orientation. There are three local stacking configurations $\rm{R_M^X}$, $\rm{R_M^M}$, and $\rm{R_X^M}$ where the atoms from the top and bottom layers are nearly aligned vertically, as shown in Fig. \ref{figs1}(a). Here M and X refer to the metal and chalcogen. In the AB stacking (H stacking), the bottom layer is rotated by $180^\circ$ with respect to the top layer. The three local stacking configurations $\rm{H_X^M}$, $\rm{H_M^M}$, and $\rm{H_X^X}$ with near interlayer alignment are shown in Fig. \ref{figs1}(b). In both cases, the atomic registry between the two different layers varies periodically due to the lattice mismatch and the three different local stacking configurations can be identified in each moir{\'e} unit cell.

\section{IV. Hartree-Fock calculations}

Under the standard Hartree-Fock approximation, the mean-field Hamiltonian reads
\begin{equation}
    H_{\rm MF} = \sum_{n,\bm{k},\tau}(E_{n,\bm{k},\tau}-\mu + V_{n,\bm{k},\tau}^{H} + V_{n,\bm{k},\tau}^{F}) c_{n,\bm{k},\tau}^\dagger c_{n,\bm{k},\tau},
\end{equation}
where the Hatree and Fock terms are
\begin{equation}
\begin{split}
    &V_{n,\bm{k},\tau}^{H}=\sum_{n',\bm{k}',\tau',\bm{q}}\frac{V_{\bm{q}}}{A}\Lambda_{n,n}^{(\tau)}(\bm{k},\bm{k},\bm{q})\Lambda_{n',n'}^{(\tau')}(\bm{k}',\bm{k}',-\bm{q}) \langle c_{n',\bm{k}',\tau'}^\dagger c_{n',\bm{k}',\tau'}\rangle, \\
    &V_{n,\bm{k},\tau}^{F}=-\sum_{n',\bm{k}',\bm{q}}\frac{V_{\bm{q}}}{A}\Lambda_{n,n'}^{(\tau)}(\bm{k},\bm{k}',\bm{q})\Lambda_{n',n}^{(\tau)}(\bm{k}',\bm{k},-\bm{q}) \langle c_{n',\bm{k}',\tau}^\dagger c_{n',\bm{k}',\tau}\rangle.
\end{split}
\end{equation}
The form factor $\Lambda_{n,n'}^{(\tau)}(\bm{k},\bm{k}',\bm{q})$ and the screened Coulomb potential $V_{\bm{q}}$ are defined in the main text. The mean-field Hamiltonian is solved self-consistently numerically. Because the bandwidth of the top valence band is larger than the gap to the next valence band, it is insufficient to consider only the top one. To guarantee the convergence of the top valence band, we retain the topmost six valence bands from each valley in the numerical calculations and  solve the mean-field Hamiltonian self-consistently at zero temperature.

\begin{figure}
\includegraphics[width=8.5 cm]{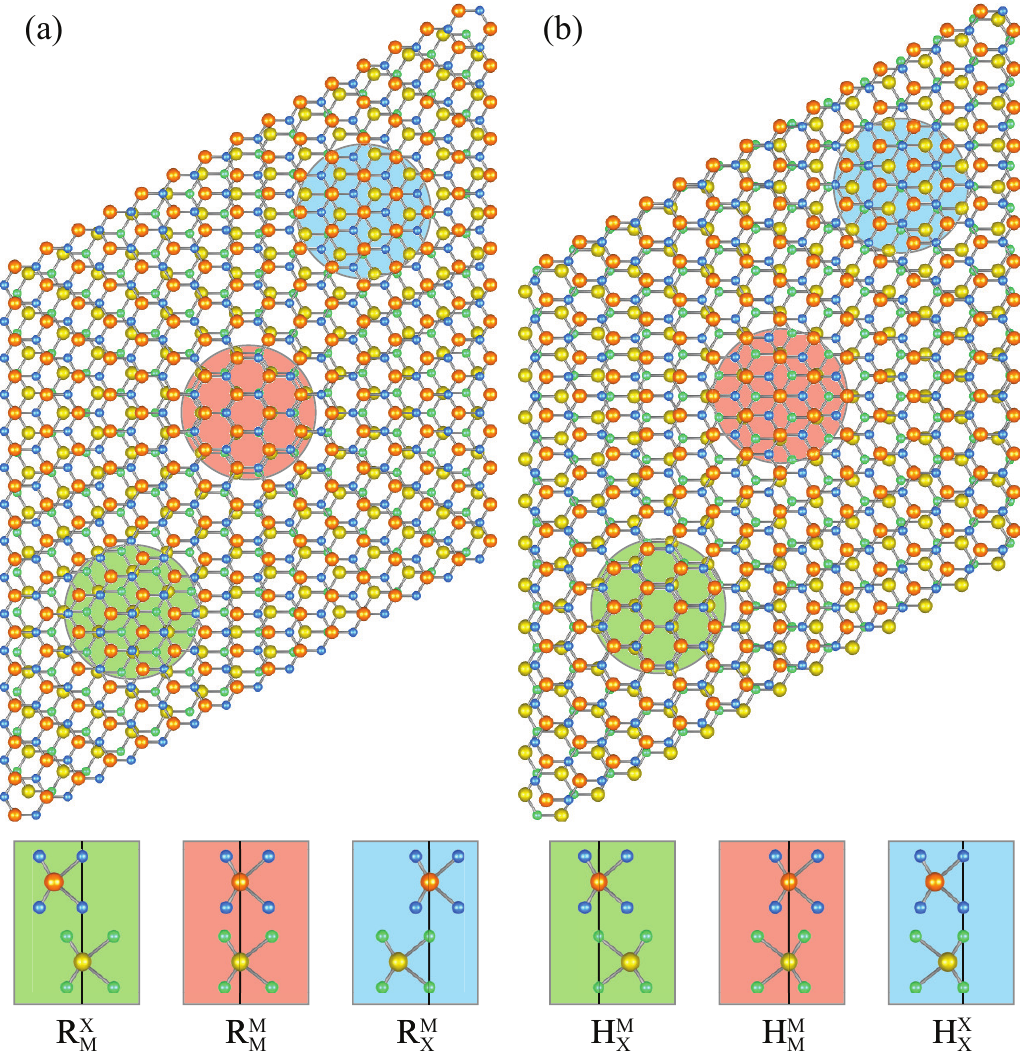}
\caption{(a) and (b) Lattice structure of AA- and AB-stacked TMD heterobilayer. The local stacking configurations $\rm{R_M^X}$, $\rm{R_M^M}$, $\rm{R_X^M}$, $\rm{H_X^M}$, $\rm{H_M^M}$, and $\rm{H_X^X}$ with near interlayer alignment are highlighted in the upper panels and the corresponding side views are shown in the lower panels. }
\label{figs1}
\end{figure}

\section{V. Topological phase protected by the $\mathcal{C}_4$ symmetry of moir\'e potential}

In the main text, we show that the topological phase can be protected by the symmetry of moir\'e potential and survives to arbitrarily larger Dirac band gap. This has been verified for the moir\'e potential whose potential minima for holes form a honeycomb lattice with $\mathcal{C}_3$ and $\mathcal{P}$ symmetries (namely the $\mathcal{C}_6$ symmetry). Now we consider a moir\'e potential 
\begin{equation}\label{square}
\begin{split}
    V(\bm{r}) = & 2V_0\left[\cos(Gx)+\cos(Gy)\right]\\
    & - 2V_1\left[ \cos(Gx+Gy) + \cos(Gx-Gy) \right],
\end{split}
\end{equation}
whose minima form a square lattice with $\mathcal{C}_4$ symmetry \cite{sun2012topologicals}, as shown in Fig. \ref{figs8} (a). Here $G=2\pi/a_M$ and  we set $V_0=8$ meV and $V_1=10$ meV (without loss of generality). There are two different sublattices that constitute a sublattice pesudospin. Then we substitute the moir\'e potential Eq. (\ref{square}) into the continuum models Eqs. (1) and (2), while all the other parameters are kept same as those used in the main text. For the free fermion coupled to the moir\'e potential, as described by Eq. (2), the $\mathcal{PT}$ and $\mathcal{C}_4$ symmetries stabilize a quadratic band-touching point at the high-symmetry point $m$ \cite{sun2012topologicals}, as shown in Fig. \ref{figs8}(b) and its insets. Here we plot only the valence bands from the $+$K valley, while those from the $-$K valley can be obtained by TRS. For the sublattice pesudospin, the $\mathcal{PT}$ symmetry prevents the appearance of a Dirac mass term that is proportional to $\sigma_z$ acting on the pesudospin space. Therefore, the band touching can occur and is further pinned at the $m$ point by the  $\mathcal{C}_4$ symmetry. The quadratic band-touching point carries a $2\pi$ Berry phase in contrast to the $\pi$ Berry phase associated with massless Dirac cone in honeycomb lattice.

When the massive Dirac fermion couples the moir\'e potential, as described by Eq. (1), the quadratic band-touching point is gapped out that gives rise to topological minibands, as shown in Fig. \ref{figs8}(b). In this case, the top valence band has a valley Chern number $C_+=-1$ and its Berry phase is shown in Fig. \ref{figs8}(c). The topologial miniband is enabled by the Dirac nature that breaks the $\mathcal{PT}$ symmetry as $\mathcal{PT}h_{\bm{k},\tau}(\mathcal{PT})^{-1}=h_{\bm{k},-\tau}$ where $h_{\bm{k},\tau}$ is the Dirac Hamiltonian in Eq. (1). Because the $\mathcal{PT}$ symmetry breaking is independent on the detailed model parameters of the Dirac Hamiltonian (as long as they are nonzero), the topological phase is protected by the $\mathcal{C}_4$ symmetry of moir\'e potential  and survives to arbitrarily large Dirac band gap.

\begin{figure*}
\includegraphics[width=14 cm]{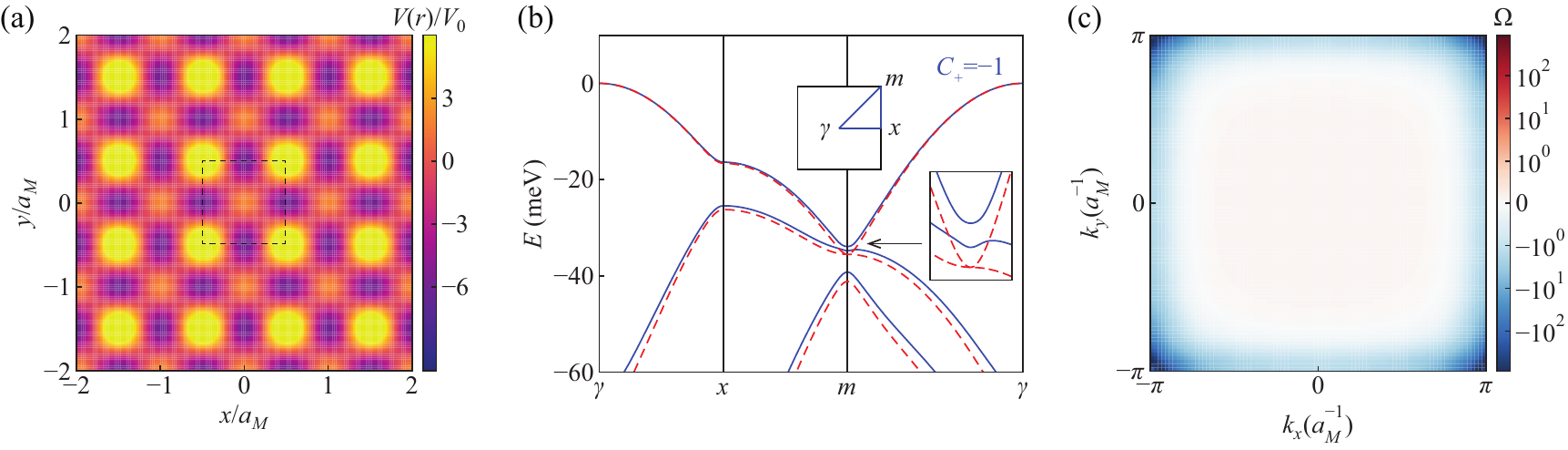}
\caption{(1) Moir\'e potential for holes as described by Eq. (\ref{square}). The potential minima form a square lattice with two sublattices. The black dashed hexagon encloses the MSL unit cell. (b) The blue solid and red dashed energy bands are from the continuum models Eqs. (1) and (2), respectively. The top blue solid band has a valley Chern number $C_+=-1$. The insets show the schematic MBZ and zoom in around the quadratic band-touching point. (c) The Berry curvature of the top blue solid band in (b).  }
\label{figs8}
\end{figure*}

\section{VI. Potential realizations of massive Dirac fermions confined in a moir\'e potential }

\begin{figure}[b]
\includegraphics[width=8.5 cm]{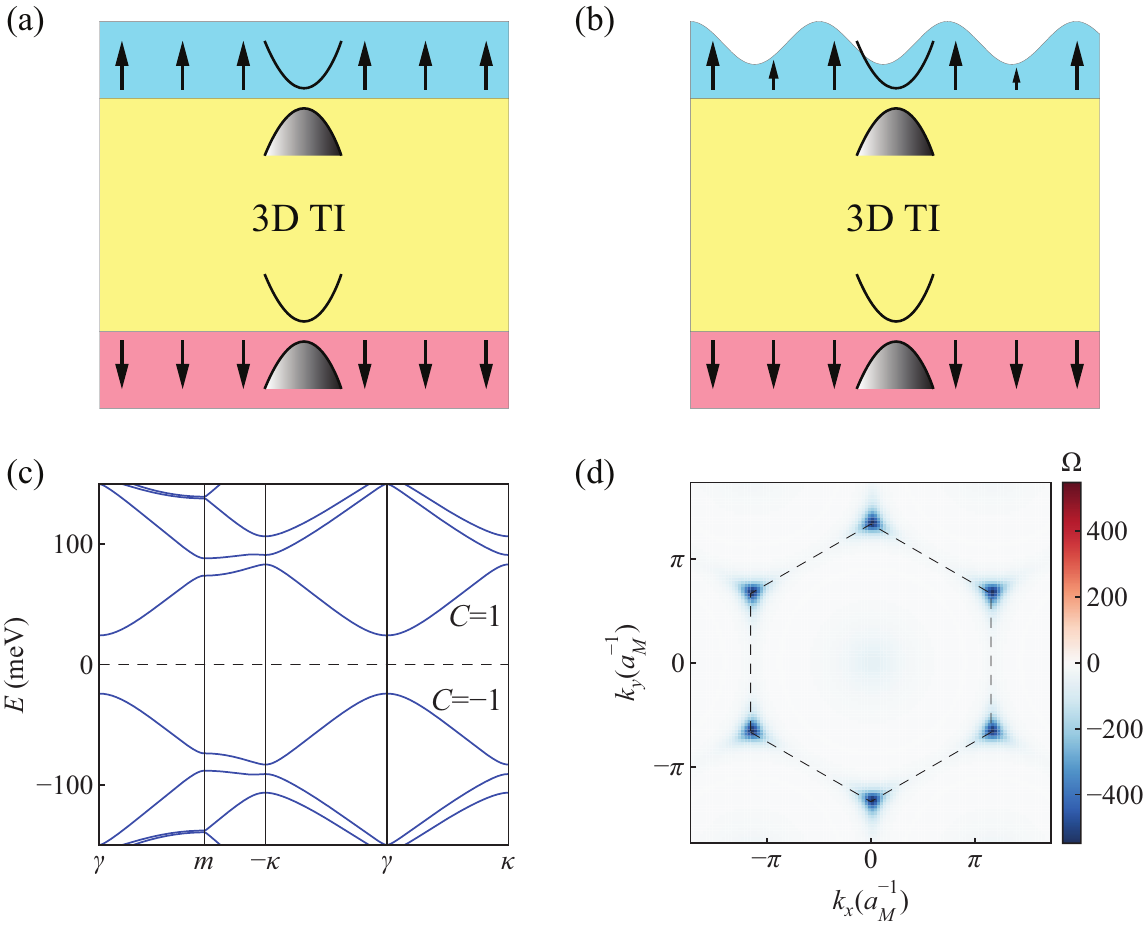}
\caption{(a) Axion insulator in a FM-3DTI-FM heterostructure. (b) Moir\'e modulation of massive Dirac surface states by a patterned FM layer. (c) Moir\'e minibands from Eq. (\ref{axion}). The Chern numbers of bottom conduction and top valence bands are $C=\pm 1$. (d) The Berry curvature of the top valence band in (c).  }
\label{figs6}
\end{figure}

{\it Axion insulator surface states under spatially periodic magnetic proximity coupling.} In 3D topological insulator (3DTI), the massless Dirac surface states are protected by the TRS. When the TRS is broken by the surface magnetization pointing in opposite directions at two opposite surfaces, the topological surface states are gapped that gives rise to the axion insulator with massive Dirac surface states \cite{Qi2008Topologicals}. The axion insulator can be realized by sandwiching a 3DTI with two antiparallel ferromagnetic (FM) layers \cite{mogi2017magnetics,Xiao2018Realizations}, as shown in Fig. \ref{figs6}(a). If the magnetization of the two FM layers is parallel, it gives rise to a Chern insulator. To couple the massive Dirac fermion with a moir\'e potential on the surface of an axion insulator, we consider a patterned FM layer with periodic modulation, as shown in Fig. \ref{figs6}(b). The patterned FM layer can be realized by spatially periodic photolithography or doping. The massive Dirac surface states under spatially periodic magnetic proximity coupling can be described by
\begin{equation}\label{axion}
    H = v_F (\bm{s}\times \bm{k})\cdot \hat{z} + \bar{M_z} s_z + M_z(\bm{r}) s_z, 
\end{equation}
where the first term describes the surface Dirac states of a 3DTI and the last two terms are the Zeeman interaction induced by the patterned FM layer on the top surface. Here $v_F=2.67$ eV$\cdot$\r{A} is the Fermi velocity, $\bm{s}$ is the vector of Pauli matrices acting in spin space, $\bar{M}_z=-25$ meV is the mean Zeeman field, and the spatial variation of Zeeman field is encoded in   
\begin{equation}
    M_z(\bm{r})=2V_0\sum_{j=1}^3 \cos(\bm{G}_j\cdot\bm{r}+\phi),
\end{equation}
in the same form as the moir\'e potential studied in  the main text. To be concrete, we consider $V_0=8$ meV, $\phi=(2n+1)\pi/3$, and the MSL constant $a_M=30a$ where $a$ is the in-plane lattice constant of the 3DTI. Note that Eq. (\ref{axion}) is different from Eq. (1) of the main text because the moir\'e potential now couples to the Pauli matrix $s_z$. Then the Hamiltonian has an extra chiral symmetry as $\mathcal{C}H\mathcal{C}^{-1} = -H$ where the chiral operator $\mathcal{C}=is_y \mathcal{K}$, that guarantees the symmetric energy bands with respect to zero energy, as shown in Fig. (\ref{figs6}). Because of the chiral symmetry, the conduction and valence bands appear in pairs with opposite Chern numbers. In Fig. \ref{figs6}(c), the bottom conduction band and top valence band have opposite Chern numbers $C=\pm1$. The Berry curvature of the top valence band in Fig. \ref{figs6}(c) is shown in Fig. \ref{figs6}(d). 

{\it Monolayer TMD under spatially periodic dielectric screening.} Recently, it has been shown that spatially periodic dielectric screening can be used to create moir\'e minibands in a monolayer TMD \cite{xu2021creations}. Because the electric field lines between charges in a monolayer TMD extend outside the 2D plane, the dielectric screening of Coulomb interaction is reduced and depends substantially on the environmental dielectric constant $\varepsilon$. By placing a monolayer WSe$_2$ on a graphene/hBN moir\'e superlattice, the spatially periodic dielectric screening induced by the substrate generates a moir\'e modulation in WSe$_2$, while the direct interlayer hybridization is suppressed by a insulating spacer \cite{xu2021creations}. The spatially periodic dielectric function $\varepsilon$ results in a spatially periodic variation of energy gap  \cite{xu2021creations}, as shown in Fig. \ref{figs7}(a),  that can be described by
\begin{equation}\label{TMD}
    H_\tau = h_{\bm{k},\tau} + V(\bm{r})\sigma_z, 
\end{equation}
where $h_{\bm{k},\tau}$ is the Dirac model in Eq. (1) and describes the massive Dirac fermions in monolayer TMD, and the last term encodes the periodic variation of energy gap. The moir\'e potential $V(\bm{r})$ takes the same form as that studied in the main text but couples to the Pauli matrix $\sigma_z$ acting in the orbital pseudospin space. Therefore, Eq. (\ref{TMD}) also has the chiral symmetry. Here we adopt the model parameters for WSe$_2$ as $v_F=3.94$ eV$\cdot$\r{A} and $\Delta=1.60$ eV.  The MSL constant $a_M=10.31$ nm for a $1^\circ$-misaligned grahene/hBN substrate. For $V_0=-8$ meV and $\phi=(2n+1)\pi/3$, the moir\'e minibands from the $+$K valley (with $\tau=+1$) are shown in Fig. \ref{figs7}(b), while those from the $-$K valley can be obtained by TRS. Here we show only the valence bands and the top one is shifted to zero energy.
The top valence band has a nonzero valley Chern number $C_+=1$ and its Berry curvature is shown in Fig. \ref{figs7}(c).

\begin{figure}
\includegraphics[width=8.5 cm]{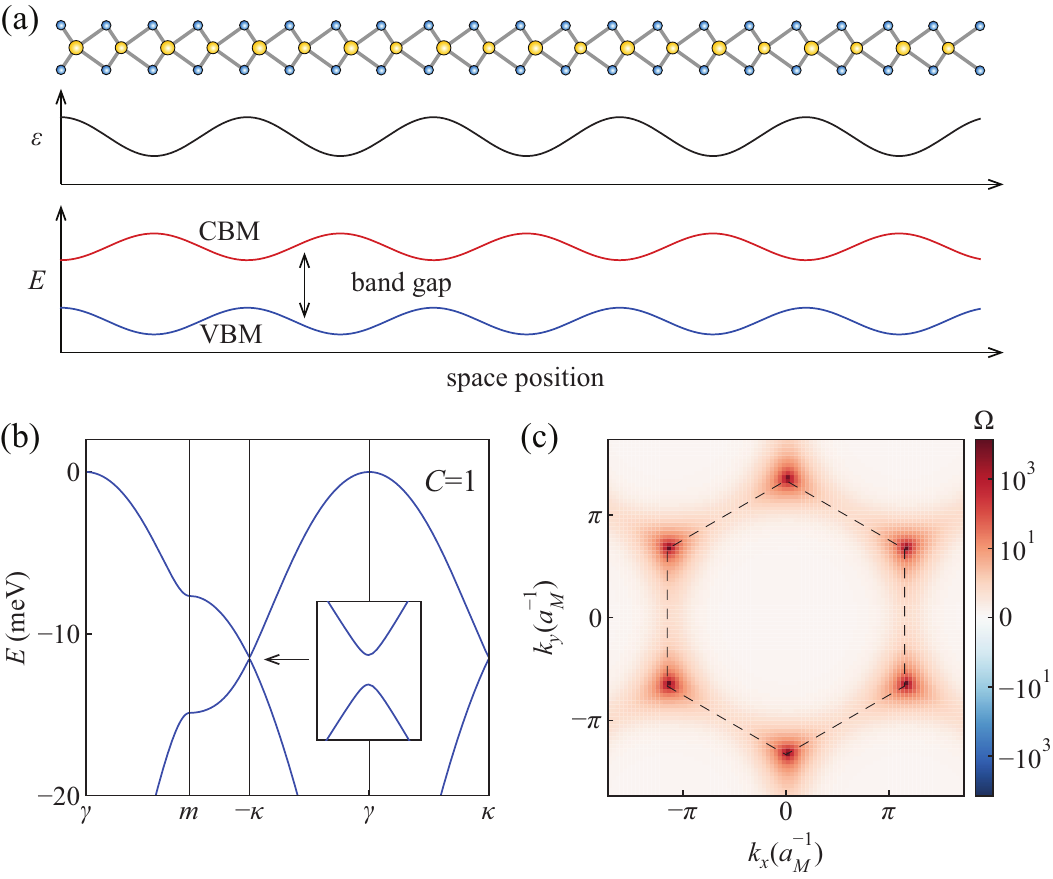}
\caption{(a) Schematic diagram for a monolayer TMD (upper panel) under a spatially periodic dielectric screening (middle panel) that results in a periodic variation of energy band gap (lower panel). Here the blue and red curves denote the valence band minimum (VBM) and conduction band maximum (CBM), respectively. (b) Moir\'e minibands from Eq. (\ref{TMD}). The valley Chern number of the top valence band is $C_+=1$. (C) The Berry curvature of the top valence band in (b).}
\label{figs7}
\end{figure}

\end{document}